\documentclass[12pt]{article}
\usepackage{amsmath,amssymb}
\usepackage{graphicx}
\newcommand{\eqdef}{\stackrel{\text{def}}{=}}
\newcommand{\n}{\nonumber\\}
\newcommand{\bm}{\boldsymbol}

\newcommand{\laprod}[2]{\prod_{#1}^{\stackrel{#2}{\longleftarrow}}}

\newcommand{\ignore}[1]{}
\numberwithin{equation}{section}
\newcommand{\Romannumeral}[1]{\uppercase\expandafter{\romannumeral#1}}
\newcommand{\I}{\text{\Romannumeral{1}}}
\newcommand{\II}{\text{\Romannumeral{2}}}
\newcommand{\III}{\text{\Romannumeral{3}}}
\newcommand{\IV}{\text{\Romannumeral{4}}}

\allowdisplaybreaks[4]

\begin{document}

\baselineskip=20pt

\newfont{\elevenmib}{cmmib10 scaled\magstep1}
\newcommand{\preprint}{
    \begin{flushright}\normalsize \sf
     DPSU-17-1\\
   \end{flushright}}
\newcommand{\Title}[1]{{\baselineskip=26pt
   \begin{center} \Large \bf #1 \\ \ \\ \end{center}}}
\newcommand{\Author}{\begin{center}
   \large \bf Satoru Odake \end{center}}
\newcommand{\Address}{\begin{center}
     Faculty of Science, Shinshu University,\\
     Matsumoto 390-8621, Japan
   \end{center}}
\newcommand{\Accepted}[1]{\begin{center}
   {\large \sf #1}\\ \vspace{1mm}{\small \sf Accepted for Publication}
   \end{center}}

\preprint
\thispagestyle{empty}

\Title{New Determinant Expressions of\\
the Multi-indexed Orthogonal Polynomials in\\
Discrete Quantum Mechanics}

\Author

\Address
\vspace{1cm}

\begin{abstract}
The multi-indexed orthogonal polynomials (the Meixner, little $q$-Jacobi
(Laguerre), ($q$-)Racah, Wilson, Askey-Wilson types) satisfying second
order difference equations were constructed in discrete quantum mechanics.
They are polynomials in the sinusoidal coordinates $\eta(x)$ ($x$ is the
coordinate of quantum system) and expressed in terms of the Casorati
determinants whose matrix elements are functions of $x$ at various points.
By using shape invariance properties, we derive various equivalent determinant
expressions, especially those whose matrix elements are functions of the
same point $x$. Except for the ($q$-)Racah case, they can be expressed in terms
of $\eta$ only, without explicit $x$-dependence.
\end{abstract}

\section{Introduction}
\label{intro}

The exceptional and multi-indexed orthogonal polynomials
$\{P_{\mathcal{D},n}(\eta)|n\in\mathbb{Z}_{\geq 0}\}$
\cite{gkm08_1}--\cite{os35}
are new type of orthogonal polynomials which
form a complete set of orthogonal basis in an appropriate Hilbert space
in spite of missing degrees.
Instead of the three term recurrence relations which characterize the
ordinary orthogonal polynomials
\cite{szego_1,szego_2,szego_3}, they satisfy the recurrence
relations with more terms \cite{stz10}--\cite{gkkm15}, and the constraints
by Bochner's theorem and its generalizations
\cite{bochner_1,bochner_2,szego_1,szego_2,szego_3} are avoided.
They satisfy second order differential or difference equations and
quantum mechanical formulation has played an important role in constructing
these new orthogonal polynomials.
The quantum mechanical systems described by the classical orthogonal
polynomials can be iso-spectrally (or almost iso-spectrally) deformed by
applying the multi-step Darboux transformations
\cite{darb_1}--\cite{os22}
and the main part of the eigenfunctions of the deformed systems are
multi-indexed orthogonal polynomials.
We distinguish the following two cases;
the set of missing degrees $\mathcal{I}=\mathbb{Z}_{\geq 0}\backslash
\{\text{deg}\,P_{\mathcal{D},n}(\eta)|n\in\mathbb{Z}_{\geq 0}\}$ is
case (1): $\mathcal{I}=\{0,1,\ldots,\ell-1\}$, or
case (2): $\mathcal{I}\neq\{0,1,\ldots,\ell-1\}$, where $\ell$ is a positive
integer. The situation of case (1) is called stable in \cite{gkm11}.
Case (1) is obtained by taking virtual states as the seed solutions of
the Darboux transformation \cite{os25,os27,os26,os35} and case (2) is
obtained by employing eigenstates or pseudo virtual states
\cite{q12b}--\cite{os30}.
In the quantum mechanical formulation, the multi-indexed orthogonal
polynomials appear as polynomials in the sinusoidal coordinate $\eta(x)$
\cite{os7_1,os7_2},
$P_{\mathcal{D},n}\bigl(\eta(x)\bigr)$, where $x$ is the
coordinate of the quantum system.

The eigenstates of the deformed system are expressed in terms of
determinants: the Wronskians for the ordinary quantum mechanics (in which
the Schr\"odinger equation is second order differential equation) and the
Casoratians for the discrete quantum mechanics (in which the Schr\"odinger
equations are second order difference equations).
In our previous paper \cite{os37}, we have presented simplified determinant
expressions of the case (1) multi-indexed Laguerre and Jacobi polynomials
in the ordinary quantum mechanics. We rewrite the Wronskians
by using the identities of the original Laguerre and Jacobi polynomials and
two simplified expressions A and B are presented. The identities used in
case A are essentially forward shift relations. For case B, the Schr\"odinger
equation is used and it is regarded as a combination of the forward and
backward shift relations, which are the consequences of the shape invariance.
Thus both A and B expressions are derived by using the properties of the
Wronskians and the shape invariance (see \S\,\ref{sec:summary} for details).
This idea is also applicable to the discrete quantum mechanics.
In this paper we consider the case (1) multi-indexed orthogonal polynomials
appearing in the discrete quantum mechanics:
the Meixner, little $q$-Laguerre, little $q$-Jacobi, Racah, $q$-Racah, Wilson
and Askey-Wilson types.
The first five appear in the discrete quantum mechanics with real shifts
(rdQM) and the last two belong to the discrete quantum mechanics with pure
imaginary shifts (idQM).
We rewrite the Casoratians by using their properties and the identities of
the original polynomials, which are the consequences of the shape invariance.
Various new determinant expressions of the multi-indexed orthogonal
polynomials are obtained.
Corresponding to the cases A and B in \cite{os37}, we present two typical
expressions explicitly.
The matrix elements of the Casoratians are functions of $x$ at various
points, $x+j-1$ for rdQM and $x+i(\frac{n+1}{2}-j)\gamma$ for idQM.
In some of new determinant expressions, the matrix elements are functions
of the same point $x$. Therefore the multi-indexed orthogonal polynomials
are expressed in terms of $\eta$ only, namely without explicit $x$-dependence.
For the Racah and $q$-Racah cases, the corresponding matrix elements for
the multi-indexed polynomials do have explicit $x$-dependence because of
the parameter ($\bm{\lambda}$) dependence of the sinusoidal coordinates,
$\eta(x;\bm{\lambda})$.
These new equivalent expressions show explicitly the constituents of the
multi-indexed orthogonal polynomials and they are helpful for deeper
understanding.

This paper is organized as follows.
We derive the new equivalent expressions for the multi-indexed Meixner,
little $q$-Laguerre, little $q$-Jacobi, Racah and $q$-Racah polynomials in
section \ref{sec:rdQM}. Those for the multi-indexed Wilson and Askey-Wilson
polynomials are presented in section \ref{sec:idQM}.
The discussions in these two sections go parallelly.
In \S\,\ref{sec:rdQM_org} (\S\,\ref{sec:idQM_org}) the discrete quantum
mechanics with real (pure imaginary) shifts is recapitulated and the data
for the original systems are presented.
The virtual state vectors (wavefunctions) and related functions are
introduced in \S\,\ref{sec:rdQM_vs} (\S\,\ref{sec:idQM_vs}).
The original systems have shape invariance which entails the well-known
forward and backward shift relations of the eigenpolynomials $P_n$.
The forward and backward shift relations for virtual state polynomials
$\xi_{\text{v}}$ are presented in \S\,\ref{sec:rdQM_si}
(\S\,\ref{sec:idQM_si}).
In \S\,\ref{sec:rdQM_miop} (\S\,\ref{sec:idQM_miop}) the definitions of the
multi-indexed orthogonal polynomials given in \cite{os26,os35} (\cite{os27})
are recapitulated.
They are expressed in terms of Casoratians.
Sections \ref{sec:rdQM_exp} and \ref{sec:idQM_exp} are the main contents
of this paper.
The Casoratians are rewritten by using their properties and the identities
stemming from the shape invariance, and new determinant expressions of the
multi-indexed orthogonal polynomials are obtained.
We present two typical examples.
Section \ref{sec:summary} is for a summary and comments.

\section{Multi-indexed Orthogonal Polynomials in rdQM}
\label{sec:rdQM}

In this section we derive various equivalent determinant expressions for
the multi-indexed orthogonal polynomials in the framework of the discrete
quantum mechanics with real shifts.
They are the multi-indexed Meixner, little $q$-Laguerre, little $q$-Jacobi,
Racah and $q$-Racah polynomials.

\subsection{Original systems}
\label{sec:rdQM_org}

The Hamiltonian of the discrete quantum mechanics with real shifts (rdQM)
\cite{os12,os34} $\mathcal{H}=(\mathcal{H}_{x,y})$ is a tri-diagonal real
symmetric (Jacobi) matrix and its rows and columns are indexed by integers
$x$ and $y$, which take values in $\{0,1,\ldots,N\}$ (finite) or
$\mathbb{Z}_{\geq 0}$ (semi-infinite),
\begin{equation}
  \mathcal{H}_{x,y}\eqdef
  -\sqrt{B(x)D(x+1)}\,\delta_{x+1,y}-\sqrt{B(x-1)D(x)}\,\delta_{x-1,y}
  +\bigl(B(x)+D(x)\bigr)\delta_{x,y},
  \label{rdQM:Hxy}
\end{equation}
where the potential functions $B(x)$ and $D(x)$ are real and positive
but vanish at the boundary, $D(0)=0$ and $B(N)=0$ for finite cases.
This Hamiltonian can be expressed in a factorized form:
\begin{align}
  &\mathcal{H}=\mathcal{A}^{\dagger}\mathcal{A},\qquad
  \mathcal{A}=(\mathcal{A}_{x,y}),
  \ \ \mathcal{A}^{\dagger}=((\mathcal{A}^{\dagger})_{x,y})
  =(\mathcal{A}_{y,x}),
  \label{factor}\\
  &\mathcal{A}_{x,y}\eqdef
  \sqrt{B(x)}\,\delta_{x,y}-\sqrt{D(x+1)}\,\delta_{x+1,y},\quad
  (\mathcal{A}^{\dagger})_{x,y}=
  \sqrt{B(x)}\,\delta_{x,y}-\sqrt{D(x)}\,\delta_{x-1,y}.
\end{align}
For simplicity in notation, we write $\mathcal{H}$, $\mathcal{A}$ and
$\mathcal{A}^{\dagger}$ as follows:
\begin{align}
  &\mathcal{H}=-\sqrt{B(x)D(x+1)}\,e^{\partial}
  -\sqrt{B(x-1)D(x)}\,e^{-\partial}+B(x)+D(x)\n
  &\phantom{\mathcal{H}}=-\sqrt{B(x)}\,e^{\partial}\sqrt{D(x)}
  -\sqrt{D(x)}\,e^{-\partial}\sqrt{B(x)}+B(x)+D(x),
  \label{rdQM:H}\\
  &\mathcal{A}=\sqrt{B(x)}-e^{\partial}\sqrt{D(x)},\quad
  \mathcal{A}^{\dagger}=\sqrt{B(x)}-\sqrt{D(x)}\,e^{-\partial},
  \label{rdQM:A,Ad}
\end{align}
where matrices $e^{\pm\partial}$ are
\begin{equation}
  e^{\pm\partial}=((e^{\pm\partial})_{x,y}),\quad
  (e^{\pm\partial})_{x,y}\eqdef\delta_{x\pm 1,y},\quad
  (e^{\partial})^{\dagger}=e^{-\partial},
  \label{partdef}
\end{equation}
and we suppress the unit matrix $\bm{1}=(\delta_{x,y})$:
$\bigl(B(x)+D(x)\bigr)\bm{1}$ in \eqref{rdQM:H},
$\sqrt{B(x)}\,\bm{1}$ in \eqref{rdQM:A,Ad}.
(The notation $f(x)Ag(x)$, where $f(x)$ and $g(x)$ are functions of $x$ and
$A$ is a matrix $A=(A_{x,y})$, stands for a matrix whose $(x,y)$-element
is $f(x)A_{x,y}g(y)$.)
Note that the matrices $e^{\partial}$ and $e^{-\partial}$ are not inverse
to each other: $e^{\pm\partial}e^{\mp\partial}\neq\bm{1}$ for finite systems
and $e^{-\partial}e^{\partial}\neq\bm{1}$ for semi-infinite systems.
The Schr\"odinger equation is the eigenvalue problem for the hermitian
matrix $\mathcal{H}$,
\begin{equation}
  \mathcal{H}\phi_n(x)=\mathcal{E}_n\phi_n(x)\quad
  (n=0,1,2,\ldots),\quad
  0=\mathcal{E}_0<\mathcal{E}_1<\mathcal{E}_2<\cdots,
  \label{Scheq}
\end{equation}
($n=0,1,\ldots,N$ for a finite case).

We consider rdQM described by the Meixner (M), little $q$-Laguerre (l$q$L),
little $q$-Jacobi (l$q$J), Racah (R) and $q$-Racah ($q$R) polynomials.
The first three are semi-infinite systems and the last two are finite systems.
Eigenvectors have the following factorized form
\begin{equation}
  \phi_n(x)=\phi_0(x)\check{P}_n(x),\quad
  \check{P}_n(x)\eqdef P_n\bigl(\eta(x)\bigr).
\end{equation}
Here the ground state eigenvector $\phi_0(x)$, which is characterized by
$\mathcal{A}\phi_0(x)=0$, is chosen as
\begin{equation}
  \phi_0(x)=\prod_{y=0}^{x-1}\sqrt{\frac{B(y)}{D(y+1)}}\,>0.
  \label{rdQM:phi0}
\end{equation}
We use the convention: $\prod\limits_{k=n}^{n-1}*=1$, which means the
normalization $\phi_0(0)=1$.
The other function $P_n(\eta)$ is a degree $n$ polynomial in $\eta$, and
$\eta(x)$ is the sinusoidal coordinate satisfying the boundary
condition $\eta(0)=0$.
We adopt the universal normalization condition \cite{os12,os34} as
\begin{equation}
  P_n(0)=1\ \bigl(\Leftrightarrow \check{P}_n(0)=1\bigr).
  \label{Pzero}
\end{equation}
Various quantities depend on a set of parameters
$\bm{\lambda}=(\lambda_1,\lambda_2,\ldots)$ and their dependence is
expressed like,
$\mathcal{H}=\mathcal{H}(\bm{\lambda})$,
$\mathcal{A}=\mathcal{A}(\bm{\lambda})$,
$\mathcal{E}_n=\mathcal{E}_n(\bm{\lambda})$,
$B(x)=B(x;\bm{\lambda})$,
$\phi_n(x)=\phi_n(x;\bm{\lambda})$,
$\check{P}_n(x)=\check{P}_n(x;\bm{\lambda})
=P_n\bigl(\eta(x;\bm{\lambda});\bm{\lambda}\bigr)$, etc.
The parameter $q$ is $0<q<1$ and $q^{\bm{\lambda}}$ stands for
$q^{(\lambda_1,\lambda_2,\ldots)}=(q^{\lambda_1},q^{\lambda_2},\ldots)$.
The symbols $(a)_n$ and $(a;q)_n$ are ($q$-)shifted factorials \cite{kls}
and $[x]$ denotes the greatest integer not exceeding $x$.

Parameters of the systems are
\begin{align}
  \text{M}:\ \ &\bm{\lambda}=(\beta,c),\quad
  \bm{\delta}=(1,0),\quad \kappa=1,\quad \beta>0,\quad 0<c<1,\\
  \text{l$q$L}:\ \ &q^{\bm{\lambda}}=a,\quad
  \bm{\delta}=1,\quad \kappa=q^{-1},\quad 0<a<q^{-1},\\
  \text{l$q$J}:\ \ &q^{\bm{\lambda}}=(a,b),\quad
  \bm{\delta}=(1,1),\quad \kappa=q^{-1},\quad 0<a<q^{-1},\quad b<q^{-1},\\
  \text{R}:\ \ &\bm{\lambda\,}=(a,b,c,d),\quad \bm{\delta}=(1,1,1,1),
  \quad\kappa=1,\\
  \text{$q$R}:\ \ &q^{\bm{\lambda}}=(a,b,c,d),\quad \bm{\delta}=(1,1,1,1),
  \quad\kappa=q^{-1},
\end{align}
and we adopt the following choice of the parameter ranges for R and $q$R:
\begin{align}
  \text{R}:\ \ &a=-N,\quad 0<d<a+b,\quad 0<c<1+d,\\
  \text{$q$R}:\ \ &a=q^{-N},\quad 0<ab<d<1,\quad qd<c<1.
\end{align}
We list the fundamental data \cite{os12}.\\
$\circ$ {\bf Meixner:}\\
We rescale the overall normalization of the Hamiltonian in \cite{os12}:
$(\mathcal{H}\text{ in \cite{os12}})\times(1-c)\to\mathcal{H}$.
\begin{align}
  &B(x;\bm{\lambda})=c(x+\beta),\quad
  D(x)=x,
  \label{MBD}\\
  &\mathcal{E}_n(\bm{\lambda})=(1-c)n,\quad\eta(x)=x,\quad
  \varphi(x)=1,
  \label{Meta}\\
  &\check{P}_n(x;\bm{\lambda})
  ={}_2F_1\Bigl(
  \genfrac{}{}{0pt}{}{-n,\,-x}{\beta}\Bigm|1-c^{-1}\Bigr)
  =M_n(x;\beta,c),
  \label{MP}\\
  &\phi_0(x;\bm{\lambda})=\sqrt{\frac{(\beta)_x\,c^x}{x!}},
  \label{Mphi0}
\end{align}
where $M_n(x;\beta,c)$ is the Meixner polynomial \cite{kls}.\\
$\circ$ {\bf little $q$-Laguerre and little $q$-Jacobi:}
\begin{align}
  &B(x;\bm{\lambda})=\left\{
  \begin{array}{ll}
  aq^{-x}&:\text{l$q$L}\\
  a(q^{-x}-bq)&:\text{l$q$J}\\
  \end{array}\right.,\quad
  D(x)=q^{-x}-1,\\[2pt]
  &\mathcal{E}_n(\bm{\lambda})=\left\{
  \begin{array}{ll}
  q^{-n}-1&:\text{l$q$L}\\
  (q^{-n}-1)(1-abq^{n+1})&:\text{l$q$J}
  \end{array}\right.,\quad
  \eta(x)=1-q^x,\quad\varphi(x)=q^x,
  \label{lqJeta}\\[2pt]
  &\check{P}_n(x;\bm{\lambda})=\left\{
  \begin{array}{ll}
  {\displaystyle
  {}_2\phi_0\Bigl(
  \genfrac{}{}{0pt}{}{q^{-n},\,q^{-x}}{-}\Bigm|q;\frac{q^x}{a}\Bigr)
  =(a^{-1}q^{-n}\,;q)_n\,{}_2\phi_1\Bigl(
  \genfrac{}{}{0pt}{}{q^{-n},\,0}{aq}\Bigm|q;q^{x+1}\Bigr)}&\\[4pt]
  {\displaystyle
  {}_3\phi_1\Bigl(
  \genfrac{}{}{0pt}{}{q^{-n},\,abq^{n+1},\,q^{-x}}{bq}\Bigm|
  q\,;\frac{q^x}{a}\Bigr)
  =\frac{(a^{-1}q^{-n};q)_n}{(bq;q)_n}\,
  {}_2\phi_1\Bigl(
  \genfrac{}{}{0pt}{}{q^{-n},\,abq^{n+1}}{aq}\Bigm|q\,;q^{x+1}\Bigr)}&
  \end{array}\right.\n
  &\phantom{\check{P}_n(x;\bm{\lambda})}
  =\left\{
  \begin{array}{ll}
  (a^{-1}q^{-n}\,;q)_n\,p_n\bigl(1-\eta(x);a|q\bigr)&:\text{l$q$L}\\[4pt]
  (bq;q)_n^{-1}(a^{-1}q^{-n};q)_n\,
  p_n\bigl(1-\eta(x);a,b|q\bigr)&:\text{l$q$J}
  \end{array}\right.,\\[2pt]
  &\phi_0(x;\bm{\lambda})=\left\{
  \begin{array}{ll}
  \sqrt{(q;q)_x^{-1}(aq)^x}&:\text{l$q$L}\\[4pt]
  \sqrt{(bq;q)_x(q;q)_x^{-1}(aq)^x}&:\text{l$q$J}
  \end{array}\right.,
\end{align}
where $p_n(y;a|q)$ and $p_n(y;a,b|q)$ are the little $q$-Laguerre and little
$q$-Jacobi polynomials in the conventional definition \cite{kls},
respectively.\\
$\circ$ {\bf Racah and $q$-Racah:}
\begin{align}
  &B(x;\bm{\lambda})=
  \left\{
  \begin{array}{ll}
  {\displaystyle
  -\frac{(x+a)(x+b)(x+c)(x+d)}{(2x+d)(2x+1+d)}}&:\text{R}\\[8pt]
  {\displaystyle-\frac{(1-aq^x)(1-bq^x)(1-cq^x)(1-dq^x)}
  {(1-dq^{2x})(1-dq^{2x+1})}}&:\text{$q$R}
  \end{array}\right.,
  \label{Bform}\\
  &D(x;\bm{\lambda})=
  \left\{
  \begin{array}{ll}
  {\displaystyle
  -\frac{(x+d-a)(x+d-b)(x+d-c)x}{(2x-1+d)(2x+d)}}&:\text{R}\\[8pt]
  {\displaystyle-\tilde{d}\,
  \frac{(1-a^{-1}dq^x)(1-b^{-1}dq^x)(1-c^{-1}dq^x)(1-q^x)}
  {(1-dq^{2x-1})(1-dq^{2x})}}&:\text{$q$R}
  \end{array}\right.,
  \label{Dform}\\
  &\mathcal{E}_n(\bm{\lambda})=
  \left\{
  \begin{array}{ll}
  n(n+\tilde{d})&:\text{R}\\[2pt]
  (q^{-n}-1)(1-\tilde{d}q^n)&:\text{$q$R}
  \end{array}\right.\!,\quad
  \eta(x;\bm{\lambda})=
  \left\{
  \begin{array}{ll}
  x(x+d)&:\text{R}\\[2pt]
  (q^{-x}-1)(1-dq^x)&:\text{$q$R}
  \end{array}\right.,
  \label{etadefs}\\
  &\varphi(x;\bm{\lambda})=
  \left\{
  \begin{array}{ll}
  {\displaystyle\frac{2x+d+1}{d+1}}&:\text{R}\\[6pt]
  {\displaystyle\frac{q^{-x}-dq^{x+1}}{1-dq}}&:\text{$q$R}
  \end{array}\right.,\quad
  \tilde{d}\eqdef
  \left\{
  \begin{array}{ll}
  a+b+c-d-1&:\text{R}\\[2pt]
  abcd^{-1}q^{-1}&:\text{$q$R}
  \end{array}\right.,
  \label{vphidef}\\
  &\check{P}_n(x;\bm{\lambda})
  =P_n\bigl(\eta(x;\bm{\lambda});\bm{\lambda}\bigr)
  =\left\{
  \begin{array}{ll}
  {\displaystyle
  {}_4F_3\Bigl(
  \genfrac{}{}{0pt}{}{-n,\,n+\tilde{d},\,-x,\,x+d}
  {a,\,b,\,c}\Bigm|1\Bigr)}\\[8pt]
  {\displaystyle
  {}_4\phi_3\Bigl(
  \genfrac{}{}{0pt}{}{q^{-n},\,\tilde{d}q^n,\,q^{-x},\,dq^x}
  {a,\,b,\,c}\Bigm|q\,;q\Bigr)}
  \end{array}\right.\n
  &\phantom{\check{P}_n(x;\bm{\lambda})
  =P_n\bigl(\eta(x;\bm{\lambda});\bm{\lambda}\bigr)}
  =\left\{
  \begin{array}{ll}
  {\displaystyle
  R_n\bigl(\eta(x;\bm{\lambda});a-1,\tilde{d}-a,c-1,d-c\bigr)}
  &:\text{R}\\[2pt]
  {\displaystyle
  R_n\bigl(1+d+\eta(x;\bm{\lambda});
  aq^{-1},\tilde{d}a^{-1},cq^{-1},dc^{-1}|q\bigr)}&:\text{$q$R}
  \end{array}\right.,
  \label{Pn=R,qR}\\[4pt]
  &\phi_0(x;\bm{\lambda})=\left\{
  \begin{array}{ll}
  {\displaystyle
  \sqrt{\frac{(a,b,c,d)_x}{(d-a+1,d-b+1,d-c+1,1)_x}\,
  \frac{2x+d}{d}}}&:\text{R}\\[12pt]
  {\displaystyle
  \sqrt{\frac{(a,b,c,d\,;q)_x}
  {(a^{-1}dq,b^{-1}dq,c^{-1}dq,q\,;q)_x\,\tilde{d}^x}\,
  \frac{1-dq^{2x}}{1-d}}}&:\text{$q$R}
  \end{array}\right.,
\end{align}
where $R_n\bigl(x(x+\gamma+\delta+1);\alpha,\beta,\gamma,\delta\bigr)$ and
$R_n(q^{-x}+\gamma\delta q^{x+1};\alpha,\beta,\gamma,\delta|q)$ are the Racah
and $q$-Racah polynomials \cite{kls}, respectively.
It should be emphasized that for these two polynomials, the sinusoidal
coordinates $\eta(x)$ \eqref{etadefs} and the auxiliary function
$\varphi(x)$ \eqref{vphidef} depend on the parameters $\eta(x;\bm{\lambda})$,
$\varphi(x;\bm{\lambda})$ in distinction with the other polynomials in this
section.

Note that the following relations are valid for all the systems discussed
in this section,
\begin{align}
  &\varphi(x;\bm{\lambda})=
  \frac{\eta(x+1;\bm{\lambda})-\eta(x;\bm{\lambda})}{\eta(1;\bm{\lambda})},\\
  &\varphi(x;\bm{\lambda})=\sqrt{\frac{B(0;\bm{\lambda})}{B(x;\bm{\lambda})}}
  \,\frac{\phi_0(x;\bm{\lambda}+\bm{\delta})}{\phi_0(x;\bm{\lambda})},\quad
  \varphi(x;\bm{\lambda})=\sqrt{\frac{B(0;\bm{\lambda})}{D(x+1;\bm{\lambda})}}
  \,\frac{\phi_0(x;\bm{\lambda}+\bm{\delta})}{\phi_0(x+1;\bm{\lambda})},
  \label{OS12(4.12,13)}\\
  &\frac{B(x;\bm{\lambda}+\bm{\delta})}{B(x+1;\bm{\lambda})}
  =\kappa^{-1}\frac{\varphi(x+1;\bm{\lambda})}{\varphi(x;\bm{\lambda})},\quad
  \frac{D(x;\bm{\lambda}+\bm{\delta})}{D(x;\bm{\lambda})}
  =\kappa^{-1}\frac{\varphi(x-1;\bm{\lambda})}{\varphi(x;\bm{\lambda})}.
  \label{OS12(5.81etc)}
\end{align}

\subsection{Virtual states}
\label{sec:rdQM_vs}

The virtual state vectors are obtained by using the discrete symmetries
of the Hamiltonian \cite{os26,os35}.
We define the twist operation $\mathfrak{t}$,
\begin{align}
  &\mathfrak{t}(\bm{\lambda})\eqdef\left\{
  \begin{array}{ll}
  (\lambda_1,\lambda_2^{-1})&:\text{M}\\
  -\lambda_1&:\text{l$q$L}\\
  (-\lambda_1,\lambda_2)&:\text{l$q$J}\\
  (\lambda_4-\lambda_1+1,\lambda_4-\lambda_2+1,\lambda_3,\lambda_4)
  &:\text{R,\,$q$R}
  \end{array}\right.,\n[2pt]
  &\text{namely}\ \ \left\{
  \begin{array}{ll}
  \mathfrak{t}(\bm{\lambda})=(\beta,c^{-1})&:\text{M}\\
  q^{\mathfrak{t}(\bm{\lambda})}=a^{-1}&:\text{l$q$L}\\
  q^{\mathfrak{t}(\bm{\lambda})}=(a^{-1},b)&:\text{l$q$J}\\
  \mathfrak{t}(\bm{\lambda})=(d-a+1,d-b+1,c,d)&:\text{R}\\
  q^{\mathfrak{t}(\bm{\lambda})}=(a^{-1}dq,b^{-1}dq,c,d)&:\text{$q$R}
  \end{array}\right.,
  \label{rdQM:twist}
\end{align}
which is an involution $\mathfrak{t}^2=\text{id}$ and satisfies
\begin{equation}
  \mathfrak{t}(\bm{\lambda})+u\bm{\delta}
  =\mathfrak{t}(\bm{\lambda}+u\tilde{\bm{\delta}})
  \ \ (\forall u\in\mathbb{R}),\quad
  \tilde{\bm{\delta}}\eqdef\left\{
  \begin{array}{ll}
  (1,0)&:\text{M}\\
  -1&:\text{l$q$L}\\
  (-1,1)&:\text{l$q$J}\\
  (0,0,1,1)&:\text{R,\,$q$R}
  \end{array}\right..
\end{equation}
Note that this twist operation does not affect $\eta(x;\bm{\lambda})$ and
$\varphi(x;\bm{\lambda})$,
\begin{equation}
  \eta\bigl(x;\mathfrak{t}(\bm{\lambda})\bigr)=\eta(x;\bm{\lambda}),\quad
  \varphi\bigl(x;\mathfrak{t}(\bm{\lambda})\bigr)=\varphi(x;\bm{\lambda}).
\end{equation}
We define two functions $B'(x;\bm{\lambda})$ and $D'(x;\bm{\lambda})$,
\begin{equation}
  B'(x;\bm{\lambda})\eqdef B\bigl(x;\mathfrak{t}(\bm{\lambda})\bigr),\quad
  D'(x;\bm{\lambda})\eqdef D\bigl(x;\mathfrak{t}(\bm{\lambda})\bigr).
  \label{B'D'}
\end{equation}
By using the twist operation, the virtual state vectors
$\tilde{\phi}_{\text{v}}(x;\bm{\lambda})$ are introduced:
\begin{align}
  \tilde{\phi}_{\text{v}}(x;\bm{\lambda})&\eqdef
  \phi_{\text{v}}\bigl(x;\mathfrak{t}(\bm{\lambda})\bigr)
  =\tilde{\phi}_0(x;\bm{\lambda})
  \check{\xi}_{\text{v}}(x;\bm{\lambda}),\quad
  \tilde{\phi}_0(x;\bm{\lambda})\eqdef
  \phi_0\bigl(x;\mathfrak{t}(\bm{\lambda})\bigr),\n
  \check{\xi}_{\text{v}}(x;\bm{\lambda})&\eqdef
  \xi_{\text{v}}\bigl(\eta(x;\bm{\lambda});\bm{\lambda}\bigr)\eqdef
  \check{P}_{\text{v}}\bigl(x;\mathfrak{t}(\bm{\lambda})\bigr)
  =P_{\text{v}}\bigl(\eta(x;\bm{\lambda});
  \mathfrak{t}(\bm{\lambda})\bigr).
  \label{rdQM:phitv}
\end{align}
The virtual state polynomials $\xi_{\text{v}}(\eta;\bm{\lambda})$
are polynomials of degree $\text{v}$ in $\eta$.
The virtual state vectors satisfy the Schr\"{o}dinger equation
$\mathcal{H}(\bm{\lambda})\tilde{\phi}_{\text{v}}(x;\bm{\lambda})
=\tilde{\mathcal{E}}_{\text{v}}(\bm{\lambda})
\tilde{\phi}_{\text{v}}(x;\bm{\lambda})$
(except for the upper boundary $x=N$ for finite cases)
with the virtual energies $\tilde{\mathcal{E}}_{\text{v}}(\bm{\lambda})$,
\begin{equation}
  \tilde{\mathcal{E}}_{\text{v}}(\bm{\lambda})=\left\{
  \begin{array}{ll}
  -(1-c)(\text{v}+\beta)&:\text{M}\\
  -(1-aq^{-\text{v}})&:\text{l$q$L}\\
  -(1-aq^{-\text{v}})(1-bq^{\text{v}+1})&:\text{l$q$J}\\
  -(c+\text{v})(a+b-d-\text{v}-1)&:\text{R}\\
  -(1-cq^{\text{v}})(1-abd^{-1}q^{-\text{v}-1})&:\text{$q$R}
  \end{array}\right..
  \label{rdQM:Etv}
\end{equation}
Other data $\alpha(\bm{\lambda})$ and $\alpha'(\bm{\lambda})$
(see \cite{os26,os35} for details) are
\begin{equation}
  \alpha(\bm{\lambda})=\left\{
  \begin{array}{ll}
  c&:\text{M}\\
  a&:\text{l$q$L,\,l$q$J}\\
  1&:\text{R}\\
  abd^{-1}q^{-1}&:\text{$q$R}
  \end{array}\right.,\quad
  \alpha^{\prime}(\bm{\lambda})=\left\{
  \begin{array}{ll}
  -(1-c)\beta&:\text{M}\\
  -(1-a)&:\text{l$q$L}\\
  -(1-a)(1-bq)&:\text{l$q$J}\\
  -c(a+b-d-1)&:\text{R}\\
  -(1-c)(1-abd^{-1}q^{-1})&:\text{$q$R}
  \end{array}\right..
\end{equation}
In order to obtain well-defined quantum systems, we have to restrict the
degree $\text{v}$ and parameters $\bm{\lambda}$ so that the conditions,
$\tilde{\mathcal{E}}_{\text{v}}(\bm{\lambda})<0$, $\alpha(\bm{\lambda})>0$,
$\alpha'(\bm{\lambda})<0$, no zeros of $\xi_{\text{v}}(\eta;\bm{\lambda})$
in the domain etc., should be satisfied, see \cite{os26,os35}.
In this paper, however, we consider algebraic properties only and we
do not bother about the ranges of $\text{v}$ and $\bm{\lambda}$.

The function $\nu(x;\bm{\lambda})$ is defined as the ratio of $\phi_0$ and
$\tilde{\phi}_0$,
\begin{equation}
  \nu(x;\bm{\lambda})\eqdef
  \frac{\phi_0(x;\bm{\lambda})}{\tilde{\phi}_0(x;\bm{\lambda})}.
  \label{rdQM:nu}
\end{equation}
Since the potential functions $B(x;\bm{\lambda})$ and $D(x;\bm{\lambda})$
are rational functions of $x$ or $q^x$, this function $\nu(x;\bm{\lambda})$
can be analytically continued into a meromorphic function of $x$ or $q^x$
through the functional relations:
\begin{equation}
  \nu(x+1;\bm{\lambda})=\frac{B(x;\bm{\lambda})}
  {\alpha(\bm{\lambda})B'(x;\bm{\lambda})}\nu(x;\bm{\lambda}),\quad
  \nu(x-1;\bm{\lambda})=\frac{D(x;\bm{\lambda})}
  {\alpha(\bm{\lambda})D'(x;\bm{\lambda})}\nu(x;\bm{\lambda}).
  \label{nurel}
\end{equation}
Explicitly it is
\begin{equation}
  \nu(x;\bm{\lambda})=\left\{
  \begin{array}{ll}
  c^x&:\text{M}\\
  a^x&:\text{l$q$L,\,l$q$J}\\[2pt]
  {\displaystyle
  \frac{\Gamma(1-a)\Gamma(x+b)\Gamma(d-a+1)\Gamma(b-d-x)}
  {\Gamma(1-a-x)\Gamma(b)\Gamma(x+d-a+1)\Gamma(b-d)}}
  &:\text{R}\\[8pt]
  {\displaystyle
  \frac{(a^{-1}q^{1-x},b,a^{-1}dq^{x+1},bd^{-1};q)_{\infty}}
  {(a^{-1}q,bq^x,a^{-1}dq,bd^{-1}q^{-x};q)_{\infty}}}
  &:\text{$q$R}
  \end{array}\right..
\end{equation}
For non-negative integer $x$, it reduces to
\begin{equation}
  \nu(x;\bm{\lambda})=\frac{(1-a-x,b)_x}{(d-a+1,b-d-x)_x}
  \ \ \text{: R},\quad
  \nu(x;\bm{\lambda})
  =\frac{(a^{-1}q^{1-x},b;q)_x}{(a^{-1}dq,bd^{-1}q^{-x};q)_x}
  \ \ \text{: $q$R}.
\end{equation}
The functions $r_j(x;\bm{\lambda},M)$ are defined as the ratio of $\nu$'s,
\begin{equation}
  r_j(x+j-1;\bm{\lambda},M)\eqdef
  \frac{\nu(x+j-1;\bm{\lambda})}
  {\nu\bigl(x;\bm{\lambda}+M\tilde{\bm{\delta}}\bigr)}
  \ \ (1\leq j\leq M+1),
  \label{rdQM:rj}
\end{equation}
whose explicit forms are
\begin{equation}
  r_j(x+j-1;\bm{\lambda},M)=\left\{
  \begin{array}{ll}
  c^{j-1}&:\text{M}\\
  a^{j-1}q^{Mx}&:\text{l$q$L,\,l$q$J}\\[2pt]
  {\displaystyle
  \frac{(x+a,x+b)_{j-1}(x+d-a+j,x+d-b+j)_{M+1-j}}
  {(d-a+1,d-b+1)_M}}&:\text{R}\\[10pt]
  {\displaystyle
  \frac{(aq^x,bq^x;q)_{j-1}(a^{-1}dq^{x+j},b^{-1}dq^{x+j};q)_{M+1-j}}
  {(abd^{-1}q^{-1})^{j-1}q^{Mx}(a^{-1}dq,b^{-1}dq;q)_M}}&:\text{$q$R}
  \end{array}\right..
\end{equation}
The auxiliary function $\varphi_M(x)$ \cite{os22} is defined by
\begin{align}
  \varphi_M(x;\bm{\lambda})&\eqdef\prod_{1\leq j<k\leq M}
  \frac{\eta(x+k-1;\bm{\lambda})-\eta(x+j-1;\bm{\lambda})}
  {\eta(k-j;\bm{\lambda})}\n
  &=\prod_{1\leq j<k\leq M}
  \varphi\bigl(x+j-1;\bm{\lambda}+(k-j-1)\bm{\delta}\bigr),
  \label{rdQM:varphiM}
\end{align}
and $\varphi_0(x)=\varphi_1(x)=1$.
For M, l$q$L and l$q$J, its explicit form is
$\varphi_M(x)=\kappa^{-\frac12M(M-1)x-\frac16M(M-1)(M-2)}$.

In the following we adopt the convention,
\begin{equation}
  \check{P}_n(x;\bm{\lambda})\eqdef 0\ \ (n\in\mathbb{Z}_{<0}),\quad
  \check{\xi}_{\text{v}}(x;\bm{\lambda})\eqdef 0
  \ \ (\text{v}\in\mathbb{Z}_{<0}).
  \label{Pn<0=0}
\end{equation}

\subsection{Shape invariance}
\label{sec:rdQM_si}

The original systems in \S\,\ref{sec:rdQM_org} are shape invariant
\cite{os12} and they satisfy the relation,
\begin{equation}
  \mathcal{A}(\bm{\lambda})\mathcal{A}(\bm{\lambda})^{\dagger}
  =\kappa\mathcal{A}(\bm{\lambda}+\bm{\delta})^{\dagger}
  \mathcal{A}(\bm{\lambda}+\bm{\delta})+\mathcal{E}_1(\bm{\lambda}),
\end{equation}
which is a sufficient condition for exact solvability.
As a consequence of the shape invariance, the action of the operators
$\mathcal{A}(\bm{\lambda})$ and $\mathcal{A}(\bm{\lambda})^{\dagger}$
on the eigenvectors is
\begin{equation}
  \mathcal{A}(\bm{\lambda})\phi_n(x;\bm{\lambda})
  =\frac{\mathcal{E}_n(\bm{\lambda})}{\sqrt{B(0;\bm{\lambda})}}\,
  \phi_{n-1}(x;\bm{\lambda}+\bm{\delta}),\quad
  \mathcal{A}(\bm{\lambda})^{\dagger}
  \phi_{n-1}(x;\bm{\lambda}+\bm{\delta})
  =\sqrt{B(0;\bm{\lambda})}\,\phi_n(x;\bm{\lambda}).
  \label{rdQM:Aphi=}
\end{equation}
These relations \eqref{rdQM:Aphi=} are equivalent to the forward and
backward shift relations of the orthogonal polynomials $P_n$ \cite{kls},
\begin{equation}
  \mathcal{F}(\bm{\lambda})\check{P}_n(x;\bm{\lambda})
  =\mathcal{E}_n(\bm{\lambda})\check{P}_{n-1}(x;\bm{\lambda}+\bm{\delta}),
  \quad
  \mathcal{B}(\bm{\lambda})\check{P}_{n-1}(x;\bm{\lambda}+\bm{\delta})
  =\check{P}_n(x;\bm{\lambda}),
  \label{rdQM:FP=}
\end{equation}
where the forward and backward shift operators $\mathcal{F}(\bm{\lambda})$
and $\mathcal{B}(\bm{\lambda})$ are defined by
\begin{align}
  \mathcal{F}(\bm{\lambda})
  &\eqdef\sqrt{B(0;\bm{\lambda})}\,\phi_0(x;\bm{\lambda}+\bm{\delta})^{-1}
  \circ\mathcal{A}(\bm{\lambda})\circ\phi_0(x;\bm{\lambda})
  =B(0;\bm{\lambda})\varphi(x;\bm{\lambda})^{-1}(1-e^{\partial}),
  \label{rdQM:Fdef}\\
  \mathcal{B}(\bm{\lambda})
  &\eqdef\frac{1}{\sqrt{B(0;\bm{\lambda})}}\,
  \phi_0(x;\bm{\lambda})^{-1}\circ\mathcal{A}(\bm{\lambda})^{\dagger}
  \circ\phi_0(x;\bm{\lambda}+\bm{\delta})\n
  &=B(0;\bm{\lambda})^{-1}
  \bigl(B(x;\bm{\lambda})-D(x;\bm{\lambda})e^{-\partial}\bigr)
  \varphi(x;\bm{\lambda}).
  \label{rdQM:Bdef}
\end{align}
(For a finite system, the first equation in \eqref{rdQM:FP=} holds for
$x\leq N-1$ as a matrix and vector equation.)
The similarity transformed Hamiltonian $\widetilde{\mathcal{H}}(\bm{\lambda})$
acting on the polynomial eigenvectors is square root free. It is defined by
\begin{align}
  &\widetilde{\mathcal{H}}(\bm{\lambda})\eqdef
  \phi_0(x;\bm{\lambda})^{-1}\circ\mathcal{H}(\bm{\lambda})
  \circ\phi_0(x;\bm{\lambda})
  =\mathcal{B}(\bm{\lambda})\mathcal{F}(\bm{\lambda})\n
  &\phantom{\widetilde{\mathcal{H}}_{\ell}(\bm{\lambda})}
  =B(x;\bm{\lambda})(1-e^{\partial})
  +D(x;\bm{\lambda})(1-e^{-\partial}),\\
  &\widetilde{\mathcal{H}}(\bm{\lambda})\check{P}_n(x;\bm{\lambda})
  =\mathcal{E}_n(\bm{\lambda})\check{P}_n(x;\bm{\lambda}).
  \label{rdQM:HtP=EP}
\end{align}
Note that the relations \eqref{rdQM:FP=} and \eqref{rdQM:HtP=EP} (after
writing down in components) are valid for any $x\in\mathbb{R}$,
because they are `polynomial' equations.

The action of $\mathcal{F}(\bm{\lambda})$ and $\mathcal{B}(\bm{\lambda})$
on $(\nu^{-1}\check{\xi}_{\text{v}})(x;\bm{\lambda})$
$\eqdef\nu(x;\bm{\lambda})^{-1}\check{\xi}_{\text{v}}(x;\bm{\lambda})$ is
\begin{equation}
  \mathcal{F}(\bm{\lambda})(\nu^{-1}\check{\xi}_{\text{v}})(x;\bm{\lambda})
  =\tilde{\mathcal{E}}_{\text{v}}(\bm{\lambda})
  (\nu^{-1}\check{\xi}_{\text{v}})(x;\bm{\lambda}+\bm{\delta}),
  \ \ \mathcal{B}(\bm{\lambda})
  (\nu^{-1}\check{\xi}_{\text{v}})(x;\bm{\lambda}+\bm{\delta})
  =(\nu^{-1}\check{\xi}_{\text{v}})(x;\bm{\lambda}).
  \label{rdQM:Fnuxi=}
\end{equation}
(For a finite system, the first equation holds for $x\leq N-1$ as a matrix
and vector equation.)
Explicitly they are equivalent to the following forward and backward shift
relations of the virtual state polynomial $\check{\xi}_{\text{v}}$ :
\begin{align}
  &A\check{\xi}_{\text{v}}(x;\bm{\lambda})
  -B\check{\xi}_{\text{v}}(x+1;\bm{\lambda})
  =C\check{\xi}_{\text{v}}(x;\bm{\lambda}+\bm{\delta}),
  \label{rdQM:xiforward}\\
  &A=\left\{
  \begin{array}{ll}
  1&\!\!:\text{M,\,l$q$L,\,l$q$J}\\
  (x+a)(x+b)&\!\!:\text{R}\\[2pt]
  (1-aq^x)(1-bq^x)&\!\!:\text{$q$R}
  \end{array}\right.\!\!\!\!,
  \ \ B=\left\{
  \begin{array}{ll}
  c^{-1}&\!\!:\text{M}\\
  a^{-1}&\!\!:\text{l$q$L,\,l$q$J}\\
  (x+d-a+1)(x+d-b+1)&\!\!:\text{R}\\[2pt]
  \frac{ab}{dq}(1-a^{-1}dq^{x+1})(1-b^{-1}dq^{x+1})&\!\!:\text{$q$R}
  \end{array}\right.\!\!,\n
  &C=\left\{
  \begin{array}{ll}
  -(c\beta)^{-1}(1-c)(\text{v}+\beta)&\!\!:\text{M}\\[2pt]
  -a^{-1}(1-aq^{-\text{v}})&\!\!:\text{l$q$L}\\[2pt]
  -a^{-1}(1-bq)^{-1}(1-aq^{-\text{v}})(1-bq^{\text{v+1}})
  &\!\!:\text{l$q$J}\\[2pt]
  c^{-1}(c+\text{v})(a+b-d-1-\text{v})(2x+d+1)&\!\!:\text{R}\\[2pt]
  (1-c)^{-1}(1-cq^{\text{v}})(1-abd^{-1}q^{-\text{v}-1})(1-dq^{2x+1})
  &\!\!:\text{$q$R}
  \end{array}\right.\!\!,
  \nonumber
\end{align}
and
\begin{align}
  &A\check{\xi}_{\text{v}}(x;\bm{\lambda}+\bm{\delta})
  -B\check{\xi}_{\text{v}}(x-1;\bm{\lambda}+\bm{\delta})
  =C\check{\xi}_{\text{v}}(x;\bm{\lambda}),
  \label{rdQM:xibackward}\\
  &A=\left\{
  \begin{array}{ll}
  x+\beta&\!\!:\text{M}\\
  1&\!\!:\text{l$q$L}\\
  1-bq^{x+1}&\!\!:\text{l$q$J}\\
  (x+c)(x+d)&\!\!:\text{R}\\
  (1-cq^x)(1-dq^x)&\!\!:\text{$q$R}
  \end{array}\right.\!\!,
  \ \ B=\left\{
  \begin{array}{ll}
  x&\!\!:\text{M}\\
  1-q^x&\!\!:\text{l$q$L,\,l$q$J}\\
  (x+d-c)x&\!\!:\text{R}\\
  c(1-c^{-1}dq^x)(1-q^x)&\!\!:\text{$q$R}
  \end{array}\right.\!\!,\n
  &C=\left\{
  \begin{array}{ll}
  \beta&\!\!:\text{M}\\
  q^x&\!\!:\text{l$q$L}\\
  (1-bq)q^x&\!\!:\text{l$q$J}\\
  c(2x+d)&\!\!:\text{R}\\
  (1-c)(1-dq^{2x})&\!\!:\text{$q$R}
  \end{array}\right.\!\!.
  \nonumber
\end{align}
Note that the relations \eqref{rdQM:Fnuxi=} (after writing down in components)
and \eqref{rdQM:xiforward}--\eqref{rdQM:xibackward} are valid for any
$x\in\mathbb{R}$, because $\check{\xi}_{\text{v}}(x)$ is a `polynomial' and
$\nu(x)$ is a meromorphic function.

\subsection{Multi-indexed orthogonal polynomials}
\label{sec:rdQM_miop}

The original systems in \S\,\ref{sec:rdQM_org} are iso-spectrally deformed
by applying multiple Darboux transformations with virtual state vectors
$\{\tilde{\phi}_{\text{v}}(x)\}$ as seed solutions.
The virtual state vectors are labeled by the degree $\text{v}$ of the
polynomial part $\xi_{\text{v}}$.
We take $M$ virtual state vectors specified by the multi-index set
$\mathcal{D}=\{d_1,\ldots,d_M\}$ (ordered set).
The deformed Hamiltonian is denoted as
$\mathcal{H}_{\mathcal{D}}(\bm{\lambda})$.
The general formula for the eigenfunctions
$\{\phi^{\text{gen}}_{\mathcal{D}\,n}(x;\bm{\lambda})\}$ of the deformed
system $\mathcal{H}_{\mathcal{D}}(\bm{\lambda})
\phi^{\text{gen}}_{\mathcal{D}\,n}(x;\bm{\lambda})=\mathcal{E}_n(\bm{\lambda})
\phi^{\text{gen}}_{\mathcal{D}\,n}(x;\bm{\lambda})$ is \cite{os26,os35}
\begin{align}
  \phi^{\text{gen}}_{\mathcal{D}\,n}(x;\bm{\lambda})&=
  \frac{\mathcal{S}_{\mathcal{D}}(\bm{\lambda})
  \sqrt{\prod_{j=1}^M\alpha(\bm{\lambda})B'(x+j-1;\bm{\lambda})}
  \,\tilde{\phi}_0(x;\bm{\lambda})\,
  \text{W}_{\text{C}}[\check{\xi}_{d_1},\ldots,\check{\xi}_{d_M},
  \nu\check{P}_n](x;\bm{\lambda})}
  {\sqrt{\text{W}_{\text{C}}[\check{\xi}_{d_1},\ldots,\check{\xi}_{d_M}]
  (x;\bm{\lambda})\,
  \text{W}_{\text{C}}[\check{\xi}_{d_1},\ldots,\check{\xi}_{d_M}]
  (x+1;\bm{\lambda})}},\n
  \mathcal{S}_{\mathcal{D}}(\bm{\lambda})&\eqdef(-1)^M
  \prod_{1\leq i<j\leq M}\text{sgn}\bigl(
  \tilde{\mathcal{E}}_{d_i}(\bm{\lambda})
  -\tilde{\mathcal{E}}_{d_j}(\bm{\lambda})\bigr),
\end{align}
where $\text{W}_{\text{C}}[f_1,\ldots,f_n](x)$ is the Casoratian
\eqref{rdQM:Wdef}.

The multi-indexed orthogonal polynomials
$\{\check{P}_{\mathcal{D},n}(x;\bm{\lambda})\}$ are the main parts
of the eigenfunctions $\{\phi_{\mathcal{D}\,n}(x;\bm{\lambda})\}$ of the
deformed system $\mathcal{H}_{\mathcal{D}}(\bm{\lambda})
\phi_{\mathcal{D}\,n}(x;\bm{\lambda})=\mathcal{E}_n(\bm{\lambda})
\phi_{\mathcal{D}\,n}(x;\bm{\lambda})$ \cite{os26,os35}:
\begin{align}
  \phi_{\mathcal{D}\,n}(x;\bm{\lambda})
  &\eqdef\psi_{\mathcal{D}}(x;\bm{\lambda})
  \check{P}_{\mathcal{D},n}(x;\bm{\lambda})
  \propto\phi^{\text{gen}}_{\mathcal{D}\,n}(x;\bm{\lambda}),\quad
  \phi_{\mathcal{D}\,n}(0;\bm{\lambda})=1,\\
  \psi_{\mathcal{D}}(x;\bm{\lambda})&\eqdef
  \sqrt{\check{\Xi}_{\mathcal{D}}(1;\bm{\lambda})}\,
  \frac{\phi_0(x;\bm{\lambda}+M\tilde{\bm{\delta}})}
  {\sqrt{\check{\Xi}_{\mathcal{D}}(x;\bm{\lambda})\,
  \check{\Xi}_{\mathcal{D}}(x+1;\bm{\lambda})}},\quad
  \psi_{\mathcal{D}}(0;\bm{\lambda})=1,\\
  \check{\Xi}_{\mathcal{D}}(x;\bm{\lambda})&\eqdef
  \text{W}_{\text{C}}[\check{\xi}_{d_1},\ldots,\check{\xi}_{d_M}]
  (x;\bm{\lambda})
  \times\mathcal{C}_{\mathcal{D}}(\bm{\lambda})^{-1}
  \varphi_M(x;\bm{\lambda})^{-1},
  \label{rdQM:cXiD}\\
  \check{P}_{\mathcal{D},n}(x;\bm{\lambda})&\eqdef
  \text{W}[\check{\xi}_{d_1},\ldots,\check{\xi}_{d_M},\nu\check{P}_n]
  (x;\bm{\lambda})
  \times\mathcal{C}_{\mathcal{D},n}(\bm{\lambda})^{-1}
  \varphi_{M+1}(x;\bm{\lambda})^{-1}
  \nu(x;\bm{\lambda}+M\tilde{\bm{\delta}})^{-1},
  \label{rdQM:cPDn}\\
  \check{P}_{\mathcal{D},n}(x;\bm{\lambda})&\eqdef
  P_{\mathcal{D},n}\bigl(\eta(x;\bm{\lambda}+M\bm{\delta});
  \bm{\lambda}\bigr),\quad
  \check{\Xi}_{\mathcal{D}}(x;\bm{\lambda})\eqdef
  \Xi_{\mathcal{D}}\bigl(\eta(x;\bm{\lambda}+(M-1)\bm{\delta});
  \bm{\lambda}\bigr).
  \label{rdQM:PDn}
\end{align}
Here the universal normalization $\check{\Xi}_{\mathcal{D}}(0;\bm{\lambda})=1$
and $\check{P}_{\mathcal{D},n}(0;\bm{\lambda})=1$ are adopted,
which determines the constants $\mathcal{C}_{\mathcal{D}}(\bm{\lambda})$
and $\mathcal{C}_{\mathcal{D},n}(\bm{\lambda})$
(convention: $\prod\limits_{1\leq j<k\leq M}\!\!\!\!\!*=1$ for $M=1$),
\begin{align}
  \mathcal{C}_{\mathcal{D}}(\bm{\lambda})&\eqdef
  \frac{1}{\varphi_M(0;\bm{\lambda})}
  \prod_{1\leq j<k\leq M}
  \frac{\tilde{\mathcal{E}}_{d_j}(\bm{\lambda})
  -\tilde{\mathcal{E}}_{d_k}(\bm{\lambda})}
  {\alpha(\bm{\lambda})B'(j-1;\bm{\lambda})},
  \label{rdQM:CD}\\
  \mathcal{C}_{\mathcal{D},n}(\bm{\lambda})&\eqdef
  (-1)^M\mathcal{C}_{\mathcal{D}}(\bm{\lambda})
  \tilde{d}_{\mathcal{D},n}(\bm{\lambda})^2,\quad
  \tilde{d}_{\mathcal{D},n}(\bm{\lambda})^2\eqdef
  \frac{\varphi_M(0;\bm{\lambda})}{\varphi_{M+1}(0;\bm{\lambda})}
  \prod_{j=1}^M\frac{\mathcal{E}_n(\bm{\lambda})
  -\tilde{\mathcal{E}}_{d_j}(\bm{\lambda})}
  {\alpha(\bm{\lambda})B'(j-1;\bm{\lambda})}.
  \label{rdQM:CDn}
\end{align}
The denominator polynomial $\Xi_{\mathcal{D}}(\eta;\bm{\lambda})$ and the
multi-indexed orthogonal polynomial $P_{\mathcal{D},n}(\eta;\bm{\lambda})$
\eqref{rdQM:PDn} are polynomials in $\eta$ and their degrees are generically
$\ell_{\mathcal{D}}$ and $\ell_{\mathcal{D}}+n$, respectively.
Here $\ell_{\mathcal{D}}$ is
$\ell_{\mathcal{D}}\eqdef\sum_{j=1}^{M}d_j-\tfrac12M(M-1)$.
Note that
$\check{P}_{\mathcal{D},0}(x;\bm{\lambda})
=\check{\Xi}_{\mathcal{D}}(x;\bm{\lambda}+\bm{\delta})$.

In \cite{os26}, the multi-indexed orthogonal polynomials
$\check{P}_{\mathcal{D},n}(x;\bm{\lambda})$ \eqref{rdQM:cPDn} for R and $q$R
are expressed as
\begin{align}
  \check{P}_{\mathcal{D},n}(x;\bm{\lambda})
  &=\mathcal{C}_{\mathcal{D},n}(\bm{\lambda})^{-1}
  \varphi_{M+1}(x;\bm{\lambda})^{-1}\n
  &\quad\times\left|
  \begin{array}{cccc}
  \check{\xi}_{d_1}(x_1)&\cdots&\check{\xi}_{d_M}(x_1)
  &r_1(x_1)\check{P}_n(x_1)\\
  \check{\xi}_{d_1}(x_2)&\cdots&\check{\xi}_{d_M}(x_2)
  &r_2(x_2)\check{P}_n(x_2)\\
  \vdots&\cdots&\vdots&\vdots\\
  \check{\xi}_{d_1}(x_{M+1})&\cdots&\check{\xi}_{d_M}(x_{M+1})
  &r_{M+1}(x_{M+1})\check{P}_n(x_{M+1})\\
  \end{array}\right|,
  \label{cPDn}
\end{align}
where $x_j\eqdef x+j-1$ and $r_j(x)=r_j(x;\bm{\lambda},M)$
($1\leq j\leq M+1$) are given by \eqref{rdQM:rj}.
This expression is also valid for M, l$q$L and l$q$J.

In the next subsection we will rewrite \eqref{rdQM:cXiD}--\eqref{rdQM:cPDn}
by using the identities implied by the shape invariance and the properties
of the Casoratians.

\subsection{New determinant expressions}
\label{sec:rdQM_exp}

\subsubsection{Casoratian}
\label{sec:rdQM_Caso}

The Casorati determinant of a set of $n$ functions $\{f_j(x)\}$ is defined by
\begin{equation}
  \text{W}_{\text{C}}[f_1,f_2,\ldots,f_n](x)
  \eqdef\det\Bigl(f_k(x+j-1)\Bigr)_{1\leq j,k\leq n},
  \label{rdQM:Wdef}
\end{equation}
(for $n=0$, we set $\text{W}_{\text{C}}[\cdot](x)=1$), which satisfies
\begin{equation}
  \text{W}_{\text{C}}[gf_1,gf_2,\ldots,gf_n](x)
  =\prod_{k=0}^{n-1}g(x+k)\cdot\text{W}_{\text{C}}[f_1,f_2,\ldots,f_n](x).
  \label{WCformula1}
\end{equation}

In the rest of this section we consider the Casoratians for $\check{P}_n(x)$,
$\check{\xi}_{\text{v}}(x)$ and $\nu(x)$.
Since they are polynomials or meromorphic functions, we can realize the
shift operator as an exponential of the differential operator,
$e^{a\frac{d}{dx}}f(x)=f(x+a)$ ($a\in\mathbb{R}$).
Contrary to the matrices $e^{\partial}$ and $e^{-\partial}$, operators
$e^{\frac{d}{dx}}$ and $e^{-\frac{d}{dx}}$ are inverse to each other.

The following determinant formula holds for any smooth functions $q_j(x)$ and
$r_j(x)$ ($j=1,2,\ldots$):
\begin{equation}
  \det\Bigl(\laprod{l=1}{j-1}\hat{D}_l\cdot f_k(x)\Bigr)_{1\leq j,k\leq n}
  =\prod_{l=1}^n\prod_{m=1}^{n-l}r_m(x+l-1)\cdot
  \text{W}_{\text{C}}[f_1,\ldots,f_n](x),
  \label{rdQM:WCid}
\end{equation}
where operators $\hat{D}_j$ ($j=1,2,\ldots$) are
\begin{equation}
  \hat{D}_j=q_j(x)+r_j(x)e^{\frac{d}{dx}},
\end{equation}
and the ordered product is
\begin{equation}
  \laprod{j=1}{n}a_j\eqdef a_n\cdots a_2a_1.
\end{equation}
This formula is shown by using the properties of the determinants
(row properties) and induction in $n$.

We define operators $\hat{\mathcal{F}}'(\bm{\lambda})$ and
$\hat{\mathcal{B}}'(\bm{\lambda})$ as
$\mathcal{F}\bigl(\mathfrak{t}(\bm{\lambda})\bigr)$ and
$\mathcal{B}\bigl(\mathfrak{t}(\bm{\lambda})\bigr)$ with the replacement
$e^{\pm\partial}\mapsto e^{\pm\frac{d}{dx}}$,
\begin{align}
  \hat{\mathcal{F}}'(\bm{\lambda})&\eqdef
  B'(0;\bm{\lambda})\varphi(x;\bm{\lambda})^{-1}(1-e^{\frac{d}{dx}}),\n
  \hat{\mathcal{B}}'(\bm{\lambda})&\eqdef
  B'(0;\bm{\lambda})^{-1}
  \bigl(B'(x;\bm{\lambda})-D'(x;\bm{\lambda})e^{-\frac{d}{dx}}\bigr)
  \varphi(x;\bm{\lambda}).
\end{align}
Then the relations \eqref{rdQM:FP=} and \eqref{rdQM:Fnuxi=} with the
replacement $\bm{\lambda}\mapsto\mathfrak{t}(\bm{\lambda})$ and
exchange $n\leftrightarrow\text{v}$ give
\begin{alignat}{2}
  \hat{\mathcal{F}}'(\bm{\lambda})\check{\xi}_{\text{v}}(x;\bm{\lambda})
  &=\mathcal{E}'_{\text{v}}(\bm{\lambda})
  \check{\xi}_{\text{v}-1}(x;\bm{\lambda}+\tilde{\bm{\delta}}),&\quad
  \hat{\mathcal{B}}'(\bm{\lambda})
  \check{\xi}_{\text{v}-1}(x;\bm{\lambda}+\tilde{\bm{\delta}})
  &=\check{\xi}_{\text{v}}(x;\bm{\lambda}),
  \label{rdQM:Fh'xiv=}\\
  \hat{\mathcal{F}}'(\bm{\lambda})(\nu\check{P}_n)(x;\bm{\lambda})
  &=\tilde{\mathcal{E}}'_n(\bm{\lambda})
  (\nu\check{P}_n)(x;\bm{\lambda}+\tilde{\bm{\delta}}),&\quad
  \hat{\mathcal{B}}'(\bm{\lambda})
  (\nu\check{P}_n)(x;\bm{\lambda}+\tilde{\bm{\delta}})
  &=(\nu\check{P}_n)(x;\bm{\lambda}),
  \label{rdQM:Fh'nuPn=}
\end{alignat}
where $(\nu\check{P}_n)(x;\bm{\lambda})\eqdef\nu(x;\bm{\lambda})
\check{P}_n(x;\bm{\lambda})$, and $\mathcal{E}'_{\text{v}}(\bm{\lambda})$ and
$\tilde{\mathcal{E}}'_n(\bm{\lambda})$ are
\begin{equation}
  \mathcal{E}'_{\text{v}}(\bm{\lambda})\eqdef
  \mathcal{E}_{\text{v}}\bigl(\mathfrak{t}(\bm{\lambda})\bigr),\quad
  \tilde{\mathcal{E}}'_n(\bm{\lambda})\eqdef
  \tilde{\mathcal{E}}_n\bigl(\mathfrak{t}(\bm{\lambda})\bigr).
\end{equation}
These relations \eqref{rdQM:Fh'xiv=}--\eqref{rdQM:Fh'nuPn=} are valid for
any $x\in\mathbb{R}$.

By taking $\hat{\mathcal{F}}'$ or $\hat{\mathcal{B}}'$ as $\hat{D}_j$
and using shape invariance properties
\eqref{rdQM:Fh'xiv=}--\eqref{rdQM:Fh'nuPn=}, the Casoratians
$\text{W}_{\text{C}}[\check{\xi}_{d_1},\ldots,\check{\xi}_{d_M}]
(x;\bm{\lambda})$ and $\text{W}_{\text{C}}[\check{\xi}_{d_1},\ldots,
\check{\xi}_{d_M},\nu\check{P}_n](x;\bm{\lambda})$
in \eqref{rdQM:cXiD}--\eqref{rdQM:cPDn} can be rewritten in various ways.
We consider two typical cases:
\begin{align}
  \text{A}:\ \,&
  \hat{D}_j=\hat{\mathcal{F}}'\bigl(\bm{\lambda}
  +(j-1)\tilde{\bm{\delta}}\bigr)\ \ (j=1,2,\ldots),
  \label{rdQM:A}\\
  \text{B}:\ \,&
  \hat{D}_{2l-1}=e^{(l-1)\frac{d}{dx}}\circ\hat{\mathcal{F}}'(\bm{\lambda})
  \circ e^{-(l-1)\frac{d}{dx}},
  \ \hat{D}_{2l}=e^{l\frac{d}{dx}}\circ\hat{\mathcal{B}}'(\bm{\lambda})
  \circ e^{-(l-1)\frac{d}{dx}}
  \ \ (l=1,2,\ldots),\!
  \label{rdQM:B}
\end{align}
which correspond to the cases A and B in \cite{os37}.

\subsubsection{case A}
\label{sec:rdQM_caseA}

Firstly we consider case A \eqref{rdQM:A}.
The functions $q_j(x)$ and $r_j(x)$ are
\begin{equation}
  -q_j(x)=r_j(x)=-B'\bigl(0;\bm{\lambda}+(j-1)\tilde{\bm{\delta}}\bigr)
  \varphi\bigl(0;\bm{\lambda}+(j-1)\bm{\delta}\bigr)^{-1}.
\end{equation}
The shape invariance properties \eqref{rdQM:Fh'xiv=}--\eqref{rdQM:Fh'nuPn=}
give ($k=0,1,\ldots$)
\begin{align}
  \laprod{l=1}{k}\hat{D}_l\cdot\check{\xi}_{\text{v}}(x;\bm{\lambda})
  &=\prod_{m=0}^{k-1}\mathcal{E}'_{\text{v}-m}
  (\bm{\lambda}+m\tilde{\bm{\delta}})\cdot
  \check{\xi}_{\text{v}-k}(x;\bm{\lambda}+k\tilde{\bm{\delta}}),\n[-2pt]
  \laprod{l=1}{k}\hat{D}_l\cdot(\nu\check{P}_n)(x;\bm{\lambda})
  &=\prod_{m=0}^{k-1}\tilde{\mathcal{E}}'_n
  (\bm{\lambda}+m\tilde{\bm{\delta}})\cdot
  (\nu\check{P}_n)(x;\bm{\lambda}+k\tilde{\bm{\delta}}).
  \label{rdQM:siA}
\end{align}

We rewrite \eqref{rdQM:cXiD}--\eqref{rdQM:cPDn} in the following way.
(\romannumeral1) By using \eqref{rdQM:WCid},
rewrite the Casoratians in \eqref{rdQM:cXiD}--\eqref{rdQM:cPDn} as
determinants $\det(a_{j,k})_{1\leq j,k\leq M'}$ ($M'=M$ for
$\check{\Xi}_{\mathcal{D}}$ and $M'=M+1$ for $\check{P}_{\mathcal{D},n}$).
(\romannumeral2) Rewrite each matrix element $a_{j,k}$ by \eqref{rdQM:siA}.
For $\check{P}_{\mathcal{D},n}$, we do the following (\romannumeral3) and
(\romannumeral4).
(\romannumeral3) Divide the $(M+1)$-th column of $\det(a_{j,k})$ by
$\nu(x;\bm{\lambda}+M\tilde{\bm{\delta}})$, and multiply the determinant
by $\nu(x;\bm{\lambda}+M\tilde{\bm{\delta}})$.
(\romannumeral4) Rewrite
$\nu\bigl(x;\bm{\lambda}+(j-1)\tilde{\bm{\delta}}\bigr)/
\nu(x;\bm{\lambda}+M\tilde{\bm{\delta}})$
in the $(M+1)$-th column of $\det(a_{j,k})$ by
\begin{equation}
  \frac{\nu\bigl(x;\bm{\lambda}+(j-1)\tilde{\bm{\delta}}\bigr)}
  {\nu(x;\bm{\lambda}+M\tilde{\bm{\delta}})}=\left\{
  \begin{array}{ll}
  1&:\text{M}\\
  q^{(M+1-j)x}=\bigl(1-\eta(x)\bigr)^{M+1-j}&:\text{l$q$L,\,l$q$J}\\[2pt]
  {\displaystyle
  \frac{(x+d-a+j,x+d-b+j)_{M+1-j}}{(d-a+j,d-b+j)_{M+1-j}}}&:\text{R}\\[10pt]
  {\displaystyle
  q^{-(M+1-j)x}\frac{(a^{-1}dq^{x+j},b^{-1}dq^{x+j};q)_{M+1-j}}
  {(a^{-1}dq^j,b^{-1}dq^j;q)_{M+1-j}}}&:\text{$q$R}
  \end{array}\right..
  \label{rdQM:A:nu/nu}
\end{equation}
Then $\check{\Xi}_{\mathcal{D}}(x;\bm{\lambda})$ and
$\check{P}_{\mathcal{D},n}(x;\bm{\lambda})$ are expressed in terms of
$\check{\xi}_{\text{v}}(x;\bm{\lambda}')$,
$\check{P}_n(x;\bm{\lambda}')$, $\varphi(x';\bm{\lambda}')$ and
\eqref{rdQM:A:nu/nu}.
Straightforward calculation shows that the factors
$\varphi(x';\bm{\lambda}')$ are canceled out.
Thus, for M, l$q$L and l$q$J cases,
$\check{\Xi}_{\mathcal{D}}(x;\bm{\lambda})$ and
$\check{P}_{\mathcal{D},n}(x;\bm{\lambda})$ are expressed in terms of
$\eta(x)$, namely
$\Xi_{\mathcal{D}}(\eta;\bm{\lambda})$ and
$P_{\mathcal{D},n}(\eta;\bm{\lambda})$ are expressed without explicit
$x$-dependence (which is trivial for M, because $\eta(x)=x$).
Their final forms are as follows.\\
$\circ$ {\bf Meixner:}
The denominator polynomial $\Xi_{\mathcal{D}}(\eta;\bm{\lambda})$ is
\begin{equation}
  \Xi_{\mathcal{D}}(\eta;\bm{\lambda})
  =\frac{(-c)^{\frac12M(M-1)}}{\mathcal{C}_{\mathcal{D}}(\bm{\lambda})
  \prod_{m=1}^{M-1}(\beta)_m}\det(a_{j,k})_{1\leq j,k\leq M},
\end{equation}
where $a_{j,k}$ are
\begin{equation}
  a_{j,k}=\prod_{m=0}^{j-2}\mathcal{E}'_{d_k-m}
  (\bm{\lambda}+m\tilde{\bm{\delta}})\cdot
  \xi_{d_k+1-j}\bigl(\eta;\bm{\lambda}+(j-1)\tilde{\bm{\delta}}\bigr)
  \ \ (j,k=1,\ldots,M),
  \label{rdQM:A:M:ajk}
\end{equation}
and the multi-indexed orthogonal polynomial
$P_{\mathcal{D},n}(\eta;\bm{\lambda})$ is
\begin{equation}
  P_{\mathcal{D},n}(\eta;\bm{\lambda})
  =\frac{(-c)^{\frac12M(M+1)}}{\mathcal{C}_{\mathcal{D},n}(\bm{\lambda})
  \prod_{m=1}^M(\beta)_m}\det(a_{j,k})_{1\leq j,k\leq M+1},
\end{equation}
where $a_{j,k}$ are
\begin{align}
  a_{j,k}&=\text{eq.\,\eqref{rdQM:A:M:ajk}}
  \ \ (j=1,\ldots,M+1;k=1,\ldots,M),\n
  a_{j,M+1}&=\prod_{m=0}^{j-2}\tilde{\mathcal{E}}'_n
  (\bm{\lambda}+m\tilde{\bm{\delta}})\cdot
  P_n\bigl(\eta;\bm{\lambda}+(j-1)\tilde{\bm{\delta}}\bigr)
  \ \ (j=1,\ldots,M+1).
\end{align}
$\circ$ {\bf little $q$-Laguerre and little $q$-Jacobi:}\\
The denominator polynomial $\Xi_{\mathcal{D}}(\eta;\bm{\lambda})$ is
\begin{equation}
  \Xi_{\mathcal{D}}(\eta;\bm{\lambda})
  =\frac{(-a)^{\frac12M(M-1)}}
  {q^{\frac16M(M-1)(M-2)}\mathcal{C}_{\mathcal{D}}(\bm{\lambda})}
  \det(a_{j,k})_{1\leq j,k\leq M}\times\left\{
  \begin{array}{ll}
  1&:\text{l$q$L}\\
  \prod_{m=1}^{M-1}(bq;q)_m^{-1}&:\text{l$q$J}
  \end{array}\right.,
\end{equation}
where $a_{j,k}$ are given by \eqref{rdQM:A:M:ajk},
and the multi-indexed orthogonal polynomial
$P_{\mathcal{D},n}(\eta;\bm{\lambda})$ is
\begin{equation}
  P_{\mathcal{D},n}(\eta;\bm{\lambda})
  =\frac{(-a)^{\frac12M(M+1)}}
  {q^{\frac16(M+1)M(M-1)}\mathcal{C}_{\mathcal{D},n}(\bm{\lambda})}
  \det(a_{j,k})_{1\leq j,k\leq M+1}\times\left\{
  \begin{array}{ll}
  1&:\text{l$q$L}\\
  \prod_{m=1}^M(bq;q)_m^{-1}&:\text{l$q$J}
  \end{array}\right.,
\end{equation}
where $a_{j,k}$ are
\begin{align}
  a_{j,k}&=\text{eq.\,\eqref{rdQM:A:M:ajk}}
  \ \ (j=1,\ldots,M+1;k=1,\ldots,M),\n
  a_{j,M+1}&=\prod_{m=0}^{j-2}\tilde{\mathcal{E}}'_n
  (\bm{\lambda}+m\tilde{\bm{\delta}})\cdot(1-\eta)^{M+1-j}
  P_n\bigl(\eta;\bm{\lambda}+(j-1)\tilde{\bm{\delta}}\bigr)
  \ \ (j=1,\ldots,M+1).
\end{align}
$\circ$ {\bf Racah and $q$-Racah:}\\
The denominator polynomial $\check{\Xi}_{\mathcal{D}}(x;\bm{\lambda})$ is
\begin{equation}
  \check{\Xi}_{\mathcal{D}}(x;\bm{\lambda})
  =\frac{(-1)^{\frac12M(M-1)}}
  {\mathcal{C}_{\mathcal{D}}(\bm{\lambda})
  \prod_{m=0}^{M-1}B'(0;\bm{\lambda}+m\tilde{\bm{\delta}})^{M-1-m}}
  \det(a_{j,k})_{1\leq j,k\leq M},
\end{equation}
where $a_{j,k}$ are
\begin{equation}
  a_{j,k}=\prod_{m=0}^{j-2}\mathcal{E}'_{d_k-m}
  (\bm{\lambda}+m\tilde{\bm{\delta}})\cdot
  \check{\xi}_{d_k+1-j}\bigl(x;\bm{\lambda}+(j-1)\tilde{\bm{\delta}}\bigr)
  \ \ (j,k=1,\ldots,M),
  \label{rdQM:A:ajk}
\end{equation}
and the multi-indexed orthogonal polynomial
$\check{P}_{\mathcal{D},n}(x;\bm{\lambda})$ is
\begin{equation}
  \check{P}_{\mathcal{D},n}(x;\bm{\lambda})
  =\frac{(-1)^{\frac12M(M+1)}}
  {\mathcal{C}_{\mathcal{D},n}(\bm{\lambda})
  \prod_{m=0}^MB'(0;\bm{\lambda}+m\tilde{\bm{\delta}})^{M-m}}
  \det(a_{j,k})_{1\leq j,k\leq M+1},
\end{equation}
where $a_{j,k}$ are
\begin{align}
  a_{j,k}&=\text{eq.\,\eqref{rdQM:A:ajk}}
  \ \ (j=1,\ldots,M+1;k=1,\ldots,M),\n
  a_{j,M+1}&=\prod_{m=0}^{j-2}\tilde{\mathcal{E}}'_n
  (\bm{\lambda}+m\tilde{\bm{\delta}})\cdot
  P_n\bigl(\eta;\bm{\lambda}+(j-1)\tilde{\bm{\delta}}\bigr)\n
  &\quad\times\left\{
  \begin{array}{ll}
  {\displaystyle
  \frac{(x+d-a+j,x+d-b+j)_{M+1-j}}{(d-a+j,d-b+j)_{M+1-j}}}&:\text{R}\\[10pt]
  {\displaystyle
  q^{-(M+1-j)x}\frac{(a^{-1}dq^{x+j},b^{-1}dq^{x+j};q)_{M+1-j}}
  {(a^{-1}dq^j,b^{-1}dq^j;q)_{M+1-j}}}&:\text{$q$R}
  \end{array}\right.
  \ \ (j=1,\ldots,M+1).
\end{align}

\subsubsection{case B}
\label{sec:rdQM_caseB}

Secondly we consider case B \eqref{rdQM:B}.
The functions $q_j(x)$ and $r_j(x)$ are ($l=1,2,\ldots$)
\begin{align}
  -q_{2l-1}(x)&=r_{2l-1}(x)=-B'(0;\bm{\lambda})
  \varphi(x+l-1;\bm{\lambda})^{-1},\n
  q_{2l}(x)&=-B'(0;\bm{\lambda})^{-1}D'(x+l;\bm{\lambda})
  \varphi(x+l-1;\bm{\lambda}),\\
  r_{2l}(x)&=B'(0;\bm{\lambda})^{-1}B'(x+l;\bm{\lambda})
  \varphi(x+l;\bm{\lambda}).
  \nonumber
\end{align}
The shape invariance properties \eqref{rdQM:Fh'xiv=}--\eqref{rdQM:Fh'nuPn=}
give ($l=1,2,\ldots$)
\begin{align}
  \laprod{m=1}{2l-2}\hat{D}_m\cdot\check{\xi}_{\text{v}}(x;\bm{\lambda})
  &=\mathcal{E}'_{\text{v}}(\bm{\lambda})^{l-1}
  \check{\xi}_{\text{v}}(x+l-1;\bm{\lambda}),\n[-2pt]
  \laprod{m=1}{2l-1}\hat{D}_m\cdot\check{\xi}_{\text{v}}(x;\bm{\lambda})
  &=\mathcal{E}'_{\text{v}}(\bm{\lambda})^l
  \check{\xi}_{\text{v}-1}(x+l-1;\bm{\lambda}+\tilde{\bm{\delta}}),\n[-2pt]
  \laprod{m=1}{2l-2}\hat{D}_m\cdot(\nu\check{P}_n)(x;\bm{\lambda})
  &=\tilde{\mathcal{E}}'_n(\bm{\lambda})^{l-1}
  (\nu\check{P}_n)(x+l-1;\bm{\lambda}),
  \label{rdQM:siB}\\
  \laprod{m=1}{2l-1}\hat{D}_m\cdot(\nu\check{P}_n)(x;\bm{\lambda})
  &=\tilde{\mathcal{E}}'_n(\bm{\lambda})^l
  (\nu\check{P}_n)(x+l-1;\bm{\lambda}+\tilde{\bm{\delta}}).
  \nonumber
\end{align}

Like case A, we rewrite \eqref{rdQM:cXiD}--\eqref{rdQM:cPDn} by the steps
(\romannumeral1)--(\romannumeral4).
In (\romannumeral4) we use
\begin{align}
  &\frac{\nu(x+l-1;\bm{\lambda})}
  {\nu(x;\bm{\lambda}+M\tilde{\bm{\delta}})}=\left\{
  \begin{array}{ll}
  c^{l-1}&:\text{M}\\
  a^{l-1}q^{Mx}&:\text{l$q$L,\,l$q$J}\\[2pt]
  {\displaystyle
  \frac{(x+d-a+l,x+d-b+l)_{M+1-l}(x+a,x+b)_{l-1}}
  {(d-a+1,d-b+1)_M}}&:\text{R}\\[10pt]
  {\displaystyle
  \frac{(a^{-1}dq^{x+l},b^{-1}dq^{x+l};q)_{M+1-l}(aq^x,bq^x;q)_{l-1}}
  {(abd^{-1}q^{-1})^{l-1}q^{Mx}
  (a^{-1}dq,b^{-1}dq;q)_M}}&:\text{$q$R}
  \end{array}\right.,\n
  &\frac{\nu(x+l-1;\bm{\lambda}+\tilde{\bm{\delta}})}
  {\nu(x;\bm{\lambda}+M\tilde{\bm{\delta}})}
  \label{rdQM:B:nu/nu}\\
  &=\left\{
  \begin{array}{ll}
  c^{l-1}&:\text{M}\\
  (aq^{-1})^{l-1}q^{(M-1)x}&:\text{l$q$L,\,l$q$J}\\[2pt]
  {\displaystyle
  \frac{(x+d-a+l+1,x+d-b+l+1)_{M-l}(x+a,x+b)_{l-1}}
  {(d-a+2,d-b+2)_{M-1}}}&:\text{R}\\[10pt]
  {\displaystyle
  \frac{(a^{-1}dq^{x+l+1},b^{-1}dq^{x+l+1};q)_{M-l}(aq^x,bq^x;q)_{l-1}}
  {(abd^{-1}q^{-2})^{l-1}q^{(M-1)x}
  (a^{-1}dq^2,b^{-1}dq^2;q)_{M-1}}}&:\text{$q$R}
  \end{array}\right..
  \nonumber
\end{align}
Contrary to case A, the determinants contain
$\check{\xi}_{\text{v}}(x';\bm{\lambda}')$ and
$\check{P}_n(x';\bm{\lambda}')$ at various points $x'$,
and the factors $\varphi(x';\bm{\lambda}')$ are not canceled out.
Their final forms are as follows.
The denominator polynomial $\check{\Xi}_{\mathcal{D}}(x;\bm{\lambda})$ is
\begin{align}
  \check{\Xi}_{\mathcal{D}}(x;\bm{\lambda})
  &=\frac{(-1)^{\frac12(1+(-1)^M)\frac{M}{2}}}
  {\mathcal{C}_{\mathcal{D}}(\bm{\lambda})B'(0;\bm{\lambda})^{[\frac{M}{2}]}}
  \prod_{m=0}^{[\frac{M-2}{2}]}\varphi(x+m;\bm{\lambda})\cdot
  \varphi_M(x;\bm{\lambda})^{-1}\n
  &\quad\times\biggl(
  \prod_{m=0}^{[\frac{M-2}{2}]}\prod_{l=1}^{M-2-2m}
  B'(x+l+m;\bm{\lambda})\biggr)^{-1}
  \det(a_{j,k})_{1\leq j,k\leq M},
\end{align}
where $a_{j,k}$ are
\begin{align}
  a_{2l-1,k}&=\mathcal{E}'_{d_k}(\bm{\lambda})^{l-1}
  \check{\xi}_{d_k}(x+l-1;\bm{\lambda})
  \ \ (l=1,\ldots,[\tfrac{M+1}{2}];k=1,\ldots,M),\n
  a_{2l,k}&=\mathcal{E}'_{d_k}(\bm{\lambda})^l
  \check{\xi}_{d_k-1}(x+l-1;\bm{\lambda}+\tilde{\bm{\delta}})
  \ \ (l=1,\ldots,[\tfrac{M}{2}];k=1,\ldots,M),
\end{align}
and the multi-indexed orthogonal polynomial
$\check{P}_{\mathcal{D},n}(x;\bm{\lambda})$ is
\begin{align}
  \check{P}_{\mathcal{D},n}(x;\bm{\lambda})
  &=\frac{(-1)^{\frac12(1-(-1)^M)\frac{M+1}{2}}}
  {\mathcal{C}_{\mathcal{D},n}(\bm{\lambda})
  B'(0;\bm{\lambda})^{[\frac{M+1}{2}]}}
  \prod_{m=0}^{[\frac{M-1}{2}]}\varphi(x+m;\bm{\lambda})\cdot
  \varphi_{M+1}(x;\bm{\lambda})^{-1}\n
  &\quad\times\biggl(
  \prod_{m=0}^{[\frac{M-1}{2}]}\prod_{l=1}^{M-1-2m}
  B'(x+l+m;\bm{\lambda})\biggr)^{-1}
  \det(a_{j,k})_{1\leq j,k\leq M+1},
\end{align}
where $a_{j,k}$ are
\begin{align}
  a_{2l-1,k}&=\mathcal{E}'_{d_k}(\bm{\lambda})^{l-1}
  \check{\xi}_{d_k}(x+l-1;\bm{\lambda})
  \ \ (l=1,\ldots,[\tfrac{M+2}{2}];k=1,\ldots,M),\n
  a_{2l,k}&=\mathcal{E}'_{d_k}(\bm{\lambda})^l
  \check{\xi}_{d_k-1}(x+l-1;\bm{\lambda}+\tilde{\bm{\delta}})
  \ \ (l=1,\ldots,[\tfrac{M+1}{2}];k=1,\ldots,M),\n
  a_{2l-1,M+1}&=\tilde{\mathcal{E}}'_n(\bm{\lambda})^{l-1}
  \check{P}_n(x+l-1;\bm{\lambda})
  \qquad(l=1,\ldots,[\tfrac{M+2}{2}])\n
  &\quad\times\left\{
  \begin{array}{ll}
  c^{l-1}&:\text{M}\\
  a^{l-1}q^{Mx}&:\text{l$q$L,\,l$q$J}\\[2pt]
  {\displaystyle
  \frac{(x+d-a+l,x+d-b+l)_{M+1-l}(x+a,x+b)_{l-1}}
  {(d-a+1,d-b+1)_M}}&:\text{R}\\[10pt]
  {\displaystyle
  \frac{(a^{-1}dq^{x+l},b^{-1}dq^{x+l};q)_{M+1-l}(aq^x,bq^x;q)_{l-1}}
  {(abd^{-1}q^{-1})^{l-1}q^{Mx}
  (a^{-1}dq,b^{-1}dq;q)_M}}&:\text{$q$R}
  \end{array}\right.,\n
  a_{2l,M+1}&=\tilde{\mathcal{E}}'_n(\bm{\lambda})^l
  \check{P}_n(x+l-1;\bm{\lambda}+\tilde{\bm{\delta}})
  \qquad(l=1,\ldots,[\tfrac{M+1}{2}])\\
  &\quad\times\left\{
  \begin{array}{ll}
  c^{l-1}&:\text{M}\\
  (aq^{-1})^{l-1}q^{(M-1)x}&:\text{l$q$L,\,l$q$J}\\[2pt]
  {\displaystyle
  \frac{(x+d-a+l+1,x+d-b+l+1)_{M-l}(x+a,x+b)_{l-1}}
  {(d-a+2,d-b+2)_{M-1}}}&:\text{R}\\[10pt]
  {\displaystyle
  \frac{(a^{-1}dq^{x+l+1},b^{-1}dq^{x+l+1};q)_{M-l}(aq^x,bq^x;q)_{l-1}}
  {(abd^{-1}q^{-2})^{l-1}q^{(M-1)x}
  (a^{-1}dq^2,b^{-1}dq^2;q)_{M-1}}}&:\text{$q$R}
  \end{array}\right..
  \nonumber
\end{align}

\section{Multi-indexed Orthogonal Polynomials in idQM}
\label{sec:idQM}

In this section we derive various equivalent expressions for the
multi-indexed Wilson and Askey-Wilson polynomials in the framework of
the discrete quantum mechanics with pure imaginary shifts.

\subsection{Original systems}
\label{sec:idQM_org}

The Hamiltonian of the discrete quantum mechanics with pure imaginary shifts
(idQM) \cite{os13_1,os13_2},
whose dynamical variables are the real coordinate
$x$ ($x_1\leq x\leq x_2$) and the conjugate momentum $p=-i\frac{d}{dx}$, is
\begin{align}
  &\mathcal{H}\eqdef\sqrt{V(x)}\,e^{\gamma p}\sqrt{V^*(x)}
  +\!\sqrt{V^*(x)}\,e^{-\gamma p}\sqrt{V(x)}
  -V(x)-V^*(x)=\mathcal{A}^{\dagger}\mathcal{A},
  \label{idQM:H}\\
  &\mathcal{A}\eqdef i\bigl(e^{\frac{\gamma}{2}p}\sqrt{V^*(x)}
  -e^{-\frac{\gamma}{2}p}\sqrt{V(x)}\,\bigr),\quad
  \mathcal{A}^{\dagger}\eqdef -i\bigl(\sqrt{V(x)}\,e^{\frac{\gamma}{2}p}
  -\sqrt{V^*(x)}\,e^{-\frac{\gamma}{2}p}\bigr).
\end{align}
Here the potential function $V(x)$ is an analytic function of $x$ and
$\gamma$ is a real constant.
The $*$-operation on an analytic function $f(x)=\sum_na_nx^n$
($a_n\in\mathbb{C}$) is defined by $f^*(x)=\sum_na_n^*x^n$, in which
$a_n^*$ is the complex conjugation of $a_n$.
Since the momentum operator appears in exponentiated forms,
the Schr\"{o}dinger equation,
\begin{equation}
  \mathcal{H}\phi_n(x)=\mathcal{E}_n\phi_n(x)
  \ \ (n=0,1,2,\ldots),\quad
  0=\mathcal{E}_0<\mathcal{E}_1<\mathcal{E}_2<\cdots,
\end{equation}
is an analytic difference equation with pure imaginary shifts.

We consider idQM described by the Wilson and Askey-Wilson polynomials.
Various parameters are
\begin{alignat}{2}
  \text{W}:\ \ &x_1=0,\ x_2=\infty,\ \gamma=1,
  &\ \ \bm{\lambda}=(a_1,a_2,a_3,a_4),
  &\ \ \bm{\delta}=(\tfrac12,\tfrac12,\tfrac12,\tfrac12),
  \ \ \kappa=1,\\
  \text{AW}:\ \ &x_1=0,\ x_2=\pi,\ \gamma=\log q,
  &\ \ q^{\bm{\lambda}}=(a_1,a_2,a_3,a_4),
  &\ \ \bm{\delta}=(\tfrac12,\tfrac12,\tfrac12,\tfrac12),
  \ \ \kappa=q^{-1},
\end{alignat}
where $0<q<1$.
The parameters are restricted by
\begin{equation}
  \{a_1^*,a_2^*,a_3^*,a_4^*\}=\{a_1,a_2,a_3,a_4\}\ \ (\text{as a set});\quad
  \text{W}:\ \text{Re}\,a_i>0,\quad
  \text{AW}:\ |a_i|<1.
  \label{rangeorg}
\end{equation}
Here are the fundamental data \cite{os13_1}:
\begin{align}
  &V(x;\bm{\lambda})=\left\{
  \begin{array}{ll}
  \bigl(2ix(2ix+1)\bigr)^{-1}\prod_{j=1}^4(a_j+ix)&:\text{W}\\[2pt]
  \bigl((1-e^{2ix})(1-qe^{2ix})\bigr)^{-1}\prod_{j=1}^4(1-a_je^{ix})
  &:\text{AW}
  \end{array}\right.,
  \label{Vform}\\
  &\eta(x)=\left\{
  \begin{array}{ll}
  x^2&:\text{W}\\
  \cos x&:\text{AW}
  \end{array}\right.,\quad
  \varphi(x)=\left\{
  \begin{array}{ll}
  2x&:\text{W}\\
  2\sin x&:\text{AW}
  \end{array}\right.,
  \label{etadef}\\
  &\mathcal{E}_n(\bm{\lambda})=\left\{
  \begin{array}{ll}
  n(n+b_1-1)&:\text{W}\\[2pt]
  (q^{-n}-1)(1-b_4q^{n-1})&:\text{AW}
  \end{array}\right.,\quad
  \begin{array}{l}
  b_1\eqdef a_1+a_2+a_3+a_4,\\
  b_4\eqdef a_1a_2a_3a_4,
  \end{array}\\
  &\phi_n(x;\bm{\lambda})
  =\phi_0(x;\bm{\lambda})\check{P}_n(x;\bm{\lambda}),
  \label{factphin}\\
  &\check{P}_n(x;\bm{\lambda})=P_n\bigl(\eta(x);\bm{\lambda}\bigr)
  =\left\{\begin{array}{ll}
  W_n\bigl(\eta(x);a_1,a_2,a_3,a_4\bigr)&:\text{W}\\[2pt]
  p_n\bigl(\eta(x);a_1,a_2,a_3,a_4|q\bigr)&:\text{AW}
  \end{array}\right.\n
  &\phantom{\check{P}_n(x;\bm{\lambda})}=\left\{\begin{array}{ll}
  {\displaystyle
  (a_1+a_2,a_1+a_3,a_1+a_4)_n}\\[2pt]
  {\displaystyle
  \quad\times
  {}_4F_3\Bigl(
  \genfrac{}{}{0pt}{}{-n,\,n+b_1-1,\,a_1+ix,\,a_1-ix}
  {a_1+a_2,\,a_1+a_3,\,a_1+a_4}\Bigm|1\Bigr)
  }&:\text{W}\\[8pt]
  {\displaystyle
  a_1^{-n}(a_1a_2,a_1a_3,a_1a_4;q)_n}\\[2pt]
  {\displaystyle
  \quad\times
  {}_4\phi_3\Bigl(\genfrac{}{}{0pt}{}{q^{-n},\,b_4q^{n-1},\,
  a_1e^{ix},\,a_1e^{-ix}}{a_1a_2,\,a_1a_3,\,a_1a_4}\!\!\Bigm|\!q\,;q\Bigr)
  }&:\text{AW}
  \end{array}\right.
  \label{Pn=W,AW}\\[2pt]
  &\phi_0(x;\bm{\lambda})=\left\{
  \begin{array}{ll}
  \sqrt{\bigl(\Gamma(2ix)\Gamma(-2ix)\bigr)^{-1}\prod_{j=1}^4
  \Gamma(a_j+ix)\Gamma(a_j-ix)}&:\text{W}\\[4pt]
  \sqrt{(e^{2ix},e^{-2ix};q)_{\infty}
  \prod_{j=1}^4(a_je^{ix},a_je^{-ix};q)_{\infty}^{-1}}
  &:\text{AW}
  \end{array}\right.,
  \label{phi0=W,AW}
\end{align}
where $W_n(\eta;a_1,a_2,a_3,a_4)$ and $p_n(\eta;a_1,a_2,a_3,a_4|q)$ are
the Wilson and Askey-Wilson polynomials \cite{kls}, respectively.
Note that
\begin{align}
  &\phi_0(x;\bm{\lambda}+\bm{\delta})
  =\varphi(x)\sqrt{V(x+i\tfrac{\gamma}{2};\bm{\lambda})}\,
  \phi_0(x+i\tfrac{\gamma}{2};\bm{\lambda}),
  \label{phi0(l+d)}\\
  &V(x;\bm{\lambda}+\bm{\delta})
  =\kappa^{-1}\frac{\varphi(x-i\gamma)}{\varphi(x)}
  V(x-i\tfrac{\gamma}{2};\bm{\lambda}).
  \label{varphiprop3}
\end{align}

\subsection{Virtual states}
\label{sec:idQM_vs}

The virtual state wavefunctions are obtained by using the discrete symmetries
of the Hamiltonian \cite{os27}.
In the following we restrict the parameters as follows:
\begin{equation}
  a_1,a_2\in\mathbb{R}\ \ \text{or}\ \ a_2^*=a_1\ ;\quad
  a_3,a_4\in\mathbb{R}\ \ \text{or}\ \ a_4^*=a_3.
\end{equation}
We define two twist operations $\mathfrak{t}$, type $\I$ and type $\II$,
\begin{equation}
  \mathfrak{t}^{\I}(\bm{\lambda})
  \eqdef(1-\lambda_1,1-\lambda_2,\lambda_3,\lambda_4),\quad
  \mathfrak{t}^{\II}(\bm{\lambda})
  \eqdef(\lambda_1,\lambda_2,1-\lambda_3,1-\lambda_4),
  \label{idQM:twist}
\end{equation}
which are involutions $\mathfrak{t}^2=\text{id}$ and satisfy
\begin{equation}
  \mathfrak{t}(\bm{\lambda})+u\bm{\delta}
  =\mathfrak{t}(\bm{\lambda}+u\tilde{\bm{\delta}})
  \ \ (\forall u\in\mathbb{R}),\quad
  \tilde{\bm{\delta}}^{\I}\eqdef(-\tfrac12,-\tfrac12,\tfrac12,\tfrac12),
  \ \ \tilde{\bm{\delta}}^{\II}\eqdef(\tfrac12,\tfrac12,-\tfrac12,-\tfrac12).
\end{equation}
By using these two types of twist operations, two types of virtual state
wavefunctions, $\tilde{\phi}_{\text{v}}^{\I}(x;\bm{\lambda})$ and
$\tilde{\phi}_{\text{v}}^{\II}(x;\bm{\lambda})$, are introduced:
\begin{align}
  \tilde{\phi}_{\text{v}}(x;\bm{\lambda})&\eqdef
  \phi_{\text{v}}\bigl(x;\mathfrak{t}(\bm{\lambda})\bigr)
  =\tilde{\phi}_0(x;\bm{\lambda})
  \check{\xi}_{\text{v}}(x;\bm{\lambda}),\quad
  \tilde{\phi}_0(x;\bm{\lambda})\eqdef
  \phi_0\bigl(x;\mathfrak{t}(\bm{\lambda})\bigr),\n
  \check{\xi}_{\text{v}}(x;\bm{\lambda})&\eqdef
  \xi_{\text{v}}\bigl(\eta(x);\bm{\lambda}\bigr)\eqdef
  \check{P}_{\text{v}}\bigl(x;\mathfrak{t}(\bm{\lambda})\bigr)
  =P_{\text{v}}\bigl(\eta(x);\mathfrak{t}(\bm{\lambda})\bigr).
  \label{idQM:phitv}
\end{align}
The virtual state polynomials $\xi_{\text{v}}(\eta;\bm{\lambda})$
(two types $\xi^{\I}_{\text{v}}$ and $\xi^{\II}_{\text{v}}$)
are polynomials of degree $\text{v}$ in $\eta$.
The virtual state wavefunctions satisfy the Schr\"{o}dinger equation
$\mathcal{H}(\bm{\lambda})\tilde{\phi}_{\text{v}}(x;\bm{\lambda})
=\tilde{\mathcal{E}}_{\text{v}}(\bm{\lambda})
\tilde{\phi}_{\text{v}}(x;\bm{\lambda})$
with the virtual energies $\tilde{\mathcal{E}}_{\text{v}}(\bm{\lambda})$,
\begin{align}
  &\tilde{\mathcal{E}}^{\I}_{\text{v}}(\bm{\lambda})=\left\{
  \begin{array}{ll}
  -(a_1+a_2-\text{v}-1)(a_3+a_4+\text{v})&:\text{W}\\
  -(1-a_1a_2q^{-\text{v}-1})(1-a_3a_4q^{\text{v}})&:\text{AW}
  \end{array}\right.,\n
  &\tilde{\mathcal{E}}^{\II}_{\text{v}}(\bm{\lambda})=\left\{
  \begin{array}{ll}
  -(a_3+a_4-\text{v}-1)(a_1+a_2+\text{v})&:\text{W}\\
  -(1-a_3a_4q^{-\text{v}-1})(1-a_1a_2q^{\text{v}})&:\text{AW}
  \end{array}\right..
  \label{idQM:Etv}
\end{align}
Other data $\alpha(\bm{\lambda})$ and $\alpha'(\bm{\lambda})$
(see \cite{os27} for details) are
\begin{align}
  &\alpha^{\I}(\bm{\lambda})=\left\{
  \begin{array}{ll}
  1&:\text{W}\\
  a_1a_2q^{-1}&:\text{AW}
  \end{array}\right.,\quad
  \alpha^{\prime\,\I}(\bm{\lambda})=\left\{
  \begin{array}{ll}
  -(a_1+a_2-1)(a_3+a_4)&:\text{W}\\
  -(1-a_1a_2q^{-1})(1-a_3a_4)&:\text{AW}
  \end{array}\right.,\n
  &\alpha^{\II}(\bm{\lambda})=\left\{
  \begin{array}{ll}
  1&:\text{W}\\
  a_3a_4q^{-1}&:\text{AW}
  \end{array}\right.,\quad
  \alpha^{\prime\,\II}(\bm{\lambda})=\left\{
  \begin{array}{ll}
  -(a_3+a_4-1)(a_1+a_2)&:\text{W}\\
  -(1-a_3a_4q^{-1})(1-a_1a_2)&:\text{AW}
  \end{array}\right..
\end{align}
In order to obtain well-defined quantum systems, we have to restrict the
degree $\text{v}$ and the parameters $a_i$ so that the conditions,
$\tilde{\mathcal{E}}_{\text{v}}(\bm{\lambda})<0$, $\alpha(\bm{\lambda})>0$,
$\alpha'(\bm{\lambda})<0$, no zeros of $\xi_{\text{v}}(\eta;\bm{\lambda})$
in the domain etc., are satisfied, see \cite{os27}.
In this paper, however, we consider algebraic properties only and we
do not bother about the ranges of $\text{v}$ and $a_i$.

The functions $\nu(x;\bm{\lambda})$ (two types $\nu^{\I}$ and $\nu^{\II}$)
are defined as the ratio of $\phi_0$ and $\tilde{\phi}_0$,
\begin{equation}
  \nu(x;\bm{\lambda})\eqdef
  \frac{\phi_0(x;\bm{\lambda})}{\tilde{\phi}_0(x;\bm{\lambda})}.
  \label{idQM:nu}
\end{equation}
The functions $r_j(x^{(M)}_j;\bm{\lambda},M)$ (two types $r^{\I}_j$ and
$r^{\II}_j$) are defined as the ratio of $\nu$'s,
\begin{equation}
  r_j(x^{(M)}_j;\bm{\lambda},M)\eqdef
  \frac{\nu(x^{(M)}_j;\bm{\lambda})}
  {\nu\bigl(x;\bm{\lambda}+(M-1)\tilde{\bm{\delta}}\bigr)}
  \ \ (1\leq j\leq M),
  \label{idQM:rj}
\end{equation}
where $x^{(M)}_j\eqdef x+i(\frac{M+1}{2}-j)\gamma$.
Their explicit forms are
\begin{align}
  r^{\I}_j(x^{(M)}_j;\bm{\lambda},M)
  &=\alpha^{\I}\bigl(\bm{\lambda}
  +(M-1)\tilde{\bm{\delta}}^{\I}\bigr)^{-\frac12(M-1)}
  \kappa^{\frac12(M-1)^2-(j-1)(M-j)}\\
  &\quad\times\left\{
  \begin{array}{ll}
  {\displaystyle
  \prod_{k=1,2}(a_k-\tfrac{M-1}{2}+ix)_{j-1}(a_k-\tfrac{M-1}{2}-ix)_{M-j}
  }&:\text{W}\\
  {\displaystyle
  e^{ix(M+1-2j)}\prod_{k=1,2}(a_kq^{-\frac{M-1}{2}}e^{ix};q)_{j-1}
  (a_kq^{-\frac{M-1}{2}}e^{-ix};q)_{M-j}
  }&:\text{AW}
  \end{array}\right.,\n
  r^{\II}_j(x^{(M)}_j;\bm{\lambda},M)
  &=\alpha^{\II}\bigl(\bm{\lambda}
  +(M-1)\tilde{\bm{\delta}}^{\II}\bigr)^{-\frac12(M-1)}
  \kappa^{\frac12(M-1)^2-(j-1)(M-j)}\\
  &\quad\times\left\{
  \begin{array}{ll}
  {\displaystyle
  \prod_{k=3,4}(a_k-\tfrac{M-1}{2}+ix)_{j-1}(a_k-\tfrac{M-1}{2}-ix)_{M-j}
  }&:\text{W}\\
  {\displaystyle
  e^{ix(M+1-2j)}\prod_{k=3,4}(a_kq^{-\frac{M-1}{2}}e^{ix};q)_{j-1}
  (a_kq^{-\frac{M-1}{2}}e^{-ix};q)_{M-j}
  }&:\text{AW}
  \end{array}\right..\nonumber
\end{align}
The auxiliary function $\varphi_M(x)$ \cite{gos} is defined by
\begin{align}
  \varphi_M(x)&\eqdef
  \varphi(x)^{[\frac{M}{2}]}\prod_{k=1}^{M-2}
  \bigl(\varphi(x-i\tfrac{k}{2}\gamma)\varphi(x+i\tfrac{k}{2}\gamma)
  \bigr)^{[\frac{M-k}{2}]}\n
  &=\prod_{1\leq j<k\leq M}
  \frac{\eta(x^{(M)}_j)-\eta(x^{(M)}_k)}
  {\varphi(i\frac{j}{2}\gamma)}
  \times\left\{
  \begin{array}{ll}
  1&:\text{W}\\
  (-2)^{\frac12M(M-1)}&:\text{AW}
  \end{array}\right.,
  \label{idQM:varphiM}
\end{align}
and $\varphi_0(x)=\varphi_1(x)=1$.

For $-M\leq k\leq M$ and $k\equiv M$ ($\text{mod}\,2$),
we can show that 
\begin{equation}
  \frac{\nu(x;\bm{\lambda}+M\bm{\delta})}{\nu(x;\bm{\lambda}+k\bm{\delta})}
  =\prod_{l=0}^{\frac{M-k}{2}-1}
  \check{U}\bigl(x;\bm{\lambda}+(k+2l)\bm{\delta}\bigr),
  \label{nuMk}
\end{equation}
where the functions $\check{U}(x;\bm{\lambda})$ (two types $\check{U}^{\I}$
and $\check{U}^{\II}$) are
\begin{align}
  \check{U}^{\I}(x;\bm{\lambda})&\eqdef\left\{
  \begin{array}{ll}
  \prod_{k=1}^2(a_k+ix)(a_k-ix)&:\text{W}\\[2pt]
  (a_1a_2)^{-1}\prod_{k=1}^2(1-a_ke^{ix})(1-a_ke^{-ix})&:\text{AW}
  \end{array}\right.,\n
  \check{U}^{\II}(x;\bm{\lambda})&\eqdef\left\{
  \begin{array}{ll}
  \prod_{k=3}^4(a_k+ix)(a_k-ix)&:\text{W}\\[2pt]
  (a_3a_4)^{-1}\prod_{k=3}^4(1-a_ke^{ix})(1-a_ke^{-ix})&:\text{AW}
  \end{array}\right..
\end{align}
In fact, these functions $\check{U}(x;\bm{\lambda})$ are polynomials
$U(\eta;\bm{\lambda})$ in $\eta$ as defined by
\begin{align}
  &U\bigl(\eta(x);\bm{\lambda}\bigr)\eqdef\check{U}(x;\bm{\lambda}),\\
  &U^{\I}(\eta;\bm{\lambda})=\left\{
  \begin{array}{ll}
  \prod_{k=1}^2(a_k^2+\eta)&:\text{W}\\[2pt]
  \prod_{k=1}^2(a_k+a_k^{-1}-2\eta)&:\text{AW}
  \end{array}\right.\!\!,
  \ \ U^{\II}(\eta;\bm{\lambda})=\left\{
  \begin{array}{ll}
  \prod_{k=3}^4(a_k^2+\eta)&:\text{W}\\[2pt]
  \prod_{k=3}^4(a_k+a_k^{-1}-2\eta)&:\text{AW}
  \end{array}\right..
  \nonumber
\end{align}
Since $\nu(x;\bm{\lambda})$ satisfies
$\nu(x;\bm{\lambda}+k\tilde{\bm{\delta}})=\nu(x;\bm{\lambda}-k\bm{\delta})$
($k\in\mathbb{Z}$), \eqref{nuMk} gives
\begin{equation}
  \frac{\nu(x;\bm{\lambda}+M\tilde{\bm{\delta}})}
  {\nu(x;\bm{\lambda}+M\bm{\delta})}
  =\prod_{l=0}^{M-1}
  \check{U}\bigl(x;\bm{\lambda}+(-M+2l)\bm{\delta}\bigr)^{-1}.
  \label{nuMtM}
\end{equation}

In the following we adopt the convention,
\begin{equation}
  \check{P}_n(x;\bm{\lambda})\eqdef 0\ \ (n\in\mathbb{Z}_{<0}),\quad
  \check{\xi}_{\text{v}}(x;\bm{\lambda})\eqdef 0
  \ \ (\text{v}\in\mathbb{Z}_{<0}).
\end{equation}

\subsection{Shape invariance}
\label{sec:idQM_si}

The original systems in \S\,\ref{sec:idQM_org} are shape invariant
\cite{os13_1,os13_2}
and they satisfy the relation
\begin{equation}
  \mathcal{A}(\bm{\lambda})\mathcal{A}(\bm{\lambda})^{\dagger}
  =\kappa\mathcal{A}(\bm{\lambda}+\bm{\delta})^{\dagger}
  \mathcal{A}(\bm{\lambda}+\bm{\delta})+\mathcal{E}_1(\bm{\lambda}),
\end{equation}
which is a sufficient condition for exact solvability.
As a consequence of the shape invariance, the action of the operators
$\mathcal{A}(\bm{\lambda})$ and $\mathcal{A}(\bm{\lambda})^{\dagger}$
on the eigenfunctions is
\begin{equation}
  \mathcal{A}(\bm{\lambda})\phi_n(x;\bm{\lambda})
  =f_n(\bm{\lambda})\phi_{n-1}(x;\bm{\lambda}+\bm{\delta}),\quad
  \mathcal{A}(\bm{\lambda})^{\dagger}
  \phi_{n-1}(x;\bm{\lambda}+\bm{\delta})
  =b_{n-1}(\bm{\lambda})\phi_n(x;\bm{\lambda}),
  \label{idQM:Aphi=}
\end{equation}
where the factors of the energy eigenvalue, $f_n(\bm{\lambda})$ and
$b_{n-1}(\bm{\lambda})$,
$\mathcal{E}_n(\bm{\lambda})=f_n(\bm{\lambda})b_{n-1}(\bm{\lambda})$,
are given by
\begin{equation}
  f_n(\bm{\lambda})\eqdef\left\{
  \begin{array}{ll}
  -n(n+b_1-1)&:\text{W}\\
  q^{\frac{n}{2}}(q^{-n}-1)(1-b_4q^{n-1})&:\text{AW}
  \end{array}\right.,
  \quad
  b_{n-1}(\bm{\lambda})\eqdef\left\{
  \begin{array}{ll}
  -1&:\text{W}\\
  q^{-\frac{n}{2}}&:\text{AW}
  \end{array}\right..
\end{equation}
The relations \eqref{idQM:Aphi=} are equivalent to the forward and
backward shift relations of the orthogonal polynomials $P_n$ \cite{kls},
\begin{equation}
  \mathcal{F}(\bm{\lambda})\check{P}_n(x;\bm{\lambda})
  =f_n(\bm{\lambda})\check{P}_{n-1}(x;\bm{\lambda}+\bm{\delta}),\quad
  \mathcal{B}(\bm{\lambda})\check{P}_{n-1}(x;\bm{\lambda}+\bm{\delta})
  =b_{n-1}(\bm{\lambda})\check{P}_n(x;\bm{\lambda}),
  \label{idQM:FP=}
\end{equation}
where the forward and backward shift operators $\mathcal{F}(\bm{\lambda})$
and $\mathcal{B}(\bm{\lambda})$ are defined by
\begin{align}
  &\mathcal{F}(\bm{\lambda})\eqdef
  \phi_0(x;\bm{\lambda}+\bm{\delta})^{-1}\circ
  \mathcal{A}(\bm{\lambda})\circ\phi_0(x;\bm{\lambda})
  =i\varphi(x)^{-1}(e^{\frac{\gamma}{2}p}-e^{-\frac{\gamma}{2}p}),
  \label{idQM:Fdef}\\
  &\mathcal{B}(\bm{\lambda})\eqdef
  \phi_0(x;\bm{\lambda})^{-1}\circ
  \mathcal{A}(\bm{\lambda})^{\dagger}
  \circ\phi_0(x;\bm{\lambda}+\bm{\delta})
  =-i\bigl(V(x;\bm{\lambda})e^{\frac{\gamma}{2}p}
  -V^*(x;\bm{\lambda})e^{-\frac{\gamma}{2}p}\bigr)\varphi(x).
  \label{idQM:Bdef}
\end{align}
Note that the forward shift operator $\mathcal{F}(\bm{\lambda})$ is parameter
independent.
The second order difference operator $\widetilde{\mathcal{H}}(\bm{\lambda})$
acting on the polynomial eigenfunctions is square root free. It is defined by
\begin{align}
  &\widetilde{\mathcal{H}}(\bm{\lambda})\eqdef
  \phi_0(x;\bm{\lambda})^{-1}\circ\mathcal{H}(\bm{\lambda})
  \circ\phi_0(x;\bm{\lambda})
  =\mathcal{B}(\bm{\lambda})\mathcal{F}(\bm{\lambda})\n
  &\phantom{\widetilde{\mathcal{H}}_{\ell}(\bm{\lambda})}
  =V(x;\bm{\lambda})(e^{\gamma p}-1)
  +V^*(x;\bm{\lambda})(e^{-\gamma p}-1),\\
  &\widetilde{\mathcal{H}}(\bm{\lambda})\check{P}_n(x;\bm{\lambda})
  =\mathcal{E}_n(\bm{\lambda})\check{P}_n(x;\bm{\lambda}).
  \label{idQM:HtP=EP}
\end{align}

The action of the operators
$\mathcal{A}(\bm{\lambda})$ and $\mathcal{A}(\bm{\lambda})^{\dagger}$
on the virtual state wavefunctions $\tilde{\phi}_{\text{v}}(x;\bm{\lambda})$
($\tilde{\phi}^{\I}_{\text{v}}$ and $\tilde{\phi}^{\II}_{\text{v}}$) is
\begin{equation}
  \mathcal{A}(\bm{\lambda})\tilde{\phi}_{\text{v}}(x;\bm{\lambda})
  =\tilde{f}_{\text{v}}(\bm{\lambda})
  \tilde{\phi}_{\text{v}}(x;\bm{\lambda}+\bm{\delta}),
  \ \ \mathcal{A}(\bm{\lambda})^{\dagger}
  \tilde{\phi}_{\text{v}}(x;\bm{\lambda}+\bm{\delta})
  =\tilde{b}_{\text{v}}(\bm{\lambda})
  \tilde{\phi}_{\text{v}}(x;\bm{\lambda}),
  \label{idQM:Atphi=}
\end{equation}
where the factors of the virtual energy eigenvalue,
$\tilde{f}_{\text{v}}(\bm{\lambda})$ and $\tilde{b}_{\text{v}}(\bm{\lambda})$,
$\tilde{\mathcal{E}}_{\text{v}}(\bm{\lambda})
=\tilde{f}_{\text{v}}(\bm{\lambda})\tilde{b}_{\text{v}}(\bm{\lambda})$,
are given by
\begin{align}
  \tilde{f}^{\I}_{\text{v}}(\bm{\lambda})&\eqdef
  \alpha^{\I}(\bm{\lambda})^{-\frac12}\times\left\{
  \begin{array}{ll}
  a_1+a_2-\text{v}-1&:\text{W}\\
  -q^{\frac{\text{v}}{2}}(1-a_1a_2q^{-\text{v}-1})&:\text{AW}
  \end{array}\right.,\n
  \tilde{b}^{\I}_{\text{v}}(\bm{\lambda})&\eqdef
  \alpha^{\I}(\bm{\lambda})^{\frac12}\times\left\{
  \begin{array}{ll}
  -(a_3+a_4+\text{v})&:\text{W}\\
  q^{-\frac{\text{v}}{2}}(1-a_3a_4q^{\text{v}})&:\text{AW}
  \end{array}\right.,\n
  \tilde{f}^{\II}_{\text{v}}(\bm{\lambda})&\eqdef
  \alpha^{\II}(\bm{\lambda})^{-\frac12}\times\left\{
  \begin{array}{ll}
  a_3+a_4-\text{v}-1&:\text{W}\\
  -q^{\frac{\text{v}}{2}}(1-a_3a_4q^{-\text{v}-1})&:\text{AW}
  \end{array}\right.,\\
  \tilde{b}^{\II}_{\text{v}}(\bm{\lambda})&\eqdef
  \alpha^{\II}(\bm{\lambda})^{\frac12}\times\left\{
  \begin{array}{ll}
  -(a_1+a_2+\text{v})&:\text{W}\\
  q^{-\frac{\text{v}}{2}}(1-a_1a_2q^{\text{v}})&:\text{AW}
  \end{array}\right..
  \nonumber
\end{align}
The relations \eqref{idQM:Atphi=} are equivalent to the forward and
backward shift relations of
$(\nu^{-1}\check{\xi}_{\text{v}})(x;\bm{\lambda})$
$\eqdef\nu(x;\bm{\lambda})^{-1}\check{\xi}_{\text{v}}(x;\bm{\lambda})$
($\nu^{\I\,-1}\check{\xi}^{\I}_{\text{v}}$ and
$\nu^{\II\,-1}\check{\xi}^{\II}_{\text{v}}$),
\begin{equation}
  \mathcal{F}(\bm{\lambda})(\nu^{-1}\check{\xi}_{\text{v}})(x;\bm{\lambda})
  =\tilde{f}_{\text{v}}(\bm{\lambda})
  (\nu^{-1}\check{\xi}_{\text{v}})(x;\bm{\lambda}+\bm{\delta}),
  \ \ \mathcal{B}(\bm{\lambda})
  (\nu^{-1}\check{\xi}_{\text{v}})(x;\bm{\lambda}+\bm{\delta})
  =\tilde{b}_{\text{v}}(\bm{\lambda})
  (\nu^{-1}\check{\xi}_{\text{v}})(x;\bm{\lambda}).
  \label{idQM:Fnuxi=}
\end{equation}
Explicitly they give square root free relations: forward and backward shift
relations of the virtual polynomials $\xi_{\text{v}}$ ($\xi^{\I}_{\text{v}}$
and $\xi^{\II}_{\text{v}}$) \cite{os27},
\begin{align}
  i\varphi(x)^{-1}\bigl(
  v_1^*(x;\bm{\lambda}+\tilde{\bm{\delta}})e^{\frac{\gamma}{2}p}
  -v_1(x;\bm{\lambda}+\tilde{\bm{\delta}})e^{-\frac{\gamma}{2}p}\bigr)
  \check{\xi}_{\text{v}}(x;\bm{\lambda})
  &=\alpha(\bm{\lambda})^{\frac12}\tilde{f}_{\text{v}}(\bm{\lambda})
  \check{\xi}_{\text{v}}(x;\bm{\lambda}+\bm{\delta}),\n
  i\varphi(x)^{-1}\bigl(
  v_2(x;\bm{\lambda})e^{\frac{\gamma}{2}p}
  -v_2^*(x;\bm{\lambda})e^{-\frac{\gamma}{2}p}\bigr)
  \check{\xi}_{\text{v}}(x;\bm{\lambda}+\bm{\delta})
  &=\alpha(\bm{\lambda})^{-\frac12}\tilde{b}_{\text{v}}(\bm{\lambda})
  \check{\xi}_{\text{v}}(x;\bm{\lambda}),
  \label{idQM:Fxi}
\end{align}
where the functions $v_1(x;\bm{\lambda})$ and $v_2(x;\bm{\lambda})$ are
\begin{equation}
  v_1(x;\bm{\lambda})\eqdef\left\{
  \begin{array}{ll}
  \!\prod_{j=1}^2(a_j+ix)&\!\!:\text{W}\\[2pt]
  \!e^{-ix}\prod_{j=1}^2(1-a_je^{ix})&\!\!:\text{AW}
  \end{array}\right.\!\!\!,
  \ \ v_2(x;\bm{\lambda})\eqdef\left\{
  \begin{array}{ll}
  \!\prod_{j=3}^4(a_j+ix)&\!\!:\text{W}\\[2pt]
  \!e^{-ix}\prod_{j=3}^4(1-a_je^{ix})&\!\!:\text{AW}
  \end{array}\right.\!\!\!.\!\!\!
  \label{v1v2}
\end{equation}

\subsection{Multi-indexed orthogonal polynomials}
\label{sec:idQM_miop}

The original systems in \S\,\ref{sec:idQM_org} are iso-spectrally deformed
by applying multiple Darboux transformations with virtual state wavefunctions
$\{\tilde{\phi}_{\text{v}}(x)\}$ as seed solutions.
The virtual states wavefunctions are labeled by ($\text{v}$, $\text{t}$):
$\text{v}$ is the degree of the polynomial part $\xi_{\text{v}}$ and
$\text{t}$ stands for type $\I$ or $\II$.
We take $M$ virtual state wavefunctions specified by the multi-index set
$\mathcal{D}=\{(d_1,\text{t}_1),\ldots,(d_M,\text{t}_M)\}$
(or shortly $\mathcal{D}=\{d_1,\ldots,d_M\}$) (ordered set),
$M=M_{\I}+M_{\II}$, $M_{\I}\eqdef\#\{d_j|(d_j,\I)\in\mathcal{D}\}$,
$M_{\II}\eqdef\#\{d_j|(d_j,\II)\in\mathcal{D}\}$.
The deformed Hamiltonian is denoted as
$\mathcal{H}_{\mathcal{D}}(\bm{\lambda})$.

The multi-indexed orthogonal polynomials
$\{\check{P}_{\mathcal{D},n}(x;\bm{\lambda})\}$ are the main parts
of the eigenfunctions $\{\phi_{\mathcal{D}\,n}(x;\bm{\lambda})\}$ of the
deformed system $\mathcal{H}_{\mathcal{D}}(\bm{\lambda})
\phi_{\mathcal{D}\,n}(x;\bm{\lambda})=\mathcal{E}_n(\bm{\lambda})
\phi_{\mathcal{D}\,n}(x;\bm{\lambda})$ \cite{os27}:
\begin{align}
  \phi_{\mathcal{D}\,n}(x;\bm{\lambda})&=\left(
  \frac{\sqrt{\prod_{j=0}^{M-1}
  V\bigl(x+i(\frac{M}{2}-j)\gamma;\bm{\lambda}\bigr)
  V^*\bigl(x-i(\frac{M}{2}-j)\gamma;\bm{\lambda}\bigr)}}
  {\text{W}_{\gamma}[\tilde{\phi}_{d_1},\ldots,\tilde{\phi}_{d_M}]
  (x-i\frac{\gamma}{2};\bm{\lambda})
  \text{W}_{\gamma}[\tilde{\phi}_{d_1},\ldots,\tilde{\phi}_{d_M}]
  (x+i\frac{\gamma}{2};\bm{\lambda})}\right)^{\frac12}\n
  &\quad\times
  \text{W}_{\gamma}[\tilde{\phi}_{d_1},\ldots,\tilde{\phi}_{d_M},\phi_n]
  (x;\bm{\lambda})\n
  &=\alpha^{\I}(\bm{\lambda}^{[M_{\I},M_{\II}]})^{\frac12M_{\I}}
  \alpha^{\II}(\bm{\lambda}^{[M_{\I},M_{\II}]})^{\frac12M_{\II}}
  \kappa^{-\frac14M_{\I}(M_{\I}+1)-\frac14M_{\II}(M_{\II}+1)
  +\frac52M_{\I}M_{\II}}\n
  &\quad\times
  \psi_{\mathcal{D}}(x;\bm{\lambda})
  \check{P}_{\mathcal{D},n}(x;\bm{\lambda}),
  \label{idQM:phiDn}\\
  \psi_{\mathcal{D}}(x;\bm{\lambda})&\eqdef
  \frac{\phi_0(x;\bm{\lambda}^{[M_{\I},M_{\II}]})}
  {\sqrt{\check{\Xi}_{\mathcal{D}}(x-i\frac{\gamma}{2};\bm{\lambda})
  \check{\Xi}_{\mathcal{D}}(x+i\frac{\gamma}{2};\bm{\lambda})}},\quad
  \bm{\lambda}^{[M_{\I},M_{\II}]}\eqdef
  \bm{\lambda}+M_{\I}\tilde{\bm{\delta}}^{\I}+M_{\II}\tilde{\bm{\delta}}^{\II},
  \label{idQM:psiD}\\
  \check{P}_{\mathcal{D},n}(x;\bm{\lambda})&\eqdef
  P_{\mathcal{D},n}\bigl(\eta(x);\bm{\lambda}\bigr),\quad
  \check{\Xi}_{\mathcal{D}}(x;\bm{\lambda})\eqdef
  \Xi_{\mathcal{D}}\bigl(\eta(x);\bm{\lambda}\bigr).
  \label{idQM:PDn}
\end{align}
Here $\text{W}_{\gamma}[f_1,\ldots,f_n](x)$ is the Casoratian
\eqref{idQM:Wdef}, and $\check{\Xi}_{\mathcal{D}}(x;\bm{\lambda})$ and
$\check{P}_{\mathcal{D},n}(x;\bm{\lambda})$ are the main polynomial part of
the Casoratians
$\text{W}_{\gamma}[\tilde{\phi}_{d_1},\ldots,\tilde{\phi}_{d_M}]
(x;\bm{\lambda})$ and
$\text{W}_{\gamma}[\tilde{\phi}_{d_1},\ldots,\tilde{\phi}_{d_M},\phi_n]
(x;\bm{\lambda})$, respectively.
The denominator polynomial $\Xi_{\mathcal{D}}(\eta;\bm{\lambda})$ and the
multi-indexed orthogonal polynomial $P_{\mathcal{D},n}(\eta;\bm{\lambda})$
are polynomials in $\eta$ and their degrees are generically
$\ell_{\mathcal{D}}$ and $\ell_{\mathcal{D}}+n$, respectively.
Here $\ell_{\mathcal{D}}$ is
$\ell_{\mathcal{D}}\eqdef\sum_{j=1}^{M}d_j-\tfrac12M(M-1)+2M_{\I}M_{\II}$.

In \cite{os27}, the denominator polynomials and the multi-indexed orthogonal
polynomials for the Wilson and Askey-Wilson types are expressed as
\begin{align}
  &\quad\check{\Xi}_{\mathcal{D}}(x;\bm{\lambda})\n
  &\eqdef i^{\frac12M(M-1)}\left|
  \begin{array}{lll}
  \vec{X}^{(M)}_{d_1}&\cdots&\vec{X}^{(M)}_{d_M}
  \end{array}\right|
  \times\varphi_M(x)^{-1}\kappa^{-\frac12M_{\I}M_{\II}(M-2)}
  \label{idQM:cXiDdef}\\
  &\quad\times\biggl(
  \prod_{l=0}^{M_{\I}-1}\check{U}^{\II}\bigl(x;\bm{\lambda}
  +(1-M+2l)\bm{\delta}\bigr)^{M_{\I}-1-l}\cdot
  \prod_{l=0}^{M_{\II}-1}\check{U}^{\I}\bigl(x;\bm{\lambda}
  +(1-M+2l)\bm{\delta}\bigr)^{M_{\II}-1-l}\biggr)^{-1},\n
  &\quad\check{P}_{\mathcal{D},n}(x;\bm{\lambda})\n
  &\eqdef i^{\frac12M(M+1)}\left|
  \begin{array}{llll}
  \vec{X}^{(M+1)}_{d_1}&\cdots&\vec{X}^{(M+1)}_{d_M}&\vec{Z}^{(M+1)}_n
  \end{array}\right|
  \times\varphi_{M+1}(x)^{-1}\kappa^{-\frac12M_{\I}M_{\II}(M+2)}
  \label{idQM:cPDndef}\\
  &\quad\times\biggl(
  \prod_{l=0}^{M_{\I}-1}\check{U}^{\II}\bigl(x;\bm{\lambda}
  +(-M+2l)\bm{\delta}\bigr)^{M_{\I}-l}\cdot
  \prod_{l=0}^{M_{\II}-1}\check{U}^{\I}\bigl(x;\bm{\lambda}
  +(-M+2l)\bm{\delta}\bigr)^{M_{\II}-l}\biggr)^{-1},
  \nonumber
\end{align}
where $\vec{X}^{(M)}_{\text{v}}$ ($\vec{X}^{(M)\,\I}_{\text{v}}$ or
$\vec{X}^{(M)\,\II}_{\text{v}}$) and $\vec{Z}^{(M)}_n$ are
\begin{align}
  &\bigl(\vec{X}^{(M)\,\text{t}}_{\text{v}}\bigr)_j
  =r_j^{\bar{\text{t}}}(x^{(M)}_j;\bm{\lambda},M)
  \check{\xi}^{\text{t}}_{\text{v}}(x^{(M)}_j;\bm{\lambda}),\quad
  \bar{\text{t}}\eqdef\left\{
  \begin{array}{ll}
  \II&:\text{t}=\I\\
  \I&:\text{t}=\II
  \end{array}\right.,\n
  &\bigl(\vec{Z}^{(M)}_n\bigr)_j
  =r^{\II}_j(x^{(M)}_j;\bm{\lambda},M)r^{\I}_j(x^{(M)}_j;\bm{\lambda},M)
  \check{P}_n(x^{(M)}_j;\bm{\lambda})
  \ \ (1\leq j\leq M).
\end{align}
By using the properties of the determinant, \eqref{dWformula1},
\eqref{idQM:rj}, \eqref{nuMk} and \eqref{nuMtM}, we can rewrite these as
follows:
\begin{align}
  &\quad\check{\Xi}_{\mathcal{D}}(x;\bm{\lambda})\n
  &=\text{W}_{\gamma}[\nu^{-1}\check{\xi}_{d_1},\ldots,
  \nu^{-1}\check{\xi}_{d_M}](x;\bm{\lambda})
  \times\varphi_M(x)^{-1}\kappa^{-\frac12M_{\I}M_{\II}(M-2)}\n
  &\quad\times
  \nu^{\I}\bigl(x;\bm{\lambda}+(M-1)\bm{\delta}\bigr)^{M_{\I}}
  \nu^{\II}\bigl(x;\bm{\lambda}+(M-1)\bm{\delta}\bigr)^{M_{\II}}
  \label{idQM:cXiD2}\\
  &\quad\times\biggl(
  \prod_{m=1}^{M_{\I}}\check{U}^{\I}\bigl(x;\bm{\lambda}
  +(M-1-2m)\bm{\delta}\bigr)^{M_{\I}-m}\cdot
  \prod_{m=1}^{M_{\II}}\check{U}^{\II}\bigl(x;\bm{\lambda}
  +(M-1-2m)\bm{\delta}\bigr)^{M_{\II}-m}\biggr)^{-1},\n
  &\quad\check{P}_{\mathcal{D},n}(x;\bm{\lambda})\n
  &=\text{W}_{\gamma}[\nu^{-1}\check{\xi}_{d_1},\ldots,
  \nu^{-1}\check{\xi}_{d_M},\check{P}_n](x;\bm{\lambda})
  \times\varphi_{M+1}(x)^{-1}\kappa^{-\frac12M_{\I}M_{\II}(M+2)}\n
  &\quad\times
  \nu^{\I}(x;\bm{\lambda}+M\bm{\delta})^{M_{\I}}
  \nu^{\II}(x;\bm{\lambda}+M\bm{\delta})^{M_{\II}}
  \label{idQM:cPDn2}\\
  &\quad\times\biggl(
  \prod_{m=1}^{M_{\I}}\check{U}^{\I}\bigl(x;\bm{\lambda}
  +(M-2m)\bm{\delta}\bigr)^{M_{\I}-m}\cdot
  \prod_{m=1}^{M_{\II}}\check{U}^{\II}\bigl(x;\bm{\lambda}
  +(M-2m)\bm{\delta}\bigr)^{M_{\II}-m}\biggr)^{-1}.
  \nonumber
\end{align}
For the special situations when the index set $\mathcal{D}$ consists type
$\I$ indices only (or type $\II$ indices only),
the above Casoratians can be simplified as
\begin{align*}
  \text{W}_{\gamma}[\nu^{-1}\check{\xi}_{d_1},\ldots,
  \nu^{-1}\check{\xi}_{d_M}](x;\bm{\lambda})
  &=\prod_{j=1}^M\nu(x^{(M)}_j;\bm{\lambda})^{-1}\cdot
  \text{W}_{\gamma}[\check{\xi}_{d_1},\ldots,\check{\xi}_{d_M}]
  (x;\bm{\lambda}),\\
  \text{W}_{\gamma}[\nu^{-1}\check{\xi}_{d_1},\ldots,
  \nu^{-1}\check{\xi}_{d_M},\check{P}_n](x;\bm{\lambda})
  &=\prod_{j=1}^{M+1}\nu(x^{(M+1)}_j;\bm{\lambda})^{-1}\cdot
  \text{W}_{\gamma}[\check{\xi}_{d_1},\ldots,\check{\xi}_{d_M},
  \nu\check{P}_n](x;\bm{\lambda}),
\end{align*}
by \eqref{dWformula1}.
However, when $\mathcal{D}$ contains both $\I$ and $\II$ types,
such rewriting is impossible due to
$\nu^{\I}(x;\bm{\lambda})\neq\nu^{\II}(x;\bm{\lambda})$.

In the next subsection we will rewrite \eqref{idQM:cXiD2}--\eqref{idQM:cPDn2}
by using the identities obtained from the shape invariance and the
Casoratian.

\subsection{New determinant expressions}
\label{sec:idQM_exp}

\subsubsection{Casoratian}
\label{sec:idQM_Caso}

The Casorati determinant of a set of $n$ functions $\{f_j(x)\}$ is defined by
\begin{equation}
  \text{W}_{\gamma}[f_1,f_2,\ldots,f_n](x)
  \eqdef i^{\frac12n(n-1)}
  \det\Bigl(f_k\bigl(x^{(n)}_j\bigr)\Bigr)_{1\leq j,k\leq n},\quad
  x_j^{(n)}\eqdef x+i(\tfrac{n+1}{2}-j)\gamma,
  \label{idQM:Wdef}
\end{equation}
(for $n=0$, we set $\text{W}_{\gamma}[\cdot](x)=1$),
which satisfies
\begin{equation}
  \text{W}_{\gamma}[gf_1,gf_2,\ldots,gf_n]
  =\prod_{j=1}^ng\bigl(x^{(n)}_j\bigr)\cdot
  \text{W}_{\gamma}[f_1,f_2,\ldots,f_n](x).
  \label{dWformula1}
\end{equation}

The following determinant formula holds for any analytic functions $q_j(x)$,
$r_j(x)$, $q'_j(x)$ and $r'_j(x)$ ($j=1,2,\ldots$):
\begin{align}
  &\quad i^{\frac12n(n-1)}\det(a_{j,k})_{1\leq j,k\leq n}\n
  &=\prod_{m=1}^{[\frac{n}{2}]}D_{n+1-2m}(x)\cdot
  \prod_{m=0}^{n-2}\prod_{l=0}^{[\frac{m-1}{2}]}
  r_{n-1-m}\bigl(x+i(\tfrac{m}{2}-l)\gamma\bigr)
  q_{n-1-m}\bigl(x-i(\tfrac{m}{2}-l)\gamma\bigr)\n
  &\quad\times\text{W}_{\gamma}[f_1,\ldots,f_n](x),
  \label{idQM:Wgid}\\
  &\left\{
  \begin{array}{ll}
  {\displaystyle
  a_{j,k}=\laprod{l=1}{n+1-2j}\hat{D}_l\cdot f_k(x)}
  &(j=1,\ldots,[\tfrac{n+1}{2}])\\
  {\displaystyle
  a_{j,k}=\hat{D}'_{2j-n-1}\laprod{l=1}{2j-n-2}\hat{D}_l\cdot f_k(x)}
  &(j=[\tfrac{n+3}{2}],\ldots,n)
  \end{array}\right.,\ \ (k=1,\ldots,n),
  \label{idQM:Wgid,ajk}
\end{align}
where operators $\hat{D}_j$, $\hat{D}'_j$ and functions $D_j(x)$
($j=1,2,\ldots$) are
\begin{align}
  &\hat{D}_j=q_j(x)e^{\frac{\gamma}{2}p}+r_j(x)e^{-\frac{\gamma}{2}p},\quad
  \hat{D}'_j=q'_j(x)e^{\frac{\gamma}{2}p}+r'_j(x)e^{-\frac{\gamma}{2}p},\\
  &D_j(x)=r_j(x)q'_j(x)-r'_j(x)q_j(x).
  \label{Dj}
\end{align}
This formula can be proven by using the properties of the determinant
(row properties) and induction in $n$ (even $n$ $\to$ odd $n+1$,
odd $n$ $\to$ even $n+1$).

By taking $\mathcal{F}$ or $\mathcal{B}$ as $\hat{D}_j$ and $\hat{D}'_j$
and using shape invariance properties \eqref{idQM:FP=} and \eqref{idQM:Fnuxi=},
the Casoratians $\text{W}_{\gamma}[\nu^{-1}\check{\xi}_{d_1},\ldots,
\nu^{-1}\check{\xi}_{d_M}](x;\bm{\lambda})$
and $\text{W}_{\gamma}[\nu^{-1}\check{\xi}_{d_1},\ldots,
\nu^{-1}\check{\xi}_{d_M},\check{P}_n](x;\bm{\lambda})$
in \eqref{idQM:cXiD2}--\eqref{idQM:cPDn2}
can be rewritten in various ways.
We consider two typical cases:
\begin{align}
  \text{A}:\ \ &
  \hat{D}_j=\mathcal{F}\bigl(\bm{\lambda}+(j-1)\bm{\delta}\bigr),
  \ \ \hat{D}'_j=\mathcal{B}\bigl(\bm{\lambda}+(j-2)\bm{\delta}\bigr)
  \ \ (j=1,2,\ldots),
  \label{idQM:A}\\
  \text{B}:\ \ &
  \hat{D}_{2l-1}=\mathcal{F}(\bm{\lambda}),
  \ \hat{D}_{2l}=\mathcal{B}(\bm{\lambda}),
  \ \hat{D}'_{2l-1}=\mathcal{B}(\bm{\lambda}-\bm{\delta}),
  \ \hat{D}'_{2l}=\mathcal{F}(\bm{\lambda}+\bm{\delta})
  \ \ (l=1,2,\ldots),
  \label{idQM:B}
\end{align}
which correspond to the cases A and B in \cite{os37}.
For the function $D_j(x)$ \eqref{Dj}, we define a function
$\check{S}(x;\bm{\lambda})$ and a polynomial $S(\eta;\bm{\lambda})$:
\begin{align}
  \check{S}(x;\bm{\lambda})&\eqdef
  -i\bigl(V(x;\bm{\lambda})\varphi(x-i\tfrac{\gamma}{2})
  -V^*(x;\bm{\lambda})\varphi(x+i\tfrac{\gamma}{2})\bigr),\n
  S(\eta;\bm{\lambda})&\eqdef\left\{
  \begin{array}{ll}
  b_1\eta-b_3&:\text{W}\\
  q^{-\frac12}\bigl(2(1-b_4)\eta+b_3-b_1\bigr)&:\text{AW}
  \end{array}\right.,\quad
  \check{S}(x;\bm{\lambda})=S\bigl(\eta(x);\bm{\lambda}\bigr),
\end{align}
where $b_3$ is $b_3\eqdef a_1a_2a_3+a_1a_2a_4+a_1a_3a_4+a_2a_3a_4$.

\subsubsection{case A}
\label{sec:idQM_caseA}

Firstly we consider case A \eqref{idQM:A}.
The functions $q_j(x)$, $r_j(x)$, $q'_j(x)$, $r'_j(x)$ and $D_j(x)$ are
\begin{align}
  q_j(x)&=i\varphi(x)^{-1},\quad r_j(x)=-i\varphi(x)^{-1},\n
  q'_j(x)&=-iV\bigl(x;\bm{\lambda}+(j-2)\bm{\delta}\bigr)
  \varphi(x-i\tfrac{\gamma}{2}),\quad
  r'_j(x)=iV^*\bigl(x;\bm{\lambda}+(j-2)\bm{\delta}\bigr)
  \varphi(x+i\tfrac{\gamma}{2}),\\
  D_j(x)&=-i\varphi(x)^{-1}
  \check{S}\bigl(x;\bm{\lambda}+(j-2)\bm{\delta}\bigr).
  \nonumber
\end{align}
The shape invariance properties \eqref{idQM:FP=} and \eqref{idQM:Fnuxi=}
give ($k=0,1,\ldots$)
\begin{align}
  &\laprod{l=1}{k}\hat{D}_l\cdot\check{P}_n(x;\bm{\lambda})
  =\prod_{m=0}^{k-1}f_{n-m}(\bm{\lambda}+m\bm{\delta})\cdot
  \check{P}_{n-k}(x;\bm{\lambda}+k\bm{\delta}),\n[-2pt]
  &\laprod{l=1}{k}\hat{D}_l\cdot
  (\nu^{-1}\check{\xi}_{\text{v}})(x;\bm{\lambda})
  =\prod_{m=0}^{k-1}\tilde{f}_{\text{v}}(\bm{\lambda}+m\bm{\delta})\cdot
  (\nu^{-1}\check{\xi}_{\text{v}})(x;\bm{\lambda}+k\bm{\delta}),
  \label{idQM:siA}\\[-2pt]
  &\hat{D}'_{k+1}\laprod{l=1}{k}\hat{D}_l\cdot\check{P}_n(x;\bm{\lambda})
  =b_{n-k}\bigl(\bm{\lambda}+(k-1)\bm{\delta}\bigr)
  \prod_{m=0}^{k-1}f_{n-m}(\bm{\lambda}+m\bm{\delta})\cdot
  \check{P}_{n+1-k}\bigl(x;\bm{\lambda}+(k-1)\bm{\delta}\bigr),\n[-2pt]
  &\hat{D}'_{k+1}\laprod{l=1}{k}\hat{D}_l\cdot
  (\nu^{-1}\check{\xi}_{\text{v}})(x;\bm{\lambda})
  =\tilde{b}_{\text{v}}\bigl(\bm{\lambda}+(k-1)\bm{\delta}\bigr)
  \prod_{m=0}^{k-1}\tilde{f}_{\text{v}}(\bm{\lambda}+m\bm{\delta})\cdot
  (\nu^{-1}\check{\xi}_{\text{v}})\bigl(x;\bm{\lambda}+(k-1)\bm{\delta}\bigr).
  \nonumber
\end{align}

We rewrite \eqref{idQM:cXiD2}--\eqref{idQM:cPDn2} in the following way.
(\romannumeral1) By using \eqref{idQM:Wgid}--\eqref{idQM:Wgid,ajk},
rewrite the Casoratians in \eqref{idQM:cXiD2}--\eqref{idQM:cPDn2} as
determinants $\det(a_{j,k})_{1\leq j,k\leq M'}$ ($M'=M$ for
$\check{\Xi}_{\mathcal{D}}$ and $M'=M+1$ for $\check{P}_{\mathcal{D},n}$).
(\romannumeral2) Rewrite each matrix element $a_{j,k}$ by \eqref{idQM:siA}.
(\romannumeral3) Multiply the $k$-th column ($1\leq k\leq M$) of
$\det(a_{j,k})$ by $\nu\bigl(x;\bm{\lambda}+(M'-1)\bm{\delta}\bigr)$
($\nu^{\I}$ or $\nu^{\II}$ depending on $d_k$), and divide the determinant by
$\nu^{\I}\bigl(x;\bm{\lambda}+(M'-1)\bm{\delta}\bigr)^{M_{\I}}
\nu^{\II}\bigl(x;\bm{\lambda}+(M'-1)\bm{\delta}\bigr)^{M_{\II}}$.
(\romannumeral4) Rewrite
$\nu\bigl(x;\bm{\lambda}+(M'-1)\bm{\delta}\bigr)/
\nu(x;\bm{\lambda}+l\bm{\delta})$ in $a_{j,k}$ by \eqref{nuMk}.
Then $\check{\Xi}_{\mathcal{D}}(x;\bm{\lambda})$ and
$\check{P}_{\mathcal{D},n}(x;\bm{\lambda})$ are expressed in terms of
$\check{\xi}_{d_k}(x;\bm{\lambda}')$,
$\check{P}_{n'}(x;\bm{\lambda}')$, $\check{U}(x;\bm{\lambda}')$,
$\check{S}(x;\bm{\lambda}')$ and $\varphi(x')$.
Straightforward calculation shows that the factors $\varphi(x')$ are
canceled out.
Thus $\check{\Xi}_{\mathcal{D}}(x;\bm{\lambda})$ and
$\check{P}_{\mathcal{D},n}(x;\bm{\lambda})$ are expressed in terms of
$\eta(x)$, namely
$\Xi_{\mathcal{D}}(\eta;\bm{\lambda})$ and
$P_{\mathcal{D},n}(\eta;\bm{\lambda})$ are expressed without explicit
$x$-dependence.
Their final forms are as follows.
The denominator polynomial $\Xi_{\mathcal{D}}(\eta;\bm{\lambda})$ is
\begin{align}
  &\quad\Xi_{\mathcal{D}}(\eta;\bm{\lambda})\n
  &=(-1)^{\frac16(M-1)M(M+1)}\kappa^{-\frac12M_{\I}M_{\II}(M-2)}
  \det(a_{j,k})_{1\leq j,k\leq M}\n
  &\quad\times\biggl(
  \prod_{l=0}^{M_{\I}-1}U^{\I}\bigl(\eta;\bm{\lambda}
  +(M_{\II}-M_{\I}-1+2l)\bm{\delta}\bigr)^l\cdot
  \prod_{l=0}^{M_{\II}-1}U^{\II}\bigl(\eta;\bm{\lambda}
  +(M_{\I}-M_{\II}-1+2l)\bm{\delta}\bigr)^l\n
  &\qquad\times\prod_{m=1}^{[\frac{M}{2}]}
  S\bigl(\eta;\bm{\lambda}+(M-1-2m)\bm{\delta}\bigr)\biggr)^{-1},
  \label{idQM:A:XiD}
\end{align}
where $a_{j,k}$ are
\begin{align}
  a_{j,k}&=\prod_{m=0}^{M-2j}\tilde{f}_{d_k}(\bm{\lambda}+m\bm{\delta})\cdot
  \prod_{l=1}^{j-1}U\bigl(\eta;\bm{\lambda}+(M-1-2l)\bm{\delta}\bigr)\cdot
  \xi_{d_k}\bigl(\eta;\bm{\lambda}+(M+1-2j)\bm{\delta}\bigr)\n
  &\qquad(j=1,\ldots,[\tfrac{M+1}{2}];k=1,\ldots,M),\n
  a_{j,k}&=\tilde{b}_{d_k}\bigl(\bm{\lambda}+(2j-M-3)\bm{\delta}\bigr)
  \prod_{m=0}^{2j-M-3}\tilde{f}_{d_k}(\bm{\lambda}+m\bm{\delta})\cdot
  \prod_{l=1}^{M+1-j}U\bigl(\eta;\bm{\lambda}+(M-1-2l)\bm{\delta}\bigr)\n
  &\quad\times
  \xi_{d_k}\bigl(\eta;\bm{\lambda}+(2j-M-3)\bm{\delta}\bigr)
  \ \ (j=[\tfrac{M+3}{2}],\ldots,M;k=1,\ldots,M).
\end{align}
The multi-indexed orthogonal polynomial
$P_{\mathcal{D},n}(\eta;\bm{\lambda})$ is
\begin{align}
  &\quad P_{\mathcal{D},n}(\eta;\bm{\lambda})\n
  &=(-1)^{\frac16M(M+1)(M+2)}\kappa^{-\frac12M_{\I}M_{\II}(M+2)}
  \det(a_{j,k})_{1\leq j,k\leq M+1}\n
  &\quad\times\biggl(
  \prod_{l=0}^{M_{\I}-1}U^{\I}\bigl(\eta;\bm{\lambda}
  +(M_{\II}-M_{\I}+2l)\bm{\delta}\bigr)^l\cdot
  \prod_{l=0}^{M_{\II}-1}U^{\II}\bigl(\eta;\bm{\lambda}
  +(M_{\I}-M_{\II}+2l)\bm{\delta}\bigr)^l\n
  &\qquad\times\prod_{m=1}^{[\frac{M+1}{2}]}
  S\bigl(\eta;\bm{\lambda}+(M-2m)\bm{\delta}\bigr)\biggr)^{-1},
  \label{idQM:A:PDn}
\end{align}
where $a_{j,k}$ are
\begin{align}
  a_{j,k}&=\prod_{m=0}^{M+1-2j}\tilde{f}_{d_k}(\bm{\lambda}+m\bm{\delta})\cdot
  \prod_{l=1}^{j-1}U\bigl(\eta;\bm{\lambda}+(M-2l)\bm{\delta}\bigr)\cdot
  \xi_{d_k}\bigl(\eta;\bm{\lambda}+(M+2-2j)\bm{\delta}\bigr)\n
  &\qquad(j=1,\ldots,[\tfrac{M+2}{2}];k=1,\ldots,M),\n
  a_{j,k}&=\tilde{b}_{d_k}\bigl(\bm{\lambda}+(2j-M-4)\bm{\delta}\bigr)
  \prod_{m=0}^{2j-M-4}\tilde{f}_{d_k}(\bm{\lambda}+m\bm{\delta})\cdot
  \prod_{l=1}^{M+2-j}U\bigl(\eta;\bm{\lambda}+(M-2l)\bm{\delta}\bigr)\n
  &\quad\times
  \xi_{d_k}\bigl(\eta;\bm{\lambda}+(2j-M-4)\bm{\delta}\bigr)
  \ \ (j=[\tfrac{M+4}{2}],\ldots,M+1;k=1,\ldots,M),\n
  a_{j,M+1}&=\prod_{m=0}^{M+1-2j}f_{n-m}(\bm{\lambda}+m\bm{\delta})\cdot
  P_{n-M-2+2j}\bigl(\eta;\bm{\lambda}+(M+2-2j)\bm{\delta}\bigr)
  \ \ (j=1,\ldots,[\tfrac{M+2}{2}]),\n
  a_{j,M+1}&=b_{n-2j+M+3}\bigl(\bm{\lambda}+(2j-M-4)\bm{\delta}\bigr)
  \prod_{m=0}^{2j-M-4}f_{n-m}(\bm{\lambda}+m\bm{\delta})\\
  &\quad\times
  P_{n-2j+M+4}\bigl(\eta;\bm{\lambda}+(2j-M-4)\bm{\delta}\bigr)
  \ \ (j=[\tfrac{M+4}{2}],\ldots,M+1).
  \nonumber
\end{align}

\subsubsection{case B}
\label{sec:idQM_caseB}

Secondly we consider case B \eqref{idQM:B}.
The functions $q_j(x)$, $r_j(x)$, $q'_j(x)$, $r'_j(x)$ and $D_j(x)$ are
($l=1,2,\ldots$)
\begin{align}
  q_{2l-1}(x)&=i\varphi(x)^{-1},\quad r_{2l-1}(x)=-i\varphi(x)^{-1},\n
  q_{2l}(x)&=-iV(x;\bm{\lambda})\varphi(x-i\tfrac{\gamma}{2}),\quad
  r_{2l}(x)=iV^*(x;\bm{\lambda})\varphi(x+i\tfrac{\gamma}{2}),\n
  q'_{2l-1}(x)&=-iV(x;\bm{\lambda}-\bm{\delta})
  \varphi(x-i\tfrac{\gamma}{2}),\quad
  r'_{2l-1}=iV^*(x;\bm{\lambda}-\bm{\delta})
  \varphi(x+i\tfrac{\gamma}{2}),\\
  q'_{2l}(x)&=i\varphi(x)^{-1},\quad r'_{2l}(x)=-i\varphi(x)^{-1},\n
  D_{2l-1}(x)&=-i\varphi(x)^{-1}\check{S}(x;\bm{\lambda}-\bm{\delta}),\quad
  D_{2l}(x)=i\varphi(x)^{-1}\check{S}(x;\bm{\lambda}).
  \nonumber
\end{align}
The shape invariance properties \eqref{idQM:FP=} and \eqref{idQM:Fnuxi=}
give ($l=1,2,\ldots$)
\begin{align}
  &\laprod{m=1}{2l-2}\hat{D}_m\cdot\check{P}_n(x;\bm{\lambda})
  =\mathcal{E}_n(\bm{\lambda})^{l-1}
  \check{P}_n(x;\bm{\lambda}),\n[-2pt]
  &\laprod{m=1}{2l-1}\hat{D}_m\cdot\check{P}_n(x;\bm{\lambda})
  =f_n(\bm{\lambda})\mathcal{E}_n(\bm{\lambda})^{l-1}
  \check{P}_{n-1}(x;\bm{\lambda}+\bm{\delta}),\n[-2pt]
  &\laprod{m=1}{2l-2}\hat{D}_m\cdot
  (\nu^{-1}\check{\xi}_{\text{v}})(x;\bm{\lambda})
  =\tilde{\mathcal{E}}_{\text{v}}(\bm{\lambda})^{l-1}
  (\nu^{-1}\check{\xi}_{\text{v}})(x;\bm{\lambda}),\n[-2pt]
  &\laprod{m=1}{2l-1}\hat{D}_m\cdot
  (\nu^{-1}\check{\xi}_{\text{v}})(x;\bm{\lambda})
  =\tilde{f}_{\text{v}}(\bm{\lambda})
  \tilde{\mathcal{E}}_{\text{v}}(\bm{\lambda})^{l-1}
  (\nu^{-1}\check{\xi}_{\text{v}})(x;\bm{\lambda}+\bm{\delta}),\\[-2pt]
  &\hat{D}'_{2l-1}\laprod{m=1}{2l-2}\hat{D}_m\cdot\check{P}_n(x;\bm{\lambda})
  =b_n(\bm{\lambda}-\bm{\delta})\mathcal{E}_n(\bm{\lambda})^{l-1}
  \check{P}_{n+1}(x;\bm{\lambda}-\bm{\delta}),\n[-2pt]
  &\hat{D}'_{2l}\laprod{m=1}{2l-1}\hat{D}_m\cdot\check{P}_n(x;\bm{\lambda})
  =f_{n-1}(\bm{\lambda}+\bm{\delta})f_n(\bm{\lambda})
  \mathcal{E}_n(\bm{\lambda})^{l-1}
  \check{P}_{n-2}(x;\bm{\lambda}+2\bm{\delta}),\n[-2pt]
  &\hat{D}'_{2l-1}\laprod{m=1}{2l-2}\hat{D}_m\cdot
  (\nu^{-1}\check{\xi}_{\text{v}})(x;\bm{\lambda})
  =\tilde{b}_{\text{v}}(\bm{\lambda}-\bm{\delta})
  \tilde{\mathcal{E}}_{\text{v}}(\bm{\lambda})^{l-1}
  (\nu^{-1}\check{\xi}_{\text{v}})(x;\bm{\lambda}-\bm{\delta}),\n[-2pt]
  &\hat{D}'_{2l}\laprod{m=1}{2l-1}\hat{D}_m\cdot
  (\nu^{-1}\check{\xi}_{\text{v}})(x;\bm{\lambda})
  =\tilde{f}_{\text{v}}(\bm{\lambda}+\bm{\delta})
  \tilde{f}_{\text{v}}(\bm{\lambda})
  \tilde{\mathcal{E}}_{\text{v}}(\bm{\lambda})^{l-1}
  (\nu^{-1}\check{\xi}_{\text{v}})(x;\bm{\lambda}+2\bm{\delta}).
  \nonumber
\end{align}

Like case A, we rewrite \eqref{idQM:cXiD2}--\eqref{idQM:cPDn2} by the steps
(\romannumeral1)--(\romannumeral4).
By using \eqref{varphiprop3} and
\begin{equation}
  V(x;\bm{\lambda})V^*(x;\bm{\lambda})
  =\alpha^{\I}(\bm{\lambda})\alpha^{\II}(\bm{\lambda})\kappa^{-1}
  \frac{\check{U}^{\I}(x;\bm{\lambda})\check{U}^{\II}(x;\bm{\lambda})}
  {\varphi(x-i\frac{\gamma}{2})\varphi(x)^2\varphi(x+i\frac{\gamma}{2})}
  \label{VV*},
\end{equation}
$\check{\Xi}_{\mathcal{D}}(x;\bm{\lambda})$ and
$\check{P}_{\mathcal{D},n}(x;\bm{\lambda})$ are expressed in terms of
$\check{\xi}_{d_k}(x;\bm{\lambda}')$,
$\check{P}_{n'}(x;\bm{\lambda}')$, $\check{U}(x;\bm{\lambda}')$,
$\check{S}(x;\bm{\lambda}')$ and $\varphi(x')$.
Straightforward calculation shows that the factors $\varphi(x')$ are
canceled out.
Thus $\check{\Xi}_{\mathcal{D}}(x;\bm{\lambda})$ and
$\check{P}_{\mathcal{D},n}(x;\bm{\lambda})$ are expressed in terms of
$\eta(x)$, namely
$\Xi_{\mathcal{D}}(\eta;\bm{\lambda})$ and
$P_{\mathcal{D},n}(\eta;\bm{\lambda})$ are expressed without explicit
$x$-dependence.
Their final forms are as follows.
The denominator polynomial $\Xi_{\mathcal{D}}(\eta;\bm{\lambda})$ is
\begin{align}
  &\quad\Xi_{\mathcal{D}}(\eta;\bm{\lambda})\n
  &=(-1)^{[\frac{M}{2}]}\kappa^{-\frac12M_{\I}M_{\II}(M-2)}
  \bigl(\alpha^{\I}(\bm{\lambda})\alpha^{\II}(\bm{\lambda})\kappa^{-1}
  \bigr)^{-\frac12[\frac{M-2}{2}][\frac{M}{2}]}
  \det(a_{j,k})_{1\leq j,k\leq M}\n
  &\quad\times\biggl(
  \prod_{m=1}^{M_{\I}}U^{\I}\bigl(\eta;\bm{\lambda}
  +(M-1-2m)\bm{\delta}\bigr)^{M_{\I}-m}\cdot
  \prod_{m=1}^{M_{\II}}U^{\II}\bigl(\eta;\bm{\lambda}
  +(M-1-2m)\bm{\delta}\bigr)^{M_{\II}-m}\n
  &\qquad\times
  \prod_{m=0}^{[\frac{M-2}{2}]}U^{\I}\bigl(\eta;\bm{\lambda}
  +(M-1-2m)\bm{\delta}\bigr)^m\cdot
  \prod_{m=0}^{[\frac{M-2}{2}]}U^{\II}\bigl(\eta;\bm{\lambda}
  +(M-1-2m)\bm{\delta}\bigr)^m\n
  &\qquad\times
  S\bigl(\eta;\bm{\lambda}-\tfrac12(1+(-1)^M)\bm{\delta}
  \bigr)^{[\frac{M}{2}]}\biggr)^{-1},
  \label{idQM:B:XiD}
\end{align}
where $a_{j,k}$ are
\begin{align}
  a_{j,k}&=\tilde{\mathcal{E}}_{d_k}(\bm{\lambda})^{\frac{M+1}{2}-j}
  \prod_{l=0}^{\frac{M-3}{2}}U(\eta;\bm{\lambda}+2l\bm{\delta})\cdot
  \xi_{d_k}(\eta;\bm{\lambda})
  \ \ (j=1,\ldots,\tfrac{M+1}{2};k=1,\ldots,M),\n
  a_{j,k}&=\tilde{f}_{d_k}(\bm{\lambda}+\bm{\delta})
  \tilde{f}_{d_k}(\bm{\lambda})
  \tilde{\mathcal{E}}_{d_k}(\bm{\lambda})^{j-\frac{M+3}{2}}
  \prod_{l=1}^{\frac{M-3}{2}}U(\eta;\bm{\lambda}+2l\bm{\delta})\cdot
  \xi_{d_k}(\eta;\bm{\lambda}+2\bm{\delta})\\
  &\qquad(j=\tfrac{M+3}{2},\ldots,M;k=1,\ldots,M),
  \nonumber
\end{align}
for odd $M$ and
\begin{align}
  a_{j,k}&=\tilde{f}_{d_k}(\bm{\lambda})
  \tilde{\mathcal{E}}_{d_k}(\bm{\lambda})^{\frac{M}{2}-j}
  \prod_{l=1}^{\frac{M-2}{2}}
  U\bigl(\eta;\bm{\lambda}+(2l-1)\bm{\delta}\bigr)\cdot
  \xi_{d_k}(\eta;\bm{\lambda}+\bm{\delta}),\n
  &\qquad(j=1,\ldots,\tfrac{M}{2};k=1,\ldots,M),\\
  a_{j,k}&=\tilde{b}_{d_k}(\bm{\lambda}-\bm{\delta})
  \tilde{\mathcal{E}}_{d_k}(\bm{\lambda})^{j-\frac{M+2}{2}}
  \prod_{l=0}^{\frac{M-2}{2}}
  U\bigl(\eta;\bm{\lambda}+(2l-1)\bm{\delta}\bigr)\cdot
  \xi_{d_k}(\eta;\bm{\lambda}-\bm{\delta})
  \ \ (j=\tfrac{M+2}{2},\ldots,M),
  \nonumber
\end{align}
for even $M$.
The multi-indexed orthogonal polynomial
$P_{\mathcal{D},n}(\eta;\bm{\lambda})$ is
\begin{align}
  &\quad P_{\mathcal{D},n}(\eta;\bm{\lambda})\n
  &=(-1)^{[\frac{M+1}{2}]}\kappa^{-\frac12M_{\I}M_{\II}(M+2)}
  \bigl(\alpha^{\I}(\bm{\lambda})\alpha^{\II}(\bm{\lambda})\kappa^{-1}
  \bigr)^{-\frac12[\frac{M-1}{2}][\frac{M+1}{2}]}
  \det(a_{j,k})_{1\leq j,k\leq M+1}\n
  &\quad\times\biggl(
  \prod_{m=1}^{M_{\I}}U^{\I}\bigl(\eta;\bm{\lambda}
  +(M-2m)\bm{\delta}\bigr)^{M_{\I}-m}\cdot
  \prod_{m=1}^{M_{\II}}U^{\II}\bigl(\eta;\bm{\lambda}
  +(M-2m)\bm{\delta}\bigr)^{M_{\II}-m}\n
  &\qquad\times
  \prod_{m=0}^{[\frac{M-1}{2}]}U^{\I}\bigl(\eta;\bm{\lambda}
  +(M-2m)\bm{\delta}\bigr)^m\cdot
  \prod_{m=0}^{[\frac{M-1}{2}]}U^{\II}\bigl(\eta;\bm{\lambda}
  +(M-2m)\bm{\delta}\bigr)^m\n
  &\qquad\times
  S\bigl(\eta;\bm{\lambda}-\tfrac12(1-(-1)^M)\bm{\delta}
  \bigr)^{[\frac{M+1}{2}]}\biggr)^{-1},
  \label{idQM:B:PDn}
\end{align}
where $a_{j,k}$ are
\begin{align}
  a_{j,k}&=\tilde{\mathcal{E}}_{d_k}(\bm{\lambda})^{\frac{M+2}{2}-j}
  \prod_{l=0}^{\frac{M}{2}-1}U(\eta;\bm{\lambda}+2l\bm{\delta})\cdot
  \xi_{d_k}(\eta;\bm{\lambda})
  \ \ (j=1,\ldots,\tfrac{M+2}{2};k=1,\ldots,M),\n
  a_{j,k}&=\tilde{f}_{d_k}(\bm{\lambda}+\bm{\delta})
  \tilde{f}_{d_k}(\bm{\lambda})
  \tilde{\mathcal{E}}_{d_k}(\bm{\lambda})^{j-\frac{M+4}{2}}
  \prod_{l=1}^{\frac{M}{2}-1}U(\eta;\bm{\lambda}+2l\bm{\delta})\cdot
  \xi_{d_k}(\eta;\bm{\lambda}+2\bm{\delta})\n
  &\qquad(j=\tfrac{M+4}{2},\ldots,M+1;k=1,\ldots,M),\\
  a_{j,M+1}&=\mathcal{E}_n(\bm{\lambda})^{\frac{M+2}{2}-j}
  P_n(\eta;\bm{\lambda})
  \ \ (j=1,\ldots,\tfrac{M+2}{2}),\n
  a_{j,M+1}&=f_{n-1}(\bm{\lambda}+\bm{\delta})f_n(\bm{\lambda})
  \mathcal{E}_n(\bm{\lambda})^{j-\frac{M+4}{2}}
  P_{n-2}(\eta;\bm{\lambda}+2\bm{\delta})
  \ \ (j=\tfrac{M+4}{2},\ldots,M+1),
  \nonumber
\end{align}
for even $M$ and
\begin{align}
  a_{j,k}&=\tilde{f}_{d_k}(\bm{\lambda})
  \tilde{\mathcal{E}}_{d_k}(\bm{\lambda})^{\frac{M+1}{2}-j}
  \prod_{l=1}^{\frac{M-1}{2}}
  U\bigl(\eta;\bm{\lambda}+(2l-1)\bm{\delta}\bigr)\cdot
  \xi_{d_k}(\eta;\bm{\lambda}+\bm{\delta}),\n
  &\qquad(j=1,\ldots,\tfrac{M+1}{2};k=1,\ldots,M),\n
  a_{j,k}&=\tilde{b}_{d_k}(\bm{\lambda}-\bm{\delta})
  \tilde{\mathcal{E}}_{d_k}(\bm{\lambda})^{j-\frac{M+3}{2}}
  \prod_{l=0}^{\frac{M-1}{2}}
  U\bigl(\eta;\bm{\lambda}+(2l-1)\bm{\delta}\bigr)\cdot
  \xi_{d_k}(\eta;\bm{\lambda}-\bm{\delta})\n
  &\qquad(j=\tfrac{M+3}{2},\ldots,M+1;k=1,\ldots,M),\\
  a_{j,M+1}&=f_n(\bm{\lambda})\mathcal{E}_n(\bm{\lambda})^{\frac{M+1}{2}-j}
  P_{n-1}(\eta;\bm{\lambda}+\bm{\delta})
  \ \ (j=1,\ldots,\tfrac{M+1}{2}),\n
  a_{j,M+1}&=b_n(\bm{\lambda}-\bm{\delta})
  \mathcal{E}_n(\bm{\lambda})^{j-\frac{M+3}{2}}
  P_{n+1}(\eta;\bm{\lambda}-\bm{\delta})
  \ \ (j=\tfrac{M+3}{2},\ldots,M+1),
  \nonumber
\end{align}
for odd $M$.

\subsubsection{single type}
\label{sec:idQM_single}

As remarked after \eqref{idQM:cPDn2}, when the index set $\mathcal{D}$
consists of type $\I$ indices only (or type $\II$ indices only),
simplifications occur.
We consider such a single type index set here.
As shown in \cite{equiv_miop,t13}, the multi-indexed polynomials with both
type indices can always be recovered from those with single type indices
$\mathcal{D}'$ (type $\I$ only or type $\II$ only) with shifted parameters
$\bm{\lambda}'$, namely
\begin{equation}
  P_{\mathcal{D},n}(\eta;\bm{\lambda})
  =\frac{c_{\mathcal{D},n}^{P}(\bm{\lambda})}
  {c_{\mathcal{D}',n}^{P}(\bm{\lambda}')}
  P_{\mathcal{D}',n}(\eta;\bm{\lambda}'),
  \label{PDn=PD'n}
\end{equation}
where $c_{\mathcal{D},n}^{P}(\bm{\lambda})$ is the coefficient of the
highest degree term of $P_{\mathcal{D},n}(\eta;\bm{\lambda})$ \cite{os27},
and $\mathcal{D}'$ and $\bm{\lambda}'$ are determined by $\mathcal{D}$ and
$\bm{\lambda}$ \cite{equiv_miop}.

Let us define $\mathcal{F}'(\bm{\lambda})$,
$\mathcal{B}'(\bm{\lambda})$, $f'_n(\bm{\lambda})$,
$\mathcal{E}'_n(\bm{\lambda})$, etc. as follows:
\begin{align}
  &\mathcal{F}'(\bm{\lambda})\eqdef
  \mathcal{F}\bigl(\mathfrak{t}(\bm{\lambda})\bigr),\quad
  \mathcal{B}'(\bm{\lambda})\eqdef
  \mathcal{B}\bigl(\mathfrak{t}(\bm{\lambda})\bigr),\n
  &f'_n(\bm{\lambda})\eqdef
  f_n\bigl(\mathfrak{t}(\bm{\lambda})\bigr),\quad
  b'_n(\bm{\lambda})\eqdef
  b_n\bigl(\mathfrak{t}(\bm{\lambda})\bigr),\quad
  \tilde{f}'_{\text{v}}(\bm{\lambda})\eqdef
  \tilde{f}_{\text{v}}\bigl(\mathfrak{t}(\bm{\lambda})\bigr),\quad
  \tilde{b}'_{\text{v}}(\bm{\lambda})\eqdef
  \tilde{b}_{\text{v}}\bigl(\mathfrak{t}(\bm{\lambda})\bigr),\\
  &\mathcal{E}'_n(\bm{\lambda})\eqdef
  \mathcal{E}_n\bigl(\mathfrak{t}(\bm{\lambda})\bigr)
  =f'_n(\bm{\lambda})b'_{n-1}(\bm{\lambda}),\quad
  \tilde{\mathcal{E}}'_{\text{v}}(\bm{\lambda})\eqdef
  \tilde{\mathcal{E}}_{\text{v}}\bigl(\mathfrak{t}(\bm{\lambda})\bigr)
  =\tilde{f}'_{\text{v}}(\bm{\lambda})\tilde{b}'_{\text{v}}(\bm{\lambda}).
  \nonumber
\end{align}
For $\mathcal{F}'(\bm{\lambda})$, $\mathcal{B}'(\bm{\lambda})$, 
$f'_n(\bm{\lambda})$, $b'_n(\bm{\lambda})$ and $\mathcal{E}'_n(\bm{\lambda})$,
the type of $\mathfrak{t}$ is $\I$ or $\II$,
e.g., $\mathcal{F}^{\prime\,\I}(\bm{\lambda})
=\mathcal{F}\bigl(\mathfrak{t}^{\I}(\bm{\lambda})\bigr)$ and
$\mathcal{F}^{\prime\,\II}(\bm{\lambda})
=\mathcal{F}\bigl(\mathfrak{t}^{\II}(\bm{\lambda})\bigr)$.
For $\tilde{f}'_{\text{v}}(\bm{\lambda})$,
$\tilde{b}'_{\text{v}}(\bm{\lambda})$ and
$\tilde{\mathcal{E}}'_{\text{v}}(\bm{\lambda})$,
the type of $\mathfrak{t}$ is matched to that of them,
e.g., $\tilde{f}^{\I\,\prime}_{\text{v}}(\bm{\lambda})
=\tilde{f}^{\I}_{\text{v}}\bigl(\mathfrak{t}^{\I}(\bm{\lambda})\bigr)$ and
$\tilde{f}^{\II\,\prime}_{\text{v}}(\bm{\lambda})
=\tilde{f}^{\II}_{\text{v}}\bigl(\mathfrak{t}^{\II}(\bm{\lambda})\bigr)$.
Then the relations \eqref{idQM:FP=} and \eqref{idQM:Fnuxi=} with the
replacement $\bm{\lambda}\mapsto\mathfrak{t}(\bm{\lambda})$
(the type of $\mathfrak{t}$ is $\I$ or $\II$ for \eqref{idQM:FP=},
and matched to $\nu^{-1}\check{\xi}_{\text{v}}$ for \eqref{idQM:Fnuxi=})
and exchange $n\leftrightarrow\text{v}$ give
\begin{alignat}{2}
  \hat{\mathcal{F}}'(\bm{\lambda})\check{\xi}_{\text{v}}(x;\bm{\lambda})
  &=f'_{\text{v}}(\bm{\lambda})
  \check{\xi}_{\text{v}-1}(x;\bm{\lambda}+\tilde{\bm{\delta}}),
  &\ \ \hat{\mathcal{B}}'(\bm{\lambda})
  \check{\xi}_{\text{v}-1}(x;\bm{\lambda}+\tilde{\bm{\delta}})
  &=b'_{\text{v}-1}(\bm{\lambda})\check{\xi}_{\text{v}}(x;\bm{\lambda}),
  \label{idQM:F'xiv=}\\
  \!\!\hat{\mathcal{F}}'(\bm{\lambda})(\nu\check{P}_n)(x;\bm{\lambda})
  &=\tilde{f}'_n(\bm{\lambda})
  (\nu\check{P}_n)(x;\bm{\lambda}+\tilde{\bm{\delta}}),
  &\ \ \hat{\mathcal{B}}'(\bm{\lambda})
  (\nu\check{P}_n)(x;\bm{\lambda}+\tilde{\bm{\delta}})
  &=\tilde{b}'_n(\bm{\lambda})(\nu\check{P}_n)(x;\bm{\lambda}),\!\!\!
  \label{idQM:F'nuPn=}
\end{alignat}
where $(\nu\check{P}_n)(x;\bm{\lambda})\eqdef\nu(x;\bm{\lambda})
\check{P}_n(x;\bm{\lambda})$.

In the following we consider the index set $\mathcal{D}$ with single type
twists. The type of unspecified $\mathfrak{t}$, $\nu(x;\bm{\lambda})$,
$\check{U}(x;\bm{\lambda})$, $\mathcal{F}'(\bm{\lambda})$,
$\mathcal{B}'(\bm{\lambda})$, $f'_n(\bm{\lambda})$,
$\mathcal{E}'_n(\bm{\lambda})$, $\tilde{\bm{\delta}}$ etc.\ is same as
the type of the indices of $\mathcal{D}$.
The expressions \eqref{idQM:cXiD2}--\eqref{idQM:cPDn2} are simplified to
\begin{align}
  \check{\Xi}_{\mathcal{D}}(x;\bm{\lambda})
  &=\text{W}_{\gamma}[\check{\xi}_{d_1},\ldots,
  \check{\xi}_{d_M}](x;\bm{\lambda})
  \times\varphi_M(x)^{-1}
  \prod_{j=1}^M\frac{\nu\bigl(x;\bm{\lambda}+(M-1)\bm{\delta}\bigr)}
  {\nu(x^{(M)}_j;\bm{\lambda})}\n
  &\quad\times\biggl(
  \prod_{m=1}^M\check{U}\bigl(x;\bm{\lambda}
  +(M-1-2m)\bm{\delta}\bigr)^{M-m}\biggr)^{-1},
  \label{idQM:cXiD3}\\
  \check{P}_{\mathcal{D},n}(x;\bm{\lambda})
  &=\text{W}_{\gamma}[\check{\xi}_{d_1},\ldots,
  \check{\xi}_{d_M},\nu\check{P}_n](x;\bm{\lambda})
  \times\varphi_{M+1}(x)^{-1}
  \prod_{j=1}^{M+1}\frac{\nu(x;\bm{\lambda}+M\bm{\delta})}
  {\nu(x^{(M+1)}_j;\bm{\lambda})}\n
  &\quad\times
  \nu(x;\bm{\lambda}+M\bm{\delta})^{-1}
  \biggl(\prod_{m=1}^M\check{U}\bigl(x;\bm{\lambda}
  +(M-2m)\bm{\delta}\bigr)^{M-m}\biggr)^{-1}.
  \label{idQM:cPDn3}
\end{align}
By taking $\mathcal{F}'$ or $\mathcal{B}'$ as $\hat{D}_j$ and $\hat{D}'_j$
and using shape invariance properties
\eqref{idQM:F'xiv=}--\eqref{idQM:F'nuPn=},
the Casoratians $\text{W}_{\gamma}[\check{\xi}_{d_1},\ldots,
\check{\xi}_{d_M}](x;\bm{\lambda})$
and $\text{W}_{\gamma}[\check{\xi}_{d_1},\ldots,
\check{\xi}_{d_M},\nu\check{P}_n](x;\bm{\lambda})$
in \eqref{idQM:cXiD3}--\eqref{idQM:cPDn3}
can be rewritten in various ways.
We consider typical two cases:
\begin{align}
  \!\text{A}:\ \,&
  \hat{D}_j=\mathcal{F}'\bigl(\bm{\lambda}+(j-1)\tilde{\bm{\delta}}\bigr),
  \ \ \hat{D}'_j=\mathcal{B}'\bigl(\bm{\lambda}+(j-2)\tilde{\bm{\delta}}\bigr)
  \ \ (j=1,2,\ldots),
  \label{idQM:singleA}\\
  \!\text{B}:\ \,&
  \hat{D}_{2l-1}=\mathcal{F}'(\bm{\lambda}),
  \,\hat{D}_{2l}=\mathcal{B}'(\bm{\lambda}),
  \,\hat{D}'_{2l-1}=\mathcal{B}'(\bm{\lambda}-\tilde{\bm{\delta}}),
  \,\hat{D}'_{2l}=\mathcal{F}'(\bm{\lambda}+\tilde{\bm{\delta}})
  \ \ (l=1,2,\ldots),\!
  \label{idQM:singleB}
\end{align}
which correspond to the cases A and B in \cite{os37}.

\medskip\noindent
{\bf case A:}\\
Firstly we consider case A \eqref{idQM:singleA}.
The functions $q_j(x)$, $r_j(x)$, $q'_j(x)$, $r'_j(x)$ and $D_j(x)$ are
\begin{align}
  q_j(x)&=i\varphi(x)^{-1},\quad r_j(x)=-i\varphi(x)^{-1},\n
  q'_j(x)&=-iV\bigl(x;\mathfrak{t}(\bm{\lambda})+(j-2)\bm{\delta}\bigr)
  \varphi(x-i\tfrac{\gamma}{2}),\quad
  r'_j(x)=iV^*\bigl(x;\mathfrak{t}(\bm{\lambda})+(j-2)\bm{\delta}\bigr)
  \varphi(x+i\tfrac{\gamma}{2}),\n
  D_j(x)&=-i\varphi(x)^{-1}
  \check{S}\bigl(x;\mathfrak{t}(\bm{\lambda})+(j-2)\bm{\delta}\bigr).
\end{align}
The shape invariance properties \eqref{idQM:F'xiv=}--\eqref{idQM:F'nuPn=}
give ($k=0,1,\ldots$)
\begin{align}
  &\laprod{l=1}{k}\hat{D}_l\cdot\check{\xi}_{\text{v}}(x;\bm{\lambda})
  =\prod_{m=0}^{k-1}f'_{\text{v}-m}(\bm{\lambda}+m\tilde{\bm{\delta}})\cdot
  \check{\xi}_{\text{v}-k}(x;\bm{\lambda}+k\tilde{\bm{\delta}}),\n[-2pt]
  &\laprod{l=1}{k}\hat{D}_l\cdot
  (\nu\check{P}_n)(x;\bm{\lambda})
  =\prod_{m=0}^{k-1}\tilde{f}'_n(\bm{\lambda}+m\tilde{\bm{\delta}})\cdot
  (\nu\check{P}_n)(x;\bm{\lambda}+k\tilde{\bm{\delta}}),
  \label{idQM:singlesiA}\\[-2pt]
  &\hat{D}'_{k+1}\laprod{l=1}{k}\hat{D}_l\cdot
  \check{\xi}_{\text{v}}(x;\bm{\lambda})
  =b'_{\text{v}-k}\bigl(\bm{\lambda}+(k-1)\tilde{\bm{\delta}}\bigr)
  \prod_{m=0}^{k-1}f'_{\text{v}-m}(\bm{\lambda}+m\tilde{\bm{\delta}})\cdot
  \check{\xi}_{\text{v}+1-k}\bigl(x;\bm{\lambda}
  +(k-1)\tilde{\bm{\delta}}\bigr),\n[-2pt]
  &\hat{D}'_{k+1}\laprod{l=1}{k}\hat{D}_l\cdot
  (\nu\check{P}_n)(x;\bm{\lambda})
  =\tilde{b}'_n\bigl(\bm{\lambda}+(k-1)\tilde{\bm{\delta}}\bigr)
  \prod_{m=0}^{k-1}\tilde{f}'_n(\bm{\lambda}+m\tilde{\bm{\delta}})\cdot
  (\nu\check{P}_n)\bigl(x;\bm{\lambda}+(k-1)\tilde{\bm{\delta}}\bigr).
  \nonumber
\end{align}

We rewrite \eqref{idQM:cXiD3}--\eqref{idQM:cPDn3} in the following way.
(\romannumeral1) By using \eqref{idQM:Wgid}--\eqref{idQM:Wgid,ajk},
rewrite the Casoratians in \eqref{idQM:cXiD3}--\eqref{idQM:cPDn3} as
determinants $\det(a_{j,k})_{1\leq j,k\leq M'}$ ($M'=M$ for
$\check{\Xi}_{\mathcal{D}}$ and $M'=M+1$ for $\check{P}_{\mathcal{D},n}$).
(\romannumeral2) Rewrite each matrix element $a_{j,k}$ by
\eqref{idQM:singlesiA}.
For $\check{P}_{\mathcal{D},n}$, we do the following
(\romannumeral3)--(\romannumeral5).
(\romannumeral3) Divide the $(M+1)$-th column of $\det(a_{j,k})$ by
$\nu(x;\bm{\lambda}+M\bm{\delta})$, and multiply the determinant
by $\nu(x;\bm{\lambda}+M\bm{\delta})$.
(\romannumeral4) Rewrite
$\nu(x;\bm{\lambda}')/\nu(x;\bm{\lambda}+M\bm{\delta})$
in the $(M+1)$-th column of $\det(a_{j,k})$ by \eqref{nuMk}.
(\romannumeral5) Multiply the $(M+1)$-th column of $\det(a_{j,k})$ by
$\prod_{m=1}^M\check{U}\bigl(x;\bm{\lambda}+(M-2m)\bm{\delta}\bigl)$,
and divide the determinant by
$\prod_{m=1}^M\check{U}\bigl(x;\bm{\lambda}+(M-2m)\bm{\delta}\bigl)$.
Then $\check{\Xi}_{\mathcal{D}}(x;\bm{\lambda})$ and
$\check{P}_{\mathcal{D},n}(x;\bm{\lambda})$ are expressed in terms of
$\check{\xi}_{\text{v}}(x;\bm{\lambda}')$,
$\check{P}_n(x;\bm{\lambda}')$, $\check{U}(x;\bm{\lambda}')$,
$\check{S}(x;\bm{\lambda}')$ and $\varphi(x')$.
Straightforward calculation shows that the factors $\varphi(x')$ are
canceled out.
Thus $\check{\Xi}_{\mathcal{D}}(x;\bm{\lambda})$ and
$\check{P}_{\mathcal{D},n}(x;\bm{\lambda})$ are expressed in terms of
$\eta(x)$, namely
$\Xi_{\mathcal{D}}(\eta;\bm{\lambda})$ and
$P_{\mathcal{D},n}(\eta;\bm{\lambda})$ are expressed without explicit
$x$-dependence.
Their final forms are as follows.
The denominator polynomial $\Xi_{\mathcal{D}}(\eta;\bm{\lambda})$ is
\begin{equation}
  \Xi_{\mathcal{D}}(\eta;\bm{\lambda})
  =(-1)^{\frac16(M-1)M(M+1)}
  \biggl(\prod_{m=1}^{[\frac{M}{2}]}
  S\bigl(\eta;\mathfrak{t}(\bm{\lambda})+(M-1-2m)\bm{\delta}\bigr)\biggr)^{-1}
  \det(a_{j,k})_{1\leq j,k\leq M},
  \label{idQM:singleA:XiD}
\end{equation}
where $a_{j,k}$ are
\begin{align}
  a_{j,k}&=\prod_{m=0}^{M-2j}f'_{d_k-m}(\bm{\lambda}+m\tilde{\bm{\delta}})
  \cdot\xi_{d_k-M-1+2j}\bigl(\eta;\bm{\lambda}
  +(M+1-2j)\tilde{\bm{\delta}}\bigr)\n
  &\qquad(j=1,\ldots,[\tfrac{M+1}{2}];k=1,\ldots,M),\n
  a_{j,k}&=b'_{d_k-2j+M+2}\bigl(\bm{\lambda}+(2j-M-3)\tilde{\bm{\delta}}\bigr)
  \prod_{m=0}^{2j-M-3}f'_{d_k-m}(\bm{\lambda}+m\tilde{\bm{\delta}})\\
  &\quad\times
  \xi_{d_k-2j+M+3}\bigl(\eta;\bm{\lambda}+(2j-M-3)\tilde{\bm{\delta}}\bigr)
  \ \ (j=[\tfrac{M+3}{2}],\ldots,M;k=1,\ldots,M).
  \nonumber
\end{align}
The multi-indexed orthogonal polynomial
$P_{\mathcal{D},n}(\eta;\bm{\lambda})$ is
\begin{equation}
  P_{\mathcal{D},n}(\eta;\bm{\lambda})
  =(-1)^{\frac16M(M+1)(M+2)}
  \biggl(\prod_{m=1}^{[\frac{M+1}{2}]}
  S\bigl(\eta;\mathfrak{t}(\bm{\lambda})+(M-2m)\bm{\delta}\bigr)\biggr)^{-1}
  \det(a_{j,k})_{1\leq j,k\leq M+1},
  \label{idQM:singleA:PDn}
\end{equation}
where $a_{j,k}$ are
\begin{align}
  a_{j,k}&=\prod_{m=0}^{M+1-2j}f'_{d_k-m}(\bm{\lambda}+m\tilde{\bm{\delta}})
  \cdot
  \xi_{d_k-M-2+2j}\bigl(\eta;\bm{\lambda}+(M+2-2j)\tilde{\bm{\delta}}\bigr)\n
  &\qquad(j=1,\ldots,[\tfrac{M+2}{2}];k=1,\ldots,M),\n
  a_{j,k}&=b'_{d_k-2j+M+3}\bigl(\bm{\lambda}+(2j-M-4)\tilde{\bm{\delta}}\bigr)
  \prod_{m=0}^{2j-M-4}f'_{d_k-m}(\bm{\lambda}+m\tilde{\bm{\delta}})\n
  &\quad\times
  \xi_{d_k-2j+M+4}\bigl(\eta;\bm{\lambda}+(2j-M-4)\tilde{\bm{\delta}}\bigr)
  \ \ (j=[\tfrac{M+4}{2}],\ldots,M+1;k=1,\ldots,M),\n
  a_{j,M+1}&=\prod_{m=0}^{M+1-2j}
  \tilde{f}'_n(\bm{\lambda}+m\tilde{\bm{\delta}})\cdot
  \prod_{l=M+2-j}^MU\bigl(\eta;\bm{\lambda}+(M-2l)\bm{\delta}\bigr)\cdot
  P_n\bigl(\eta;\bm{\lambda}+(M+2-2j)\tilde{\bm{\delta}}\bigr)\n
  &\qquad(j=1,\ldots,[\tfrac{M+2}{2}]),\\
  a_{j,M+1}&=\tilde{b}'_n\bigl(\bm{\lambda}+(2j-M-4)\tilde{\bm{\delta}}\bigr)
  \prod_{m=0}^{2j-M-4}\tilde{f}'_n(\bm{\lambda}+m\tilde{\bm{\delta}})\cdot
  \prod_{l=j-1}^MU\bigl(\eta;\bm{\lambda}+(M-2l)\bm{\delta}\bigr)\n
  &\quad\times
  P_n\bigl(\eta;\bm{\lambda}+(2j-M-4)\tilde{\bm{\delta}}\bigr)
  \ \ (j=[\tfrac{M+4}{2}],\ldots,M+1).
  \nonumber
\end{align}

\medskip\noindent
{\bf case B:}\\
Secondly we consider case B \eqref{idQM:singleB}.
The functions $q_j(x)$, $r_j(x)$, $q'_j(x)$, $r'_j(x)$ and $D_j(x)$ are
($l=1,2,\ldots$)
\begin{align}
  q_{2l-1}(x)&=i\varphi(x)^{-1},\quad r_{2l-1}(x)=-i\varphi(x)^{-1},\n
  q_{2l}(x)&=-iV\bigl(x;\mathfrak{t}(\bm{\lambda})\bigr)
  \varphi(x-i\tfrac{\gamma}{2}),\quad
  r_{2l}(x)=iV^*\bigl(x;\mathfrak{t}(\bm{\lambda})\bigr)
  \varphi(x+i\tfrac{\gamma}{2}),\n
  q'_{2l-1}(x)&=-iV\bigl(x;\mathfrak{t}(\bm{\lambda})-\bm{\delta}\bigr)
  \varphi(x-i\tfrac{\gamma}{2}),\quad
  r'_{2l-1}=iV^*\bigl(x;\mathfrak{t}(\bm{\lambda})-\bm{\delta}\bigr)
  \varphi(x+i\tfrac{\gamma}{2}),\\
  q'_{2l}(x)&=i\varphi(x)^{-1},\quad r'_{2l}(x)=-i\varphi(x)^{-1},\n
  D_{2l-1}(x)&=-i\varphi(x)^{-1}
  \check{S}\bigl(x;\mathfrak{t}(\bm{\lambda})-\bm{\delta}\bigr),\quad
  D_{2l}(x)=i\varphi(x)^{-1}
  \check{S}\bigl(x;\mathfrak{t}(\bm{\lambda})\bigr).
  \nonumber
\end{align}
The shape invariance properties \eqref{idQM:F'xiv=}--\eqref{idQM:F'nuPn=}
give ($l=1,2,\ldots$)
\begin{align}
  &\laprod{m=1}{2l-2}\hat{D}_m\cdot\check{\xi}_{\text{v}}(x;\bm{\lambda})
  =\mathcal{E}'_{\text{v}}(\bm{\lambda})^{l-1}
  \check{\xi}_{\text{v}}(x;\bm{\lambda}),\n[-2pt]
  &\laprod{m=1}{2l-1}\hat{D}_m\cdot\check{\xi}_{\text{v}}(x;\bm{\lambda})
  =f'_{\text{v}}(\bm{\lambda})\mathcal{E}'_{\text{v}}(\bm{\lambda})^{l-1}
  \check{\xi}_{\text{v}-1}(x;\bm{\lambda}+\tilde{\bm{\delta}}),\n[-2pt]
  &\laprod{m=1}{2l-2}\hat{D}_m\cdot
  (\nu\check{P}_n)(x;\bm{\lambda})
  =\tilde{\mathcal{E}}'_n(\bm{\lambda})^{l-1}
  (\nu\check{P}_n)(x;\bm{\lambda}),\n[-2pt]
  &\laprod{m=1}{2l-1}\hat{D}_m\cdot
  (\nu\check{P}_n)(x;\bm{\lambda})
  =\tilde{f}'_n(\bm{\lambda})
  \tilde{\mathcal{E}}'_n(\bm{\lambda})^{l-1}
  (\nu\check{P}_n)(x;\bm{\lambda}+\tilde{\bm{\delta}}),\\[-2pt]
  &\hat{D}'_{2l-1}\laprod{m=1}{2l-2}\hat{D}_m\cdot
  \check{\xi}_{\text{v}}(x;\bm{\lambda})
  =b'_{\text{v}}(\bm{\lambda}-\tilde{\bm{\delta}})
  \mathcal{E}'_{\text{v}}(\bm{\lambda})^{l-1}
  \check{\xi}_{\text{v}+1}(x;\bm{\lambda}-\tilde{\bm{\delta}}),\n[-2pt]
  &\hat{D}'_{2l}\laprod{m=1}{2l-1}\hat{D}_m\cdot
  \check{\xi}_{\text{v}}(x;\bm{\lambda})
  =f'_{\text{v}-1}(\bm{\lambda}+\tilde{\bm{\delta}})f'_{\text{v}}(\bm{\lambda})
  \mathcal{E}'_{\text{v}}(\bm{\lambda})^{l-1}
  \check{\xi}_{\text{v}-2}(x;\bm{\lambda}+2\tilde{\bm{\delta}}),\n[-2pt]
  &\hat{D}'_{2l-1}\laprod{m=1}{2l-2}\hat{D}_m\cdot
  (\nu\check{P}_n)(x;\bm{\lambda})
  =\tilde{b}'_n(\bm{\lambda}-\tilde{\bm{\delta}})
  \tilde{\mathcal{E}}'_n(\bm{\lambda})^{l-1}
  (\nu\check{P}_n)(x;\bm{\lambda}-\tilde{\bm{\delta}}),\n[-2pt]
  &\hat{D}'_{2l}\laprod{m=1}{2l-1}\hat{D}_m\cdot
  (\nu\check{P}_n)(x;\bm{\lambda})
  =\tilde{f}'_n(\bm{\lambda}+\tilde{\bm{\delta}})
  \tilde{f}'_n(\bm{\lambda})
  \tilde{\mathcal{E}}'_n(\bm{\lambda})^{l-1}
  (\nu\check{P}_n)(x;\bm{\lambda}+2\tilde{\bm{\delta}}).
  \nonumber
\end{align}

Like case A, we rewrite \eqref{idQM:cXiD3}--\eqref{idQM:cPDn3} by the steps
(\romannumeral1)--(\romannumeral5).
In (\romannumeral5), multiply and divide by
$\prod_{l=1}^{[\frac{M+2}{2}]}
\check{U}\bigl(x;\bm{\lambda}+(M-2l)\bm{\delta}\bigr)$.
By using \eqref{varphiprop3} and \eqref{VV*},
$\check{\Xi}_{\mathcal{D}}(x;\bm{\lambda})$ and
$\check{P}_{\mathcal{D},n}(x;\bm{\lambda})$ are expressed in terms of
$\check{\xi}_{\text{v}}(x;\bm{\lambda}')$,
$\check{P}_n(x;\bm{\lambda}')$, $\check{U}(x;\bm{\lambda}')$,
$\check{S}(x;\bm{\lambda}')$ and $\varphi(x')$.
Straightforward calculation shows that the factors $\varphi(x')$ are
canceled out.
Thus $\check{\Xi}_{\mathcal{D}}(x;\bm{\lambda})$ and
$\check{P}_{\mathcal{D},n}(x;\bm{\lambda})$ are expressed in terms of
$\eta(x)$, namely
$\Xi_{\mathcal{D}}(\eta;\bm{\lambda})$ and
$P_{\mathcal{D},n}(\eta;\bm{\lambda})$ are expressed without explicit
$x$-dependence.
Their final forms are as follows.
The denominator polynomial $\Xi_{\mathcal{D}}(\eta;\bm{\lambda})$ is
\begin{align}
  \Xi_{\mathcal{D}}(\eta;\bm{\lambda})
  &=(-1)^{[\frac{M}{2}]}
  \bigl(\alpha^{\text{t}}(\bm{\lambda})
  \alpha^{\bar{\text{t}}}(\bm{\lambda})^{-1}\kappa
  \bigr)^{\frac12[\frac{M-2}{2}][\frac{M}{2}]}
  \det(a_{j,k})_{1\leq j,k\leq M}\n
  &\quad\times\biggl(
  \prod_{m=[\frac{M+3}{2}]}^MU\bigl(\eta;\bm{\lambda}
  +(M-1-2m)\bm{\delta}\bigr)^{M-m}\cdot
  \prod_{m=0}^{[\frac{M-2}{2}]}U^{\bar{\text{t}}}
  \bigl(\eta;\bm{\lambda}+(M-1-2m)\bm{\delta}\bigr)^m\n
  &\qquad\times
  S\bigl(\eta;\mathfrak{t}(\bm{\lambda})-\tfrac12(1+(-1)^M)\bm{\delta}
  \bigr)^{[\frac{M}{2}]}\biggr)^{-1},
  \label{idQM:singleB:XiD}
\end{align}
where $\text{t}$ is the type of the indices of $\mathcal{D}$ and $a_{j,k}$ are
\begin{align}
  a_{j,k}&=\mathcal{E}'_{d_k}(\bm{\lambda})^{\frac{M+1}{2}-j}
  \xi_{d_k}(\eta;\bm{\lambda})
  \ \ (j=1,\ldots,\tfrac{M+1}{2};k=1,\ldots,M),\\
  a_{j,k}&=f'_{d_k-1}(\bm{\lambda}+\tilde{\bm{\delta}})
  f'_{d_k}(\bm{\lambda})
  \mathcal{E}'_{d_k}(\bm{\lambda})^{j-\frac{M+3}{2}}
  \xi_{d_k-2}(\eta;\bm{\lambda}+2\tilde{\bm{\delta}})
  \ \ (j=\tfrac{M+3}{2},\ldots,M;k=1,\ldots,M),
  \nonumber
\end{align}
for odd $M$ and
\begin{align}
  a_{j,k}&=f'_{d_k}(\bm{\lambda})
  \mathcal{E}'_{d_k}(\bm{\lambda})^{\frac{M}{2}-j}
  \xi_{d_k-1}(\eta;\bm{\lambda}+\tilde{\bm{\delta}})
  \ \ (j=1,\ldots,\tfrac{M}{2};k=1,\ldots,M),\n
  a_{j,k}&=b'_{d_k}(\bm{\lambda}-\tilde{\bm{\delta}})
  \mathcal{E}'_{d_k}(\bm{\lambda})^{j-\frac{M+2}{2}}
  \xi_{d_k+1}(\eta;\bm{\lambda}-\tilde{\bm{\delta}})
  \ \ (j=\tfrac{M+2}{2},\ldots,M;k=1,\ldots,M),
\end{align}
for even $M$.
The multi-indexed orthogonal polynomial
$P_{\mathcal{D},n}(\eta;\bm{\lambda})$ is
\begin{align}
  P_{\mathcal{D},n}(\eta;\bm{\lambda})
  &=(-1)^{[\frac{M+1}{2}]}
  \bigl(\alpha^{\text{t}}(\bm{\lambda})
  \alpha^{\bar{\text{t}}}(\bm{\lambda})^{-1}\kappa
  \bigr)^{\frac12[\frac{M-1}{2}][\frac{M+1}{2}]}
  \det(a_{j,k})_{1\leq j,k\leq M+1}\n
  &\quad\times\biggl(
  \prod_{m=[\frac{M+4}{2}]}^MU\bigl(\eta;\bm{\lambda}
  +(M-2m)\bm{\delta}\bigr)^{M-m}\cdot
  \prod_{m=0}^{[\frac{M-1}{2}]}U^{\bar{\text{t}}}
  \bigl(\eta;\bm{\lambda}
  +(M-2m)\bm{\delta}\bigr)^m\n
  &\qquad\times
  S\bigl(\eta;\mathfrak{t}(\bm{\lambda})-\tfrac12(1-(-1)^M)\bm{\delta}
  \bigr)^{[\frac{M+1}{2}]}\biggr)^{-1},
  \label{idQM:singleB:PDn}
\end{align}
where $\text{t}$ is the type of the indices of $\mathcal{D}$ and $a_{j,k}$ are
\begin{align}
  a_{j,k}&=\mathcal{E}'_{d_k}(\bm{\lambda})^{\frac{M+2}{2}-j}
  \xi_{d_k}(\eta;\bm{\lambda})
  \ \ (j=1,\ldots,\tfrac{M+2}{2};k=1,\ldots,M),\n
  a_{j,k}&=f'_{d_k-1}(\bm{\lambda}+\tilde{\bm{\delta}})
  f'_{d_k}(\bm{\lambda})
  \mathcal{E}'_{d_k}(\bm{\lambda})^{j-\frac{M+4}{2}}
  \xi_{d_k-2}(\eta;\bm{\lambda}+2\tilde{\bm{\delta}})\n
  &\qquad(j=\tfrac{M+4}{2},\ldots,M+1;k=1,\ldots,M),\n
  a_{j,M+1}&=\tilde{\mathcal{E}}'_n(\bm{\lambda})^{\frac{M+2}{2}-j}
  U(\eta;\bm{\lambda}-2\bm{\delta})P_n(\eta;\bm{\lambda})
  \ \ (j=1,\ldots,\tfrac{M+2}{2}),\\
  a_{j,M+1}&=\tilde{f}'_n(\bm{\lambda}+\tilde{\bm{\delta}})
  \tilde{f}'_n(\bm{\lambda})
  \tilde{\mathcal{E}}'_n(\bm{\lambda})^{j-\frac{M+4}{2}}
  P_n(\eta;\bm{\lambda}+2\tilde{\bm{\delta}})
  \ \ (j=\tfrac{M+4}{2},\ldots,M+1),
  \nonumber
\end{align}
for even $M$ and
\begin{align}
  a_{j,k}&=f'_{d_k}(\bm{\lambda})
  \mathcal{E}'_{d_k}(\bm{\lambda})^{\frac{M+1}{2}-j}
  \xi_{d_k-1}(\eta;\bm{\lambda}+\tilde{\bm{\delta}})
  \ \ (j=1,\ldots,\tfrac{M+1}{2};k=1,\ldots,M),\n
  a_{j,k}&=b'_{d_k}(\bm{\lambda}-\tilde{\bm{\delta}})
  \mathcal{E}'_{d_k}(\bm{\lambda})^{j-\frac{M+3}{2}}
  \xi_{d_k+1}(\eta;\bm{\lambda}-\tilde{\bm{\delta}})
  \ \ (j=\tfrac{M+3}{2},\ldots,M+1;k=1,\ldots,M),\n
  a_{j,M+1}&=\tilde{f}'_n(\bm{\lambda})
  \tilde{\mathcal{E}}'_n(\bm{\lambda})^{\frac{M+1}{2}-j}
  P_n(\eta;\bm{\lambda}+\tilde{\bm{\delta}})
  \ \ (j=1,\ldots,\tfrac{M+1}{2}),\\
  a_{j,M+1}&=\tilde{b}'_n(\bm{\lambda}-\tilde{\bm{\delta}})
  \tilde{\mathcal{E}}'_n(\bm{\lambda})^{j-\frac{M+3}{2}}
  U(\eta;\bm{\lambda}-\bm{\delta})
  P_n(\eta;\bm{\lambda}-\tilde{\bm{\delta}})
  \ \ (j=\tfrac{M+3}{2},\ldots,M+1),
  \nonumber
\end{align}
for odd $M$.

\section{Summary and Comments}
\label{sec:summary}

The multi-indexed orthogonal polynomials (the Meixner, little $q$-Laguerre,
little $q$-Jacobi, Racah, $q$-Racah, Wilson and Askey-Wilson types)
introduced in the framework of the discrete quantum mechanics are expressed
in terms of Casoratians.
Various new determinant expressions of the multi-indexed orthogonal
polynomials are derived by using the properties of the Casoratians and
shape invariance.
Two typical cases are presented explicitly.
For the Meixner, little $q$-Laguerre, little $q$-Jacobi, Wilson and
Askey-Wilson cases, the new expressions for the multi-indexed polynomials
$P_{\mathcal{D},n}(\eta)$ do not have explicit dependence on $x$,
which is the coordinate of the quantum system.
We hope that these new expressions are helpful for deeper understanding of
the multi-indexed orthogonal polynomials.

In \cite{os37}, simplified expressions of the multi-indexed Laguerre
and Jacobi polynomials are derived.
From the viewpoint adopted in this paper, the calculation in \cite{os37}
is regarded as follows.
The eigenfunctions of the deformed systems, namely the multi-indexed orthogonal
polynomials are expressed in terms of the Wronskian.
The Wronskian of a set of $n$ functions $\{f_j(x)\}$ is defined by
\begin{equation}
  \text{W}[f_1,f_2,\ldots,f_n](x)\eqdef
  \det\Bigl(\frac{d^{j-1}f_k(x)}{dx^{j-1}}\Bigr)_{1\leq j,k\leq n},
\end{equation}
(for $n=0$, we set $\text{W}\,[\cdot](x)=1$), and the following formula
holds for any smooth functions $q_j(x)$ ($j=1,2,\ldots$):
\begin{equation}
  \det\Bigl(\laprod{l=1}{j-1}\hat{D}_l\cdot f_k(x)
  \Bigr)_{1\leq j,k\leq n}
  =\text{W}[f_1,\ldots,f_n](x),
  \ \ \hat{D}_j=\frac{d}{dx}-q_j(x).
  \label{QM:Wid}
\end{equation}
The original systems have shape invariance
$\mathcal{A}(\bm{\lambda})\mathcal{A}(\bm{\lambda})^{\dagger}
=\mathcal{A}(\bm{\lambda}+\bm{\delta})^{\dagger}
\mathcal{A}(\bm{\lambda}+\bm{\delta})
+\mathcal{E}_1(\bm{\lambda})$, where $\mathcal{A}(\bm{\lambda})
=\frac{d}{dx}-\frac{d}{dx}\log\phi_0(x;\bm{\lambda})$ and
$\mathcal{A}(\bm{\lambda})^{\dagger}=
-\frac{d}{dx}-\frac{d}{dx}\log\phi_0(x;\bm{\lambda})$.
These operators act on the eigenstates and virtual states as follows:
\begin{alignat}{2}
  \mathcal{A}(\bm{\lambda})\phi_n(x;\bm{\lambda})
  &=f_n(\bm{\lambda})\phi_{n-1}(x;\bm{\lambda}+\bm{\delta}),&\quad
  \mathcal{A}(\bm{\lambda})^{\dagger}\phi_{n-1}(x;\bm{\lambda}+\bm{\delta})
  &=b_{n-1}(\bm{\lambda})\phi_n(x;\bm{\lambda}),
  \label{QM:Aphi=}\\
  \mathcal{A}(\bm{\lambda})\tilde{\phi}_{\text{v}}(x;\bm{\lambda})
  &=\tilde{f}_{\text{v}}(\bm{\lambda})
  \tilde{\phi}_{\text{v}}(x;\bm{\lambda}+\bm{\delta}),&\quad
  \mathcal{A}(\bm{\lambda})^{\dagger}
  \tilde{\phi}_{\text{v}}(x;\bm{\lambda}+\bm{\delta})
  &=\tilde{b}_{\text{v}}(\bm{\lambda})
  \tilde{\phi}_{\text{v}}(x;\bm{\lambda}),
  \label{QM:Aphit=}
\end{alignat}
where
\begin{alignat}{2}
  f_n(\bm{\lambda})&=\left\{
  \begin{array}{ll}
  -2&:\text{L}\\
  -2(n+g+h)&:\text{J}
  \end{array}\right.,&\quad
  b_{n-1}(\bm{\lambda})&=-2n:\text{L,\,J}\\
  \tilde{f}_{\text{v}}(\bm{\lambda})&=\left\{
  \begin{array}{ll}
  2&:\text{L,\,$\I$}\\
  -2(g-\frac12-\text{v})&:\text{L,\,$\II$}\\
  2(h-\frac12-\text{v})&:\text{J,\,$\I$}\\
  -2(g-\frac12-\text{v})&:\text{J,\,$\II$}
  \end{array}\right.,&\quad
  \tilde{b}_{\text{v}}(\bm{\lambda})&=\left\{
  \begin{array}{ll}
  -2(g+\frac12+\text{v})&:\text{L,\,$\I$}\\
  2&:\text{L,\,$\II$}\\
  -2(g+\frac12+\text{v})&:\text{J,\,$\I$}\\
  2(h+\frac12+\text{v})&:\text{J,\,$\II$}
  \end{array}\right..
\end{alignat}
By taking $\mathcal{A}$ or $\mathcal{A}^{\dagger}$ as $\hat{D}_j$ in
\eqref{QM:Wid} and using shape invariant properties
\eqref{QM:Aphi=}--\eqref{QM:Aphit=}, the Wronskians can be rewritten
in various ways.
Typical two cases are studied in \cite{os37}:
\begin{align}
  \text{A}:\ \ &
  \hat{D}_j=\mathcal{A}\bigl(\bm{\lambda}+(j-1)\bm{\delta}\bigr),
  \ \ (j=1,2,\ldots),
  \label{QM:A}\\
  \text{B}:\ \ &
  \hat{D}_{2l-1}=\mathcal{A}(\bm{\lambda}),
  \ \hat{D}_{2l}=-\mathcal{A}(\bm{\lambda})^{\dagger},
  \ \ (l=1,2,\ldots).
  \label{QM:B}
\end{align}
Like \S\,\ref{sec:idQM_single}, the equivalence property \eqref{PDn=PD'n}
holds \cite{equiv_miop,t13} and simplifications occur for the index set
with a single type. This calculation is left to readers as an exercise.

\section*{Acknowledgments}

I thank R.\,Sasaki for discussion and reading of the manuscript.
I am supported in part by Grant-in-Aid for Scientific Research
from the Ministry of Education, Culture, Sports, Science and Technology
(MEXT), No.25400395.


\end{document}